\documentclass[numbook,natbib,final,runningheads]{svjour3}
\usepackage{graphicx,epsfig,amssymb,mathptmx} 
\usepackage{makeidx} 
\makeindex

\include{macros_new}

\begin{document}

\title{Properties of Energetic Ions in the Solar Atmosphere from $\gamma$-Ray and Neutron Observations} 

\author{N. Vilmer$^1$,
        A.~L. MacKinnon$^2$, and
        G.~J. Hurford$^3$
}


\institute{$^1$
              LESIA-CNRS, Observatoire de Paris, 5 Place Jules Janssen, Meudon, France \\
              \email{nicole.vilmer@obspm.fr}  \\         %  \\
           $^2$
           DACE/Physics and Astronomy, University of Glasgow, Glasgow, G12 8QQ, UK   \\
             \email{a.mackinnon@educ.gla.ac.uk}\\
           $^3$
            Space Sciences Laboratory, University of California, Berkeley, CA 94720-7450, California, USA \\
            \email{ghurford@ssl.berkeley.edu}
}

\date{Received: date / Accepted: date}
\authorrunning{Vilmer et al.}
\titlerunning{Gamma Rays and Neutrons}
\maketitle

\begin{abstract}
Gamma-rays and neutrons are the only sources of information on energetic ions present during solar flares and on 
properties of these ions when they interact in the solar atmosphere. The production of $\gamma$-rays and neutrons results from convolution of 
the nuclear cross-sections with the ion distribution functions in the atmosphere. The observed $\gamma$-ray 
and neutron fluxes
thus provide useful diagnostics for the properties of energetic ions, yielding strong constraints on acceleration mechanisms as 
well as properties of the interaction sites. The problem of ion transport between the accelerating and interaction sites must also be 
addressed to infer as much information as possible on the properties of the primary ion accelerator. 
In the last couple of decades, both theoretical and observational developments have led to substantial progress in understanding the 
origin of solar $\gamma$-rays and neutrons. 
This chapter reviews recent developments in the study of solar $\gamma$-rays and of solar neutrons at the time of the \textit{RHESSI} era.\index{RHESSI@\textit{RHESSI}}\index{eras!\textit{RHESSI}}
The unprecedented quality of the \textit{RHESSI} data reveals $\gamma$-ray line shapes for the first time and provides $\gamma$-ray images. Our previous understanding 
of the properties of energetic ions based on measurements from the former solar cycles is also summarized. The new results -- 
obtained owing both to the gain in spectral resolution (both with \textit{RHESSI} and with the non solar-dedicated \textit{INTEGRAL}/SPI instrument) and 
to the pioneering imaging technique in the $\gamma$-ray domain --
are presented in the context of this previous knowledge. Still open 
questions are emphasized in the last section of the chapter and future perspectives on this field are briefly discussed. 

\end{abstract}

\setcounter{tocdepth}{3}
\tableofcontents

\section{Introduction}

\begin{figure*}
\centering
\includegraphics[width=0.70\textwidth,angle=0.5]{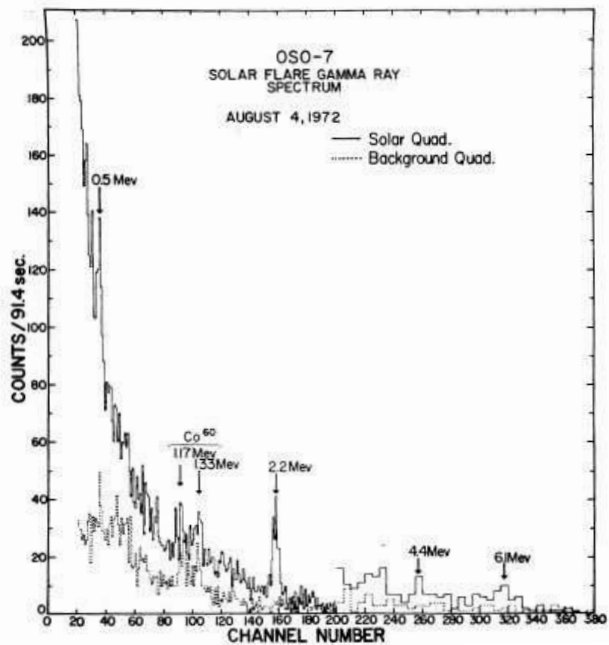}
\caption{Time-integrated $\gamma$-ray spectrum observed aboard \textit{OSO-7} in the solar and background quadrants for the discovery event, SOL1972-08-04. 
The integration time is from 06:24 to 06:33~UT. The solar quadrant 
shows significant enhancement up to 7 MeV as well as line emissions at 0.5~MeV, 2.2~MeV, 4.4~MeV and 6.1~MeV. The lines at 1.17~and 1.33~MeV are from the on-board calibration source (from Suri et al. 1973).
}        
\label{fig:Chupp73}       
\index{flare (individual)!SOL1972-08-04 (pre-\textit{GOES})!illustration}
\end{figure*}

The first evidence of acceleration of charged particles (electrons and ions) to 
relativistic energies in association with solar flares was found in 1942 with the 
discovery of large increases in the count rates of several ground level cosmic-ray 
intensity monitors\index{ground-level events}\index{cosmic rays!ground-level events}\index{acceleration!discovery of solar relativistic particles}\index{flares!and relativistic particles}\index{ions!gamma@$\gamma$-ray production}.
These 
observations told us, more than fifty years ago, that solar flares are capable of accelerating 
protons to GeV energies. While \cite{1951ZNatA...6...47B} 
had predicted that relativistic protons accelerated during a solar flare could 
produce a flux of high-energy neutrons observable at the Earth (as well as some 
$\gamma$-ray emission), \cite{1958cora.conf..305M} published the first prediction of 
$\gamma$-ray fluxes expected from several celestial objects including solar flares. 
\index{gamma-rays!2.223~MeV line!prediction}
These first estimates were at the base of former detailed theoretical studies of 
high-energy neutral emissions which were initiated by \citet {1967henr.book...99L}.
They provided a detailed treatment of the production of nuclear reactions in the solar 
atmosphere and calculated 
the yield of neutrons, positrons and $\gamma$-ray lines us\nocite{1975SoPh...43..415S}ing normalized rigidity 
spectra for the accelerated ions. 
The first observational evidence for $\gamma$-rays came 
from the observations of the flare SOL1972-08-04 with the \textit{OSO-7}\footnote{Orbiting Solar Observatory-7.}
$\gamma$-ray spectrometer \citep[][see Figure~\ref{fig:Chupp73}]{1973Natur.241..333C}, which confirmed the prediction by Morrison that nuclear reactions in the solar atmosphere could produce $\gamma$-ray lines.
\index{flare (individual)!SOL1972-08-04 (pre-\textit{GOES})}\index{satellites!OSO-7@\textit{OSO-7}}\index{reactions!nuclear}
Pioneering observations of 
solar $\gamma$-ray lines were also obtained for later flares of August 1972 by spectrometers on \textit{Prognoz-6} \citep{1975IAUS...68..315T} and for the flares 
SOL1978-07-11T10:54 
by \textit{HEAO~1} \citep{1980ApJ...236L..91H}
and SOL1979-11-09T03:20
by \textit{HEAO~3} \citep{1982ApJ...255L..81P}\index{satellites!HEAO-1@\textit{HEAO-1}}\index{satellites!HEAO-3@\textit{HEAO-3}}\index{satellites!Prognoz-6@\textit{Prognoz-6}}\index{flare (individual)!SOL1978-07-11T10:54 (X3)}\index{flare (individual)!SOL1979-11-09T03:20 (M9)}.
Since then solar $\gamma$-ray astronomy has provided a new 
window for the study of ion acceleration in solar flares\index{flare (individual)!SOL1972-08-04 (pre-\textit{GOES})!$\gamma$-ray discovery}.

\nocite{1975SoPh...43..415S}

Energetic ions interacting with the solar  atmosphere produce a wealth of $\gamma$-ray emissions\index{acceleration!gamma@$\gamma$-ray diagnostics}.
A complete  $\gamma$-ray line spectrum is produced through interactions of ions in the range
$\sim$1-100~MeV/nucleon and consists of several  
nuclear de-excitation lines\index{gamma-rays!gamma@$\gamma$-ray lines!nuclear de-excitation}, neutron capture\index{gamma-rays!gamma@$\gamma$-ray lines!neutron capture} and positron annihilation\index{gamma-rays!gamma@$\gamma$-ray lines!positron annihilation} lines. 
See, for example, \cite{1984ARA&A..22..359C}, \cite{1986psun....2..291R}, \cite{1996AIPC..374....3C}, \citet{1997LNP...489..219T}, \cite{2000ASPC..206..377S},\cite{2003LNP...612..127V}, and \cite{2006GMS...165..177S}. 
The temporal and  
spectral characteristics of all these radiations provide strong constraints on 
acceleration timescales, ion energy spectra and numbers as  
well as energetic ion abundances. If more energetic ions are present,
over a few hundred MeV/nucleon,
nuclear interactions with the ambient medium produce secondary pions whose decay products
lead to a broad-band continuum at photon energies above 10~MeV (with a broad peak around 70~MeV 
from neutral pion\index{gamma-rays!gamma@$\gamma$-ray continuum!pi@$\pi^0$-decay} radiation) \citep{1970Ap&SS...6..377S,1987ApJS...63..721M} and also secondary 
energetic neutrons\index{neutrons!energetic} which, if energetic enough, may escape from the Sun and be directly detected 
in interplanetary space ($\geq$~10~MeV neutrons at 
1~AU) or at ground level ($\geq$~200~MeV neutrons) \citep[see, e.g.,][]{2000ASPC..206..400C,2009RAA.....9...11C}. 
The importance of 
observing neutrons was pointed out by \citet{1951ZNatA...6...47B} and by 
\citet{1965JGR....70.4077L, 1965JGR....70.4087L}.
Lower-energy neutrons (kinetic energies below $ 100\,{\rm MeV} $) can only 
be observed in space because they are strongly attenuated in the Earth's atmosphere and cannot 
reach the ground.
Neutrons with kinetic energies higher than $ 100\,{\rm MeV} $ can be observed on the 
ground and hence with simultaneous observations in space and on the ground, it is 
possible to obtain the energy spectrum of solar neutrons and of accelerated particles, in a wide 
energy range.
\index{continuum!pi@$\pi^0$-decay}
\index{pions!gamma@$\gamma$-ray production}
\index{ions!secondary neutron production}
\index{neutrons!ground-level detection}
\index{acceleration!and neutron production}
\index{ground-level events}

The first detection of solar high-energy neutrons was provided by the observations of the 
\textit{SMM}/GRS\footnote{\textit{Solar Maximum Mission}/Gamma Ray Spectrometer.}
spectrometer for the flare SOL1980-06-21T02:00 (X2.6) which showed not only the emission from
\index{flare (individual)!SOL1980-06-21T02:00 (X2.6)!neutrons}
\index{neutrons!production in flares} 
ultra-relativistic electrons at the time of the flare but also a later signature 
from high-energy neutrons \citep {1982ApJ...263L..95C}. 
This same experiment provided the first observations of 
pion-decay radiation in solar flares (high energy radiation produced 
by ultra-relativistic ions) \citep {1985ICRC....4..146F} for the event SOL1982-06-03T13:26 (X8.0). 
\index{flare (individual)!SOL1982-06-03T13:26 (X8.0)!neutrons}
For this event, solar neutrons were also detected for the first time at ground-level by 
the IGY-type neutron monitor installed at Jungfraujoch, Switzerland, as well as 
by {\it SMM}/GRS \citep{1987ApJ...318..913C}\index{neutron monitors}\index{observatories!Jungfraujoch}.
With these observations, 
all the components predicted by \citet{1967henr.book...99L} had been detected 
\citep[see, e.g.,][for a detailed review of early $\gamma$-ray observations of solar flares]{1996AIPC..374....3C}. Since these first 
observations, solar energetic neutrons have been 
detected with ground-based detectors on a handful of occasions. 10-120 MeV neutrons 
were also detected in space with COMPTEL 
on the {\it Compton Gamma-Ray Observatory (CGRO)} mission, whose mode of operation enabled direct 
determination of their energies 
\citep{1993AdSpR..13..255R,1998A&A...340..257K}. We consider these
various neutron observations in more detail below (Section~\ref{neu}). 
\nocite{lange-forbush1995}

Before the launch of \textit{RHESSI} and \textit{INTEGRAL} in 2002, high energy observations of solar 
flares had been obtained for two solar cycles respectively with \textit{SMM}/GRS as well as on
 \textit{Hinotori} for cycle~21\index{solar cycles!cycle 21 $\gamma$-ray observations} and 
by the PHEBUS and SIGMA experiments aboard \textit{Granat}, the wide-band spectrometer on \textit{Yohkoh}, 
\textit{GAMMA-1} and \textit{CGRO} for cycle~22 
(see, e.g., Chupp, 1984; Chupp, 1996; Ramaty, 1986; Trottet \& Vilmer, 1997; Vilmer \& MacKinnon, 2003, for reviews).\nocite{1984ARA&A..22..359C,1996AIPC..374....3C,1997LNP...489..219T,2003LNP...612..127V1986psun....2..291R}
\index{solar cycles!cycle 22 $\gamma$-ray observations}\index{satellites!Hinotori@\textit{Hinotori}}\index{satellites!RHESSI@\textit{RHESSI}!launch}\index{satellites!SMM@\textit{SMM}}\index{satellites!GRANAT@\textit{GRANAT}}\index{satellites!GRANAT@\textit{GRANAT}!SIGMA}\index{satellites!GRANAT@\textit{GRANAT}!PHEBUS}\index{satellites!CGRO@\textit{CGRO}}\index{satellites!Yohkoh@\textit{Yohkoh}}\index{satellites!GAMMA-1@\textit{GAMMA-1}}Before the advent of \textit{RHESSI} observations, the 
solar electromagnetic radiation above 100~keV (including the $\gamma$-ray line 
region) was one of the last spectral domains with no spatially resolved 
observations, so that no information was available on the location of the 
interaction sites of the energetic ions (nor of electrons above 100~keV) on the Sun. 
$\gamma$-ray line emission from ions had been observed in many flares but the 
quantitative constraints from $\gamma$-ray line spectroscopy, which provides information on the bulk of the flare fast ions, whose presence had been deduced from observations with limited spectral resolution. 
Higher energy radiations and neutrons have been observed in fewer events, but
they provide crucial information on the extremes of particle acceleration.  

Since the launch of the solar-dedicated \textit{RHESSI} mission \citep{2002SoPh..210....3L} 
and of \textit{INTEGRAL} in 2002 \citep{2003A&A...411L..63V}, both equipped 
with high-resolution germanium detectors, several $\gamma$-ray line flares produced strong 
enough fluxes so that a detailed $\gamma$-ray line shape analysis could be achieved for a few 
events\index{satellites!INTEGRAL@\textit{INTEGRAL}}\index{satellites!RHESSI@\textit{RHESSI}}.
Furthermore, since February 2002, \textit{RHESSI} has observed a total of 29 events above 
300~keV with 18 clearly showing $\gamma$-ray line emission \citep{2009ApJ...698L.152S}. 
Pioneering results were 
obtained by  \textit{RHESSI} on a few of these events on the localization of the interaction sites of 
energetic ions. 
We shall review below the results provided by \textit{RHESSI} and also by \textit{INTEGRAL} on these 
issues.

\section{$\gamma$-ray line observations of energetic ions}

\subsection{General features of a $\gamma$-ray spectrum}
\index{spectrum!gamma@$\gamma$-rays}

\begin{figure*}
\centering
\includegraphics[width=0.70\textwidth]{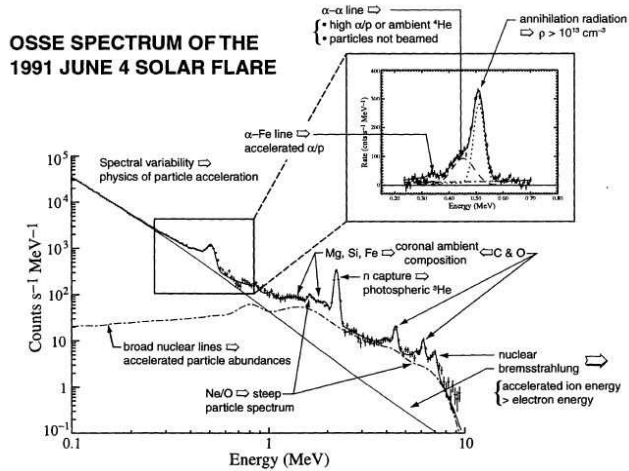}
\caption{OSSE spectrum of the flare SOL1991-06-04T03:37 (X12.0) summarizing the different components and the 
physics from $\gamma$-ray line spectroscopy \citep[from][]{2000ASPC..206..377S}.}
\index{flare (individual)!SOL1991-06-04T03:37 (X12.0)!illustration}
\label{fig:Share00}      
\end{figure*}

Many studies performed before the launch of \textit{RHESSI} have been devoted to the 
study of the complete $\gamma$-ray line spectrum which is produced in the 
solar atmosphere through the interactions of ions in the $\sim$1-100~MeV/nucleon range. 
In contrast to the monotonic electron bremsstrahlung continuum, the nuclear $\gamma$-ray 
spectrum displays a rich variety of features: broad and narrow lines and various sorts of continuum arising from several mechanisms.\index{bremsstrahlung!gamma@$\gamma$-ray continuum}\index{continuum!gamma@$\gamma$-ray bremsstrahlung}\index{continuum!gamma@$\gamma$-ray broad-line component}The bombardment of ambient nuclei of the solar atmosphere by accelerated 
protons and $\alpha$ particles produces a whole spectrum of de-excitation $\gamma$-ray 
lines (prompt narrow lines), almost all of them with rest energies lying between a 
few hundred~keV and 8~MeV. 
Inverse reactions in which accelerated C, O and
heavier nuclei collide on ambient~H and~He are the origin of prompt broad 
lines centered at the same line energy. Since the emitting nuclei retain 
significant momentum after the exciting collisions, these lines are 
substantially Doppler-broadened and/or shifted, to the extent that they are not individually 
resolvable.\index{reactions!nuclear!and broad-line component}
These broad lines merge with many other weaker lines 
from nuclei heavier than oxygen\index{gamma-rays!unresolved continuum}\index{gamma-rays!narrow-line component}\index{gamma-rays!broad-line component} to form an unresolved continuum (see Section~\ref{broad}) \citep{2009ApJS..183..142M}.
Figure~\ref{fig:Share00} from \citet {2000ASPC..206..377S}, based on the \textit{HEAO-1}/OSSE\footnote{\textit{High Energy Astrophysical Observatory-1}/Oriented Scintillation Spectrometer Experiment.} 
observations of the large flare SOL1991-06-04T03:37 (X12.0), shows the major features of\index{flare (individual)!SOL1991-06-04T03:37 (X12.0)}\index{gamma-rays!list of strong de-excitation lines} the $\gamma$-ray line spectrum: strong narrow de-excitation lines are found at 6.129~MeV 
from $^{16}$O, 4.438~MeV from $^{12}$C, 1.779~MeV from $^{28}$Si, 1.634~MeV 
from $^{20}$Ne, 1.369~MeV from $^{24}$Mg and 0.847~MeV from $^{56}$Fe. 
The broad lines merge into a quasi continuum dominating over the bremsstrahlung 
emission from electrons in the $\sim$1-8~MeV range. 
The single strongest line is the extremely narrow 2.223~MeV deuterium formation line. 
\index{gamma-rays!2.223~MeV line!deuterium formation}
Also present are photons of flare origin at 0.511~MeV from positron annihilation and the 
broad feature at 0.4-0.5~MeV formed from the two lines, at 0.429~and 0.47~MeV, that accompany formation of $^7$Li and $^7$Be in $\alpha$-$^4$He fusion reactions.\index{ions!nuclear reactions!alpha@$\alpha$/$\alpha$}\index{spectrum!gamma@$\gamma$-rays}\index{ions!nuclear reactions!fusion}\index{reactions!fusion}

\subsection{Magnetic loop transport and interaction model}
\subsubsection{Gamma-ray line production including transport}
\label{shmu}
\index{transport!ions}
\index{magnetic structures!gamma@$\gamma$-ray models}
\index{loops!ion transport models}

Information derived from $\gamma$-ray line emissions on the characteristics of the accelerated ions requires a comprehensive understanding of the 
transport of these energetic ions between the acceleration site, presumably located in the low corona, and the interaction and emitting sites in the 
denser layers of the solar atmosphere. This requires a model for ion acceleration, transport in flare loops and interaction. 
Several models have been developed since the 1980s on this topic, a couple of which we describe. 
The recent model described by \citet{2007ApJS..168..167M}
addesses particle transport and interaction and includes energy losses due to Coulomb collisions, removal by nuclear reactions, 
magnetic mirroring in the converging magnetic flux and MHD pitch-angle scattering in the corona, treated via the quasilinear formalism\index{magnetic structures!mirror geometry!and $\gamma$-ray production}.\index{magnetic structures!mirror geometry!Zweibel-Haber parametrization}\index{chromospheric density model!and magnetic field}\index{magnetic field!in chromosphere}\index{scattering!pitch-angle}
The loop formalism is the one previously developed by \citet{1989ApJ...341..516H}. 
The loop consists of a semicircular coronal portion of length 
{\it{L}} and two straight portions extending vertically from the transition region through the chromosphere into the photosphere. 
Below the transition region, the magnetic field strength is assumed proportional to a power $\delta$ of the pressure \citep{1983ApJ...264..648Z} taken here as $\delta$ $\simeq 0.2$.
\index{magnetic structures!mirror geometry!Zweibel-Haber parametrization} 
Such a converging magnetic field results in mirroring of accelerated particles\index{magnetic structures!below transition region}. 
Pitch-angle scattering is characterized by $\Lambda$, the mean free path required for an arbitrary initial angular distribution to relax to an isotropic distribution\index{ions!pitch-angle scattering!model}\index{pitch-angle scattering}\index{scattering!pitch-angle}.
The dependence of $\Lambda$ on particle energy is expected to be weak \citep[see discussion by][]{1989ApJ...341..516H} and is assumed to be independent of particle energy. 
In the model, the level of pitch-angle scattering of the energetic particles is simply characterized by $\lambda$, 
the ratio of $\Lambda$ to the loop half-length $L_c = L/2$. 
Magnetic convergence and pitch-angle scattering determine the angular distribution of 
the accelerated particles when they interact with the ambient medium, which is crucial to several aspects of the observable emissions. 
Several height-density profiles  ($n(h)$) for the solar atmosphere are also assumed. 

\subsubsection{Time history of the nuclear interaction rate}

\citet{2007ApJS..168..167M} studied the effects of transport on temporal behavior by assuming instantaneous release of all particles at $t=0$.
Actual observed time profiles represent the convolution of this behavior with the time profile of the particle accelerator. 
In the absence of magnetic convergence, ions do not mirror and the nuclear interaction time history depends only on the energy loss rate 
in the lower chromosphere and upper photosphere where the density is greatest and most of the interactions occur. 

\begin{figure}
\begin{center}
\hspace*{-5mm} \includegraphics[scale=0.70]{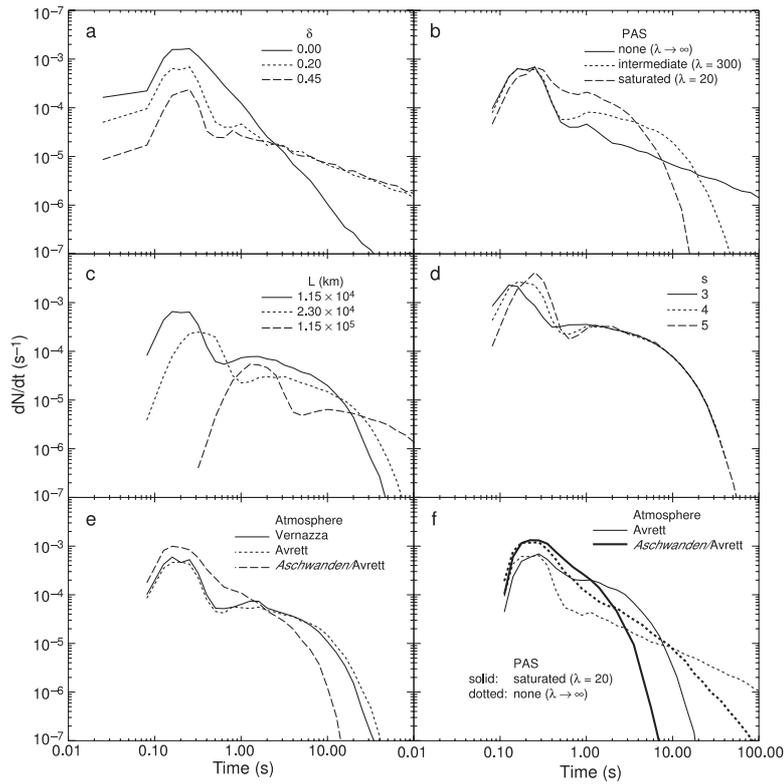}
\end{center}
\caption{Nuclear interaction rate as a function of the different parameters: (a) magnetic convergence, (b) pitch-angle scattering, (c) loop length, (d) spectral index, (e) atmopsheric model, (f) combined effect of the pitch-angle scattering and of the atmospheric model. The yields are normalized to one proton of energy greater than 30~MeV \citep[from][]{2007ApJS..168..167M}.}
\label{Murphy0207}
\end{figure}

In the presence of magnetic convergence, particles behave differently depending on pitch-angle $\alpha$. Increase of the magnetic field and  of the density towards the deeper atmosphere leads to 
partial particle trapping and to the formation of a loss cone of angular half-width ($\alpha_{0}$).\index{trapping}
Ions with $\alpha < \alpha_0$ precipitate in a loop transit time, behaving as described above. 
Ions outside the loss cone mirror and lose energy much more slowly as they traverse the 
low-density corona. 
\index{trapping!coronal}
These interactions thus occur on longer time scales, with the time scale increasing with increasing convergence. 
\index{ions!pitch-angle scattering}\index{scattering!pitch-angle!ions}
Pitch-angle scattering causes the loss cone to be continuously repopulated, and in fact is essential if observed, impulsive time profiles
are to be understood \citep{1983ApJ...264..648Z,1989ApJ...341..516H}. As a result, with increasing pitch-angle scattering rate (i.e., decreasing 
$\lambda$) the nuclear interaction rate increases at early times and correspondingly decreases at later times. 
However, the time history is no longer affected by increasing pitch-angle scattering\index{ions!pitch-angle scattering} when the rate of loss-cone replenishment exceeds the rate of nuclear reactions in the loss cone. 
Increasing the loop length increases the time scale of the interaction rate since mirroring particles spend more time at lower coronal densities where nuclear reactions are less likely. 
All these effects are illustrated in Figure~\ref{Murphy0207} from \citet{2007ApJS..168..167M}. 
On short timescales ($<$1~s) the effects of velocity dispersion may be seen, as higher energy particles transit the loop and encounter the deeper regions more rapidly\index{ions!velocity dispersion}.
On longer timescales, trapping and precipitation, via pitch-angle scattering, dominate the temporal behavior.

The results of these detailed calculations are summarized and put to work in the ``trap-plus-precipitation'' model, in which the flaring atmosphere is divided into a coronal, magnetic trap characterized by a single, uniform density $n_0$, and a high density, chromosphere-photosphere region\index{precipitation}\index{models!trap-plus-precipitation}.
The escape of particles from the trap to the deep atmosphere is summarized in an escape rate. 
The saturation of escape rate with increasing scattering rate, 
described above, may be recognized as the ``strong pitch-angle scattering'' limit\index{pitch-angle scattering!strong} identified by \citet{1966JGR....71....1K}, in which case the coronal pitch-angle distribution is always more or less isotropic and the escape time from the trap is given by 
\citep[see, e.g.,][]{1989A&A...213..383H}
\begin{equation}
\mu_{prep}={\frac{\alpha_0^2}{2}}{\frac{V}{L}},
\label{prec}
\end{equation}
where $\alpha_0$ and $L$ are as above and $V$ is  the particle velocity. 
\index{pitch-angle scattering!strong}\index{scattering!pitch-angle!strong limit}
\index{coronal sources}
For scattering rates below the strong limit, the resulting long coronal 
particle lifetimes become inconsistent with the short duration usually observed in solar $\gamma$-ray bursts \citep {1983ApJ...264..648Z}. 
The trap-plus-precipitation model has been applied to observations, with results
described below (Section \ref{kp}).

\index{loops!ion transport models!illustration}

\subsubsection{Angular distribution of the interacting ions}
\index{accelerated particles!pitch-angle distribution!ions}
\index{ions!pitch-angle distributions}

In the presence of magnetic convergence, particles encounter the greatest densities at their mirror points when they also have the greatest pitch angles. 
Observations (e.g., of line shapes) reveal the density-weighted, source-integrated pitch-angle distributions and these may consequently
take a ``fan-beam'' form  \citep[e.g.,][]{1987SoPh..107..351H}. 
As pitch-angle scattering approaches the strong limit ($\lambda \approx 20$), however, more and more particles precipitate in the loss cone, at small pitch angles, and the density-weighted distribution becomes more strongly beamed downward.
So the overall pitch-angle distribution reflects a combination of the initial pitch-angle distribution, the degree of magnetic convergence in the loop, and the pitch-angle scattering rate (see Murphy et al. 2007 for details).
\index{loops!converging field}\nocite{2007ApJS..168..167M} 

\subsubsection{Depth distribution of the interaction site}

In the absence of magnetic convergence, there is no mirroring and the depth distribution directly reflects the grammage required for the accelerated ions to interact as they move downward through the solar atmosphere. In the presence of magnetic convergence, mirroring results in interactions occurring at higher elevations (and therefore lower densities) as more and more particles are prevented from penetrating the lower atmosphere. 
Even with minimal convergence and in case of no pitch-angle scattering, a significant fraction of the interactions occurs at low densities (almost 20\% of the interactions occur at densities less than 10$^{10}$ cm$^{-3}$). As the scattering is increased, more particles 
are able to precipitate and the bulk of the interactions move deeper, to higher densities. 
Because higher-energy ions tend to interact farther along their paths, harder particle spectra result in interactions occurring at higher densities since more higher-energy ions are producing the interactions. 
Similarly, interactions whose cross-sections have higher threshold energies tend to occur preferentially at higher densities\index{cross-sections!thresholds}.
This dependence of the depth distribution on the interaction cross-section also explains why the depth distribution is affected by the accelerated $\alpha$/proton ratio\index{ions!alpha@$\alpha/p$ ratio}.
Alpha-particle interactions generally have lower threshold energies and, as just discussed, such interactions occur at lower densities. 
When the $\alpha$/proton ratio is high, a larger fraction of the line yield is due to such reactions and the depth distribution shifts to lower densities \citep[see][for details]{2007ApJS..168..167M}.\index{reactions!nuclear!depth distribution}

\subsubsection{Effective energies of the accelerated ions producing $\gamma$-ray lines}
\label{eff}
\index{spectrum!ions}
\index{ions!spectrum}

The main factor determining the effective energy range for a nuclear reaction is the cross-section, but the shape of the accelerated-ion spectrum can be important. 
For thin-target interactions, only ions with initial energies where the cross-section is significant contribute\index{thick-target model!gamma@$\gamma$-rays}.\index{thin target!ions}
For thick target interactions, however, even ions with higher initial energies also contribute, since they can lose energy and then interact at energies where the cross-section is significant.

\begin{figure}[tbp]
\begin{center}
\hspace*{-5mm} \includegraphics[scale=0.60]{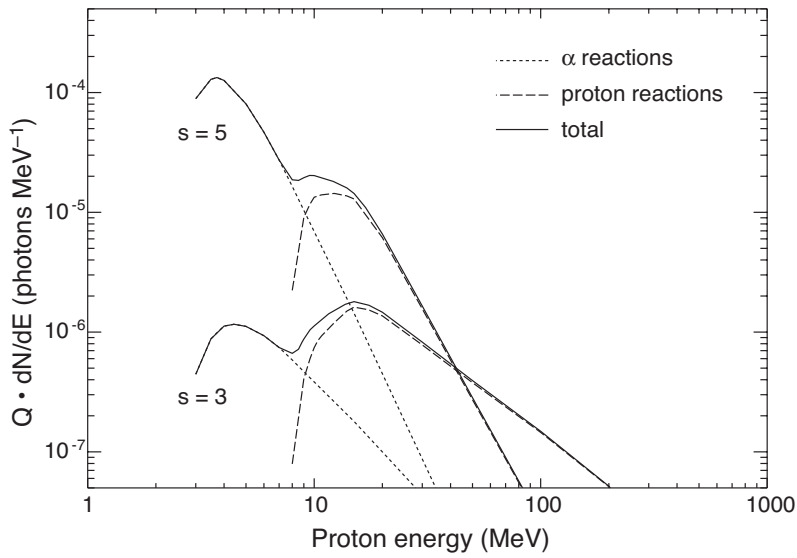}
\end{center}
\caption{Yield of the 6.129~MeV $^{16}$O line weighted by accelerated ion power-law spectra with indices $s$ = 3 and 5. The contributions to the line from accelerated protons and $\alpha$ particles are separately shown \citep[from][]{2007ApJS..168..167M}.}
\label{61_ener}
\end{figure}

\citet{2007ApJS..168..167M} determined the effective ion energies to produce a $\gamma$-ray line for a particular ion spectrum by weighting yields from monoenergetic ions with the spectrum. The ambient abundances are assumed to be coronal \citep {1996AIPC..374..172R}\index{abundances!and $\gamma$-ray production}.
As an example, Figure \ref{61_ener} from \citet {2007ApJS..168..167M} shows the differential yield of the 1.634 MeV $^{20}$Ne line as a function of accelerated ion energy for power-law spectra with indices of 3 and 5 and accelerated $\alpha$/proton = 0.5. These energies are the energies of the ions leaving the acceleration region.\index{acceleration region!escape from}
Separate contributions to the line from accelerated protons (dashed curves) and $\alpha$~particles (dotted curves) are plotted. For soft spectra, the $\alpha$-particle contributions dominate due to their lower threshold energies, and the most effective ion energies are around a few MeV/nucleon. 
For hard spectra, the proton interactions become important due to their higher  threshold energies and because the effective energies are higher. 
An effective ion energy range for producing the line is defined as follows: the ion energy where the yield is maximum is  determined and the effective range is the one for which the yield has fallen to 50\% on each side of the maximum. We note that if the effective energy distribution is very broad, the arbitrary value of 50\% could be misleading.

\begin{figure}[tbp]
\begin{center}
\hspace*{-5mm} \includegraphics[scale=0.60]{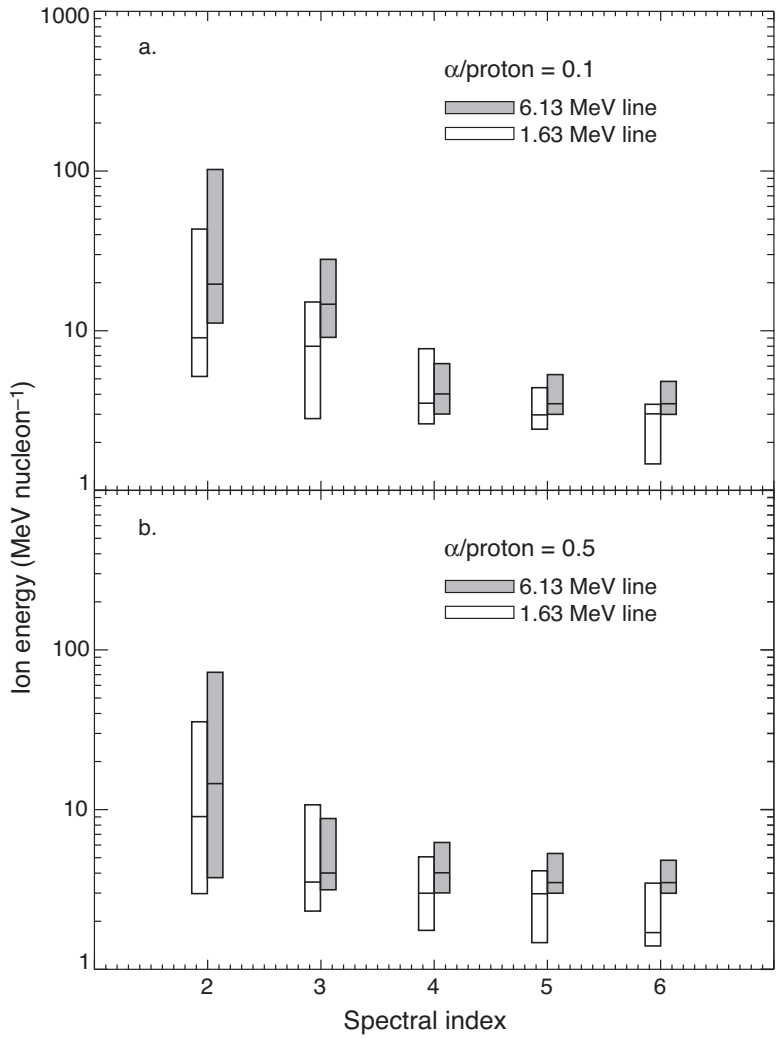}
\end{center}
\caption{Effective accelerated ion energies for production of the 1.634~MeV $^{20}$Ne line (white boxes) and 6.129~MeV $^{16}$O line (grey boxes) as a function of spectral index. 
The effective energy ranges are the 50\% yield range as defined in the text. Panel~(a) is for accelerated $\alpha$/proton = 0.1 and panel~(b) is for 0.5 (from Murphy et al. 2007).}
\label{16_61}
\end{figure}

The fluence ratio of the 6.129~MeV $^{16}$O and 1.634 MeV $^{20}$Ne lines has been frequently used as a measure of the accelerated-ion spectral index.\index{gamma-rays!6.129 MeV $^{16}$O line}\index{gamma-rays!1.634 MeV $^{20}$Ne line}
Figure~\ref{16_61} shows the effective accelerated-ion energy ranges for producing the 1.634~MeV $^{20}$Ne line (white boxes) and the 6.129~MeV $^{16}$O line (grey boxes) for several power-law spectral indices and for accelerated $\alpha$/proton = 0.1 (panel a) and 0.5 (panel b). The horizontal line within each box corresponds to the peak of the distribution. 
The Figure shows that the effective energy range for producing the $^{16}$O line is usually shifted to energies higher than for the $^{20}$Ne line. 
The extent of the separation of the effective energy ranges for the two lines determines the sensitivity of the ratio to the spectral index\index{ions!energy range for $\gamma$-ray production}.
Because the separation for these two lines is not large, the ratio is not very sensitive. 
For very hard spectra, the upper range of the effective ion energies can be $\sim$100 MeV/nucleon, but for most (softer) spectra it is less than 10~MeV/nucleon. 
The lower range is typically a few MeV/nucleon but for very soft spectra (particularly when the $\alpha/p$ ratio has the commonly assumed value of 0.5) it can be less than 2~MeV/nucleon. 
As the spectrum hardens, the higher-energy proton reactions begin to contribute and the effective ion energy range shifts to higher energies and becomes broader as interactions of both $\alpha$ particles and protons contribute. 
The interpretation of this particular energy-distribution diagnostic is discussed further below (Section~\ref{atmos}), where the roles of ambient chemical abundances and the relative contributions of protons and $\alpha$-particles are discussed.

\nocite{2007ApJS..168..167M}

\begin{figure}[tbp]
\begin{center}
\hspace*{-5mm} \includegraphics[scale=0.60]{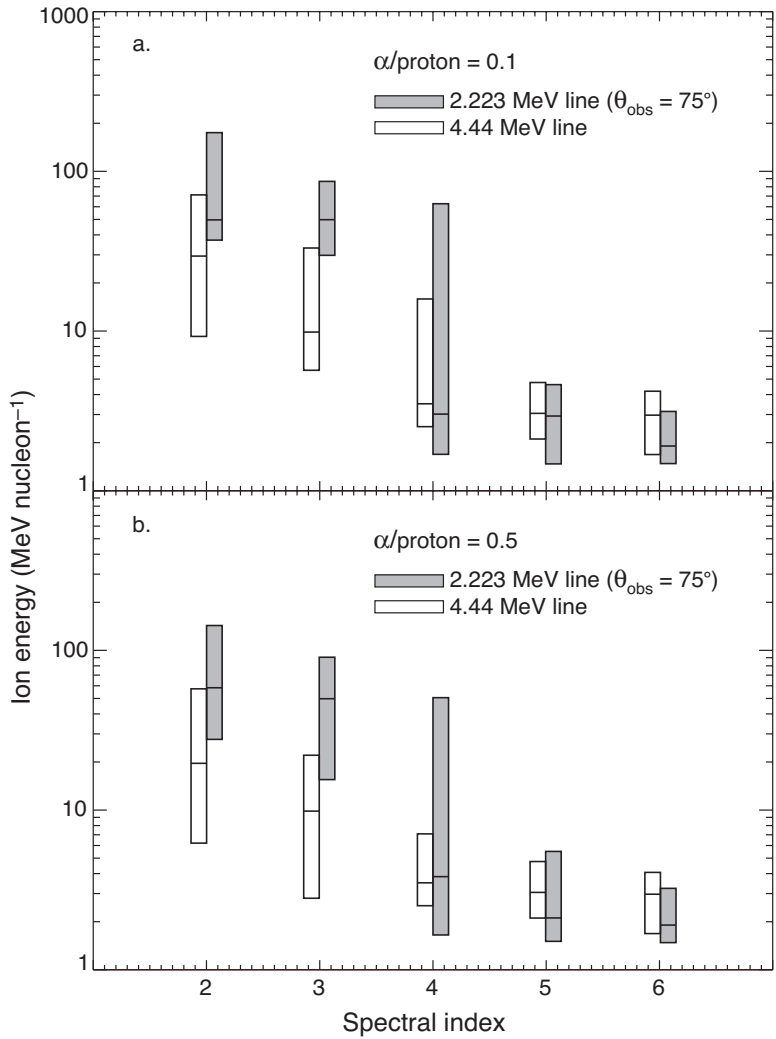}
\end{center}
\caption{Effective accelerated ion energies for production of the 4.438 MeV $^{12}$C line (white boxes) and 2.223~MeV neutron capture line (grey boxes) as a function of spectral index. The effective energy ranges are the 50\% yield range as defined in the text. Panel a is for accelerated $\alpha$/proton = 0.1 and panel b is for 0.5. For the neutron capture line, $\lambda$ = 300, $\delta$ = 0.2, $L$ = 11,500 km, and $\theta_{\rm{obs}}$ = 75$^\circ$  (from Murphy et al. 2007).}
\label{44_22}
\end{figure}

The fluence ratio of the 4.438~MeV $^{12}$C de-excitation line\index{gamma-rays!4.438~MeV $^{12}$C line} and the 2.223~MeV line\index{gamma-rays!2.223~MeV line} has also been used as a measure of the accelerated-ion spectral index. 
Figure \ref{44_22} shows the effective accelerated ion energy ranges for the 4.44~MeV line (white boxes) and the 2.223~MeV line (grey boxes) for several power-law spectral indices and for accelerated $\alpha$/proton = 0.1 (panel a) and 0.5 (panel~b). 
For hard spectra (spectral index $s <$4), the effective ion energies for the neutron capture line extend to very high energies, up to $\sim$100~MeV/nucleon, and are generally much higher than those producing the $^{12}$C line. 
However, for soft spectra ($s >$~4), the effective ion energies for the neutron capture line can be very low ($<$7~MeV/nucleon) due to neutron production by the exothermic ($\alpha,n$) reactions on heavy elements\index{ions!nuclear reactions!alpha@($\alpha,n$)}.\index{reactions!nuclear!exothermic}
For such soft spectra the effective ion energies for the neutron capture line are even less than those for the $^{12}$C line. 
As the flare location moves from disk center to limb, the effective energies shift to slightly lower energies since the neutrons from higher energy reactions are generally produced deeper and are subsequently more attenuated by Compton scattering.\index{Compton scattering!and attenuation of 2.223~MeV line}\index{gamma-rays!2.223~MeV line!Compton scattering}\index{scattering!Compton}
When a spectral index is derived using the ratio of these two lines, the relevant ion energies cover a broad range of energies. For hard spectra, the relevant ion energy range extends from a few MeV/nucleon up to and greater than 100 MeV/nucleon. For soft spectra ($s >$4), the relevant ion energy range is much narrower, from around 1 to a few MeV/nucleon. For very soft spectra, the lack of separation of the effective ion energies producing the two lines reduces the sensitivity of the ratio to the spectral index\index{gamma-rays!2.223~MeV line!effective primary energies}.

\subsubsection{Loop parameters and time development of $\gamma$-radiation}
\label{kp}
\index{loops!ion transport models!and time development}

\cite{2007ApJS..168..167M} discuss an integrated approach to the interpretation
of $\gamma$-ray measurements, distinguishing in particular between the deduction
of quantities describing the accelerated particles (``acceleration parameters'') 
and quantities describing the atmospheric environment in which the particles
evolve and radiate (``physical parameters'' -- loop lengths, target abundances, 
parameters describing the state of plasma turbulence in the corona).\index{acceleration region!turbulence}\index{turbulence}\index{plasma turbulence}
$\gamma$-ray measurements in turn yield various quantities, each of which contains information on some combination of acceleration and/or physical parameters. 
De-excitation line fluences\index{gamma-rays!de-excitation lines!fluences}, for example, are determined by acceleration parameters and otherwise only by source region chemical abundances.\index{pitch-angle scattering}\index{scattering!pitch-angle}
Line widths and centroid energies, on the other hand, are determined by the 
energy distribution and composition of energetic ions, but also by the amount of magnetic convergence in the loop and the degree of pitch-angle scattering, i.e., in principle, the energy distribution over wavenumbers of coronal MHD turbulence, but typically parametrized in terms of the ion mean free path $\lambda$ (normalized to the loop half-length). 
Below we concentrate on deduction of loop parameters from timing measurements.

Peak time delays, i.e., differences between the times of peak flux measured at different photon energies, have been reported 
in a few flares between hard X-rays and $\gamma$-ray line emissions
\citep[e.g.,][]{1989A&A...213..383H,1989SSRv...51...85Y}. Their interpretation was focused on the deduction of loop parameters\index{hard X-rays!timing relative to $\gamma$-rays}.
The relative timing of 
prompt $\gamma$-ray line emissions and of hard X-rays was studied in this context 
by Hulot and collaborators. 
These models were again used to interpret the delays of the hard X-ray and $\gamma$-ray line emissions observed for two flares 
by \textit{ISEE-3} and \textit{SMM}/GRS \citep{1992A&A...256..273H} and also applied to one of the $\gamma$-ray line flares observed with 
\textit{RHESSI} \citep{2007A&A...468..289D}.

\begin{figure}[tbp]
\begin{center}
%\hspace*{-5mm} \includegraphics[scale=0.70]{time_delay.eps}
\hspace*{-5mm} \includegraphics[scale=0.4]{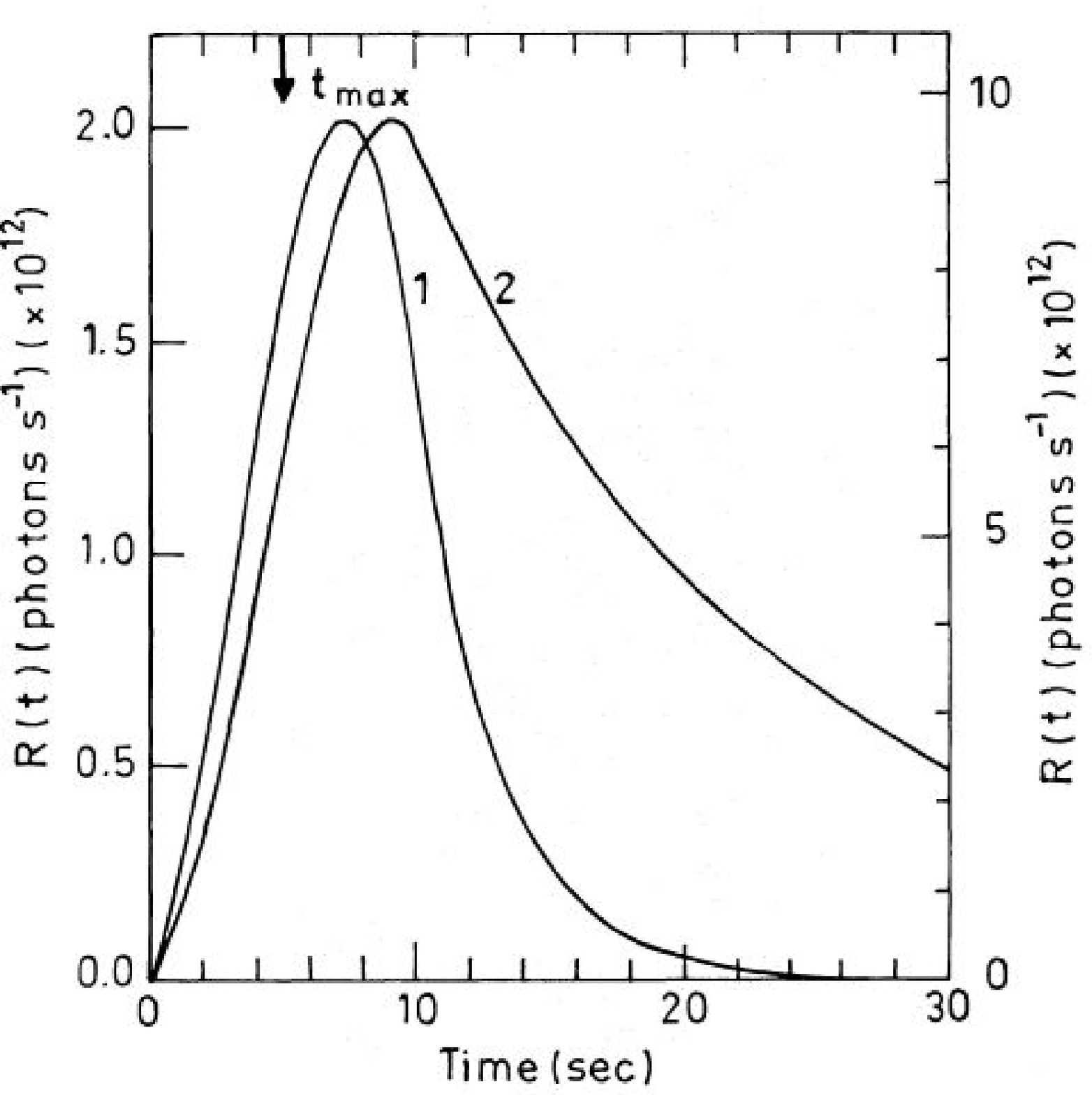}
\end{center}
\caption{Time profile of the production of the line at 4.4 MeV in the case of the trap-plus-precipitation model developed by Hulot et al. (1989) in the case of a short injection time $t_{max}$ and for two values of the precipitation rate $\alpha_0^2$/L = $10^{-5}$ km$^{-1}$ and $10^{-6}$ km$^{-1}$ (from Hulot et al. 1989).
}
\label{time_delay.eps}
\end{figure}
\nocite{1989A&A...213..383H}

\begin{figure}[h!]
   \centering
   \includegraphics[width=8.5cm,height=11cm]{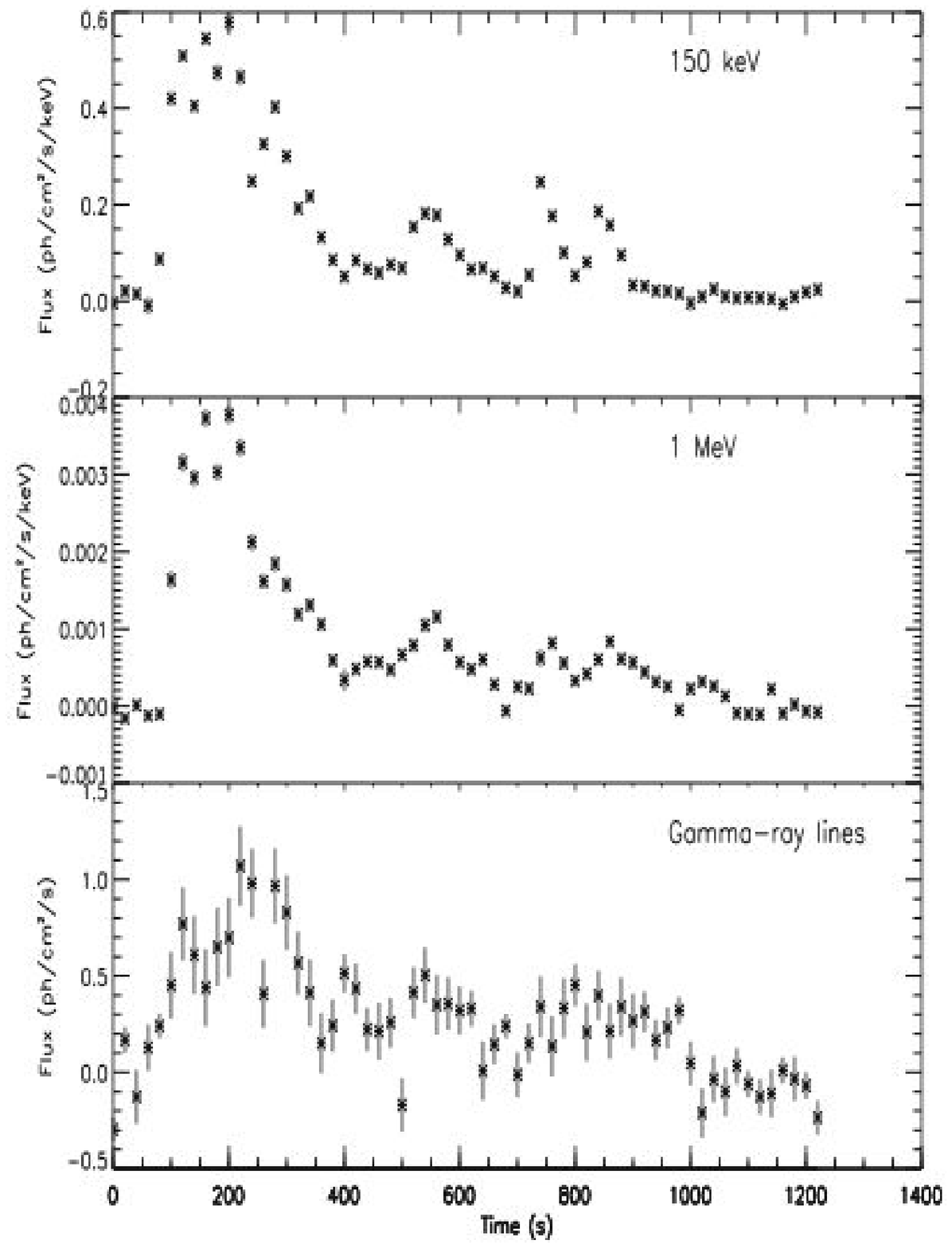}
      \caption{Time evolution of the bremsstrahlung flux at 150~keV and 1~MeV and of the prompt nuclear de-excitation lines observed with a time integration of 20~s with \textit{RHESSI} for SOL2002-07-23T00:35 (X4.8). 
      The starting time of the figure is 00:26~UT \citep{2007A&A...468..289D}, adapted from \cite{2003ApJ...595L..85S}.}
          \label{time_profile_23july.ps}
          \index{flare (individual)!SOL2002-07-23T00:35 (X4.8)!illustration}
   \end{figure}

For the study of temporal behavior, the basic characteristics of the hard X-ray and $\gamma$-ray trap-plus-precipitation models\index{models!trap-plus-precipitation} are summarized above, and described in 
\cite{1982A&A...108..306V} and \cite{1989A&A...213..383H}.  
The principle of the model is the following:
electrons and ions are supposed to be continuously injected over a finite injection time in a coronal loop where they experience trapping and pitch-angle scattering as above.\index{trapping!modelization} 
While particles trapped in the loops produce X-ray and $\gamma$-ray thin-target emissions, electrons and protons escaping from the trap propagate downwards to the denser chromospheric layers where they produce hard X-ray and $\gamma$-ray line emissions in a thick-target approximation.
\index{flare models!trap-plus-precipitation}\index{loops!trapping}
\index{thick-target model!gamma@$\gamma$-rays}
In these works, the precipitation rate is supposed to be in the ``strong diffusion limit'' given in Equation~(\ref{prec}). The free, ``physical'' parameters of the model describing the loop thus reduce to the precipitation rate (Section~\ref{prec}) of the trapped particles into the chromosphere and the density of the ambient medium.
\index{precipitation}
\index{atmospheric models!chromospheric density}
Additional parameters describe the production of energetic ions: the spectral index and the time 
evolution of the injected particles which in this model are not instantaneously, but continuously released. 
Energetic electrons, protons and alphas are assumed injected in the corona at a rate given by
\begin{equation}
q(E,t)=q_0f(E)f(t),
\end{equation}
where $q_0$ is the amplitude of the injection and $f(t)$ is given by a simple parametric form with rise-and-fall behavior:
\begin{equation}
 \begin{array}{ll}
f(t)=(t- t_0)(2t_{max}-(t-t_0)) \ \mathrm{for} \ t_0 \leq t \leq 2t_{max}+t_0; \\
f(t)=0 \ \mathrm{elsewhere},
\end{array}
\end{equation}
where $t_{max}$ and $t_0$ are, respectively, the peak time and the starting time of the injection.
Here  $f(E)$ represents the energy spectrum of injected electrons and ions;
 $f(E)$ is taken as a power law in energy with a spectral parameter $\delta$ defined by
\begin{equation}
f(E)=E^{-\delta}.
\end{equation}

These models compute electron bremsstrahlung radiation and the strong $\gamma$-ray lines at 6.129~MeV from $^{16}$O and 4.438~MeV from the first excited state of $^{12}$C, populated directly, and by the spallation reactions on $^{16}$O, and 1.779~MeV from $^{28}$Si, 1.634~MeV 
from $^{20}$Ne, 1.369~MeV from $^{24}$Mg  (with the photospheric target abundance for $\gamma$-ray lines).
\index{spectrum!gamma@$\gamma$-ray model}\index{ions!nuclear reactions!spallation}\index{electrons!relativistic}\index{reactions!nuclear!spallation}
The free parameters are finally the 
parameter $\alpha_0^2/L$ linked to the precipitation rate, the density, the starting and the peak time $t_{max}$ of the injection, the amplitude of the injection and the spectral index of the energy spectrum of injected particles.
Figure~\ref{time_delay.eps} shows the effect of the precipitation rate on the delay with respect to the injection time and on the time decay of the production of the prompt $\gamma$-ray line. Decreasing the value of the precipitation rate has a similar effect on the decay time as decreasing the pitch-angle scattering in the previous model.\index{precipitation}\index{scattering!pitch-angle}

 \cite{1989A&A...213..383H}, by using a trap-plus-precipitation model, were the first to calculate the relative time delay of $\gamma$-ray flux with respect to the hard X-ray flux.\index{models!trap-plus-precipitation}
They showed that such delays strongly depend on both the trapping time of the accelerated particles and the density of the loop. 
In particular, they showed that the time delays of the hard X-ray and $\gamma$-ray line emissions observed by \textit{ISEE-3}\footnote{International Sun-Earth Explorer-3.} and \textit{SMM}/GRS during the SOL1980-06-07T03:22 and SOL1981-04-27T09:45 events could be explained by this model and that parameters of the particles and of the ambient model could be obtained\index{satellites!SMM@\textit{SMM}}\index{satellites!ISEE-3@\textit{ISEE-3}}\index{flare (individual)!SOL1981-04-27T09:45 (X5.5)!modeling}\index{flare (individual)!SOL1980-06-07T03:22 (M7.3)!modeling}.
However, these studies were performed using the time profile of the 4-8~MeV range as a proxy for the time profiles of the nuclear lines, and no images at X-rays or $\gamma$-rays were used as additional constraints on the model. 

\begin{figure}[h!]
   \centering
    \hspace*{-5mm} 
      \includegraphics[scale=0.50]{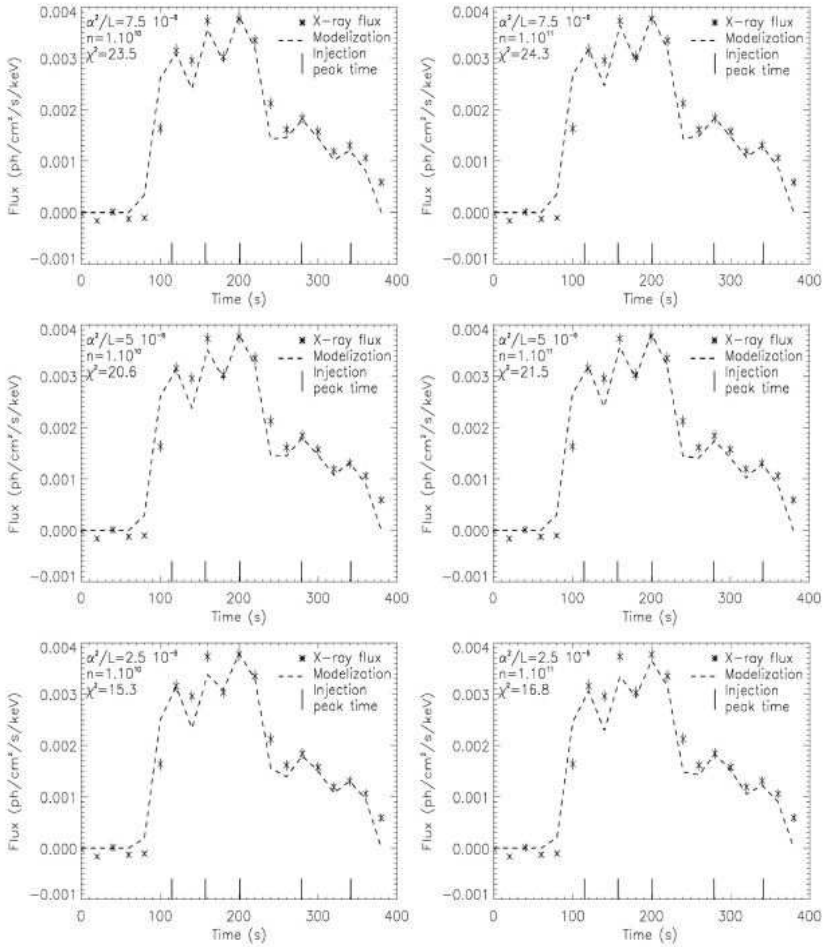}
      \caption{Time evolution of the observed and computed $\gamma$-ray flux derived for different parameters of the trap plus precipitation model \cite[from][]{2007A&A...468..289D}.}
          \label{dauphin_vilmer.ps}
   \end{figure}

A similar analysis was performed by \cite{2007A&A...468..289D} on the first $\gamma$-ray line event observed by \textit{RHESSI}, SOL2002-07-23T00:35 (X4.8).
\index{flare (individual)!SOL2002-07-23T00:35 (X4.8)!modeling}
Indeed, the analysis of the hard X-ray and $\gamma$-ray flux time evolution \citep{2003ApJ...595L..69L,2003ApJ...595L..85S} showed that there is a slight delay between hard X-ray emission at 150 keV and $\gamma$-ray line emission. 
\index{hard X-rays!timing relative to $\gamma$-rays}
\index{gamma-rays!timing relative to hard X-rays}
Figure~\ref {time_profile_23july.ps} 
shows the temporal evolution of the HXR emissions at 150~keV and 1~MeV and of the $\gamma$-ray line time profile. 
These time profiles are obtained from the spectroscopic analysis presented in \citet{2003ApJ...595L..85S}. 
The spectral analysis is performed for successive time intervals of 20~s, and results in the fitting to observed count spectra of a model photon spectrum including a double power law for the bremsstrahlung continuum, a nuclear $\gamma$-ray line function made of 15 narrow and broad Gaussians \citep{2003ApJ...595L..81S}, a neutron-capture line \citep{2003ApJ...595..L93M}, an $\alpha-^4$He fusion line complex between $\sim$400 and 500 keV \citep{2003ApJ...595L..85S}, 
and the solar annihilation line at 511~keV and its positronium continuum.
\index{continuum!positronium}
Figure~\ref{time_profile_23july.ps} -- adapted from \citet{2003ApJ...595L..85S} -- shows the results of this fit as a function of time: the bremsstrahlung flux at 150~keV and 1~MeV and the nuclear de-excitation line flux. 
This figure shows that the hard X-ray and $\gamma$-ray time evolutions are roughly similar, indicating a common origin of the accelerated electrons and ions, but that for the first main peak (from 0~s to 400~s) a time delay of around 12~s is observed between the time profile at 150~keV 
and the $\gamma$-ray line time profile. 
There is, however, no significant delay between the time profile at 150~keV and at 1~MeV. 

\cite{2007A&A...468..289D} investigated whether this delay could be reproduced by the trap-plus-precipitation model\index{models!trap-plus-precipitation}, given the additional constraint provided by the X-ray and $\gamma$-ray images. 
They found that the time profiles of both HXR and prompt $\gamma$-ray line fluxes could be reproduced (given that electrons and ions are injected and partially trapped in different coronal loop systems with slightly different characteristics discussed below) but they also found that the detailed X-ray and 
$\gamma$-ray time profiles could be well reproduced if the ratio of energetic electrons with respect to ions slightly varies from peak to peak during the flare. 
Such a variability of electron-to-ion production from peak to peak had been observed previously in flares based on spectral analysis 
\citep[e.g.,][]{1993A&A...275..602C} and \citet{2007A&A...468..289D} showed here that a similar result can be found from time-delay analysis. 
They finally found, given this variability in the electron-to-ion ratio, that a good reproduction of the evolution of both X-ray and $\gamma$-ray 
line time profiles (see Figure~\ref{dauphin_vilmer.ps})  was obtained if compared to electrons -- ions were injected in a system of coronal loops of lower density and with a 
slightly larger mirror ratio leading to a slightly more efficient trapping.\index{loops!mirror ratio}
If we consider a similar value for the loss cone for the two coronal loops, we find that ions should propagate in slightly longer loop lengths, which is consistent with the imaging observations (see Section~\ref{img}). 

The emphasis above has been on inter-comparison of X-ray and $\gamma$-ray time profiles. 
As discussed in detail by \citet{2007ApJS..168..167M}, 
analysis of $\gamma$-ray measurements on their own 
can also provide constraints on the loop parameters, e.g., from the flare $\gamma$-ray line shapes and the fluence and time evolution of the 2.223~MeV neutron-capture line.

\subsection {Characteristics of the solar atmosphere during flares from $\gamma$-ray line measurements}
\label{atmos}
\subsubsection{Atmospheric abundances: deduction from and consequences for $\gamma$-ray lines}
\label{abundances}
\index{abundances!gamma@$\gamma$-ray constraints}

The fluence in any particular de-excitation line depends on the number and energy distribution of fast particles, and on the abundance of the species that can contribute to it (either directly or via spallation reactions)\index{ions!nuclear reactions!spallation}.
A set of measured fluences in these lines thus includes information on both source chemical abundances, and the numbers and energy distributions of the exciting particles. If the energy distributions of the fast particles were known, we could deduce a set of source relative abundances from the measured fluences\index{abundances!and $\gamma$-ray spectra}\index{gamma-rays!abundance determinations}.
Solar photospheric abundance determinations are fundamental to an enormously wide range of astrophysical topics. 
Determinations in the
$\gamma$-ray source region would complement these, extending knowledge of the outer solar atmosphere.
\index{abundances!and model atmospheres}
This could be particularly important at a time when 3-D radiative-transfer modeling has produced several significant revisions of solar heavy element abundances \citep{2005ASPC..336...25A}, with the caveat that exceptional conditions might occur in the flaring atmosphere\index{caveats!exceptional conditions in flares}.

If, on the other hand, the source chemical abundances were known, we could deduce the energy distributions of protons and $\alpha$ particles in principle.
This depends on the extent to which the various cross-sections have different dependences on fast particle energy \citep{2004SoPh..223..155T}. 
In practice neither of these sets of variables is known with much reliability, and both the abundances and the parameters of the energy distributions should be deduced together\index{satellites!SMM@\textit{SMM}}.
To date, the \textit{SMM}/GRS work of \cite{1991ApJ...371..793M} remains the only published study where source abundances and particle energy distributions are determined together. 
Other studies have determined particle distributions using, e.g., coronal or 
photospheric abundances \citep{1995ApJ...455L.193R}, further deciding between these possibilities using goodness-of-fit tests.
Trends in line fluence ratios across many events have been used to make at least qualitative statements \citep{1995ApJ...452..933S};
with many events one can attempt to disentangle the separate influences of particle distributions and abundances. 
Target abundances may be determined non-parametrically or as a choice among several alternative sets\index{ions!distribution functions}.
Ion energy distributions are always represented by assumed analytical forms, however (e.g., power-law or Bessel function $K_2$ in momentum) whose parameters 
are determined as best fits to data.  
From these various studies, certain general statements may be made (see below) -- always noting that these might be subject to change via future revisions of nuclear cross-sections or 
refinements of data analysis techniques. 
We now survey some results of these studies.

Abundances of several species are enhanced compared to standard photospheric values\index{abundances!and FIP effect}\index{first ionization potential (FIP) effect}.
As is found using other measures of coronal abundances \citep[e.g.,][]{1993AdSpR..13..377M}, enhancements may be ordered by first ionization potential (FIP) of the species. 
Low-FIP elements (Mg, Si, Fe) generally show enhancements by an amount that varies from 
one event to the next but may be as great as 3-4. 
There is a definite ``FIP effect,'' with low-FIP elements showing more 
variability than high. 
At least in \textit{SMM}/GRS data, the 1.63 MeV line from $^{20}$Ne needs both a $^{20}$Ne enhancement (Ne/O~=~0.25, rather than the photospheric value of~0.15) and a steeply falling ion energy distribution that can exploit its low threshold for excitation by protons \citep{1995ApJ...452..933S}.
\index{abundances!and helioseismology}
Although an enhanced Ne abundance in the solar interior might have helped with certain problems in understanding helioseismology data \citep{2005Natur.436..525D}, there is no convincing evidence for such an enhancement at other wavelengths \citep{2005ApJ...634L.197S}. 
It now seems likely that we are identifying some peculiarity of the flaring atmosphere rather than the normal state of affairs in the non-flaring Sun.

Rather than a complete fit of the whole spectrum, one often tries to make progress using ratios of pairs of line fluences, for instance the ratio mentioned in Section~\ref{eff} of fluences in the 1.63 and 6.13 MeV lines. 
Let $\Phi_{\epsilon}$ stand for the measured fluence (cm$^{-2}$) in a line at photon energy $\epsilon$ (keV). 
Then the assumed abundances of the emitting species play a crucial role. 
With a photospheric or a coronal abundance ratio for neon and oxygen, one can only accommodate the observed values of $\Phi_{6.13}$, $\Phi_{1.63}$
and $\Phi_{0.4-0.5}$ by allowing protons and $\alpha$s to have significantly different energy distributions \citep{tonerphd,2004SoPh..223..155T}.
\index{ions!energy distributions!$\alpha$ particles}
Because the true degree of enhancement of $^{20}$Ne is uncertain, the proton and $\alpha$ energy distributions are correspondingly uncertain. 
This uncertainty adds to that introduced by the different energy-dependence of the cross-sections for excitation of the lines by protons and $\alpha$s.
\index{flare (individual)!SOL1981-04-27T09:45 (X5.5)!abundances}
For example, assuming $\alpha/p = 0.1$ among the fast particles,  $\Phi_{6.13}/\Phi_{1.63}$
for SOL1981-04-27T09:45 (X5.5) implies a power-law energy spectral index $\delta$ roughly in the range 4.3 - 4.8 if Ne/O = 0.15,
but 3.6 - 4.1 if Ne/O = 0.25 \citep{1995ApJ...455L.193R}, the former value believed appropriate to the photosphere \citep{2005ASPC..336...25A}. 
Ratios involving the 2.223~MeV line give a valuable further constraint
in the face of this uncertainty. All target species can contribute to free neutron production so some uncertainty in the abundance of any one has much less impact on deductions. 
On the other hand, the intensity of the 2.223~MeV line depends on other quantities that must be 
constrained, if at all, by other means: pitch-angle distribution and magnetic geometry in which the fast ions move\index{pitch-angle distributions!and 2.223~MeV radiation}.

Usually one needs to integrate over the whole of a flare to get a $\gamma$-ray spectrum of adequate statistical quality\index{gamma-rays!time-resolved spectra}\index{gamma-rays!statistical significance}.
There have, however, been a few attempts to study the time development of the $\gamma$-ray spectrum\index{spectrum!gamma@$\gamma$-ray!variability}.
\cite{1997ApJ...490..883M} found evidence for an increase with time of the abundances of low-FIP elements relative to high-FIP ones, from the \textit{CGRO}/OSSE data of SOL1991-06-04T03:37 (X12.0).
\index{flare (individual)!SOL1991-06-04T03:37 (X12.0)!abundance variations}
Evidence for a similar trend was found in \textit{RHESSI} data for the 
SOL2002-07-23T00:35 (X4.8) by \cite{2003AGUFMSH11D1132S}. \cite{2006GMS...165..177S} additionally show such a trend in the flare of SOL2003-11-03T09:55 (X3.9)\index{flare (individual)!SOL2002-07-23T00:35 (X4.8)!abundance variations}\index{flare (individual)!SOL2003-11-03T09:55 (X3.9)!abundance variations}.
It is interesting that a similar trend is found in data from three different flares, involving two completely different instruments, although the ubiquity or otherwise of this effect remains to be shown.
Assuming, unjustifiably, that this effect proves commonplace we may make some preliminary comments. 
We can imagine at least three reasons why $\gamma$-ray source region abundances might change \citep[see also][]{2006GMS...165..157M}. 
First, abundances might vary with position in an active region, and the regions of fast ion precipitation change as the flare progresses. 
\index{precipitation!and abundance variations}
It is not at first easy to see why this would produce a systematic change in the ratio of low-FIP to high-FIP elements, however. 
Second, abundances may well vary with height in the atmosphere. 
Even if the atmosphere remains static as the flare progresses, an ion energy distribution evolving in time (e.g., hardening) would sample an evolving range of atmospheric heights and thus relative
abundances. 
Third, relative abundances might indeed evolve in the source region during the flare \citep[e.g.,][]{1989ApJ...343..511W}.

\cite{2006A&A...445..725K} studied the \textit{INTEGRAL} data from the SOL2003-10-28T11:10 (X17.2) flare, noting a change in time of the relative intensities of the 4.44~MeV $^{12}$C and 6.13~MeV $^{16}$O lines\index{flare (individual)!SOL2003-10-28T11:10 (X17.2)!gamma@$\gamma$-rays}\index{satellites!INTEGRAL@\textit{INTEGRAL}}.
With line shape data providing additional constraints, their preferred explanation involves a change in time of the $\alpha/p$ fast ion abundance ratio, rather than the abundances of the 
target species. 
This example shows the importance of the extra information obtainable from line shapes.

\begin{figure}[tbp]
\begin{center}
\hspace*{-5mm} \includegraphics[scale=0.60]{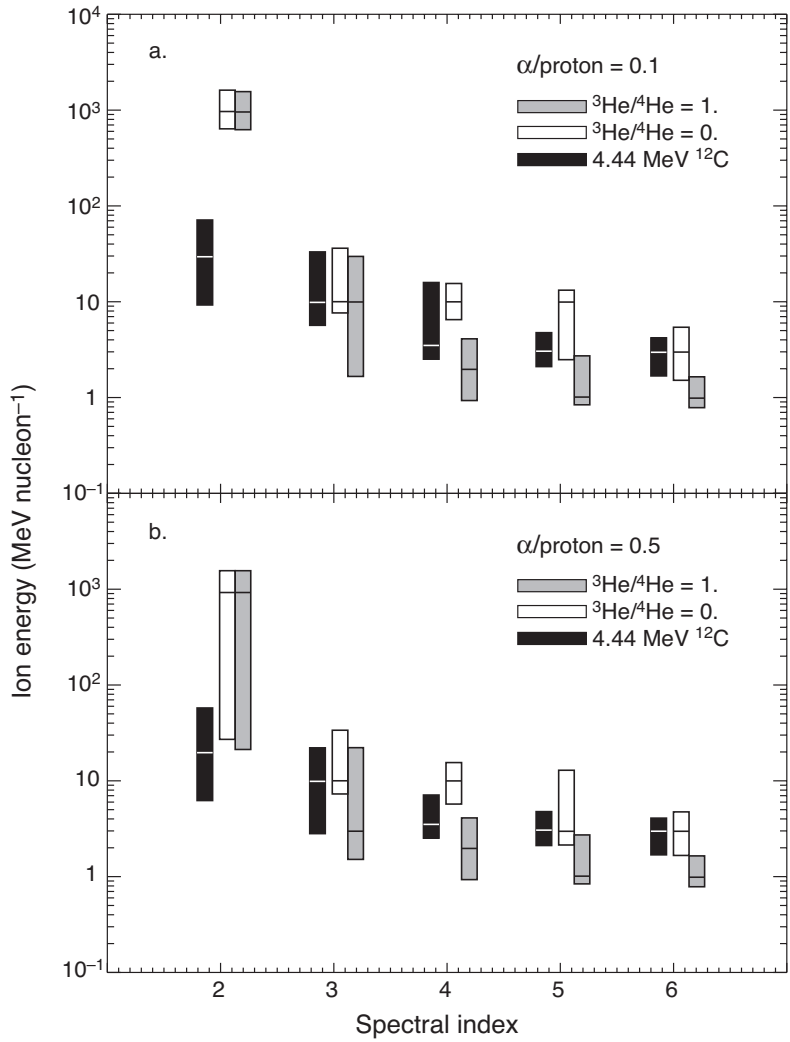}
\end{center}
\caption{Effective accelerated ion energies for production of positrons as a function of spectral index for accelerated $^3$He/$^4$He = 0 (white boxes) and 1 (grey boxes). 
The effective energy ranges are the 50\% yield range as defined in Section \ref{eff}. Panel~(a) is for accelerated $\alpha$/proton~=~0.1 and panel~(b) for~0.5. 
Also shown are the effective energies for the 4.438~MeV $^{12}$C line (black boxes) from Figure \ref{44_22} (from Murphy et al. 2007).}
\index{positrons!effective ion energies!illustration}
\label{44_posi}
\end{figure}

The time development of the 2.223~MeV line gives a quite different sort of constraint on the abundance of $^3$He\index{abundances!$^3$He}.
Indeed, after thermalization (typical duration around 100~s), the free neutrons produced in nuclear reactions are captured in the solar photosphere in two competitive processes: radiative 
capture on protons to form deuterium (and the 2.2~MeV line) and non-radiative capture on $^3$He ($^3$He(n,p)$^3$H).\index{reactions!nuclear!neutron-producing}
This method was first described and employed by \cite{1974SoPh...36..129W}.
They showed that the greater the $^3$He abundance, the more rapid the decay of the 2.2 MeV line. 
The determination of $^3$He/H  was thereafter done on a few flares observed by \textit{HEAO-1} \citep{1980ApJ...236L..91H}, \textit{SMM} \citep{1981ApJ...244L.171C,1987SoPh..113..229H}, \textit{GRANAT}/PHEBUS \citep{1993A&AS...97..337T}, \textit{CGRO}/OSSE \citep{1998ApJ...508..876S}, \textit{CGRO}/EGRET \citep{1999SoPh..187...45D,1999ICRC....6....9D}, 
\textit{CGRO}/COMPTEL \citep{2001A&A...378.1046R} and \textit{Yohkoh}/GRS \citep{1999ICRC....6....5Y}; see, e.g., \cite{2000ASPC..206...64M} for a review of observations before the \textit{RHESSI} era.
\index{eras!RHESSI@\textit{RHESSI}}
An upper limit on the photospheric $^3$He/H of $1.6\pm0.1 x 10 ^{-5}$ was found in the review of \cite{2000ASPC..206...64M}.       
The analysis of the time profile of the 2.223 MeV line was also achieved for SOL2003-07-23T00:35 (X4.8) observed with \textit{RHESSI} \citep{2003ApJ...595..L93M} 
\index{flare (individual)!SOL2002-07-23T00:35 (X4.8)!2.2 MeV time profile}
\index{flare (individual)!SOL2002-07-23T00:35 (X4.8)!$^3$He abundance}
to get a measurement of $^3$He/H. 
It is one of the factors included in the analysis of \cite{2008AGUFMSH31B1670N}.

\subsubsection{Conditions in the positron annihilation region}
\index{gamma-rays!511 keV line!formation conditions}

\begin{figure}
\begin{center}
\hspace*{-5mm} \includegraphics{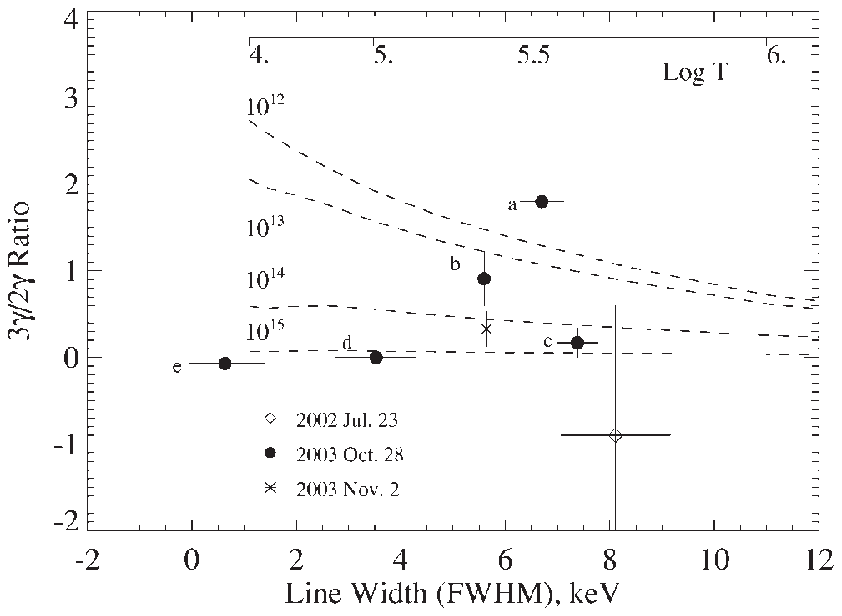}
\end{center}
\caption{\textit{RHESSI} measurements of the 3$\gamma$/2$\gamma$ ratio vs. 511~keV line width for three flares.
Time intervals for SOL2003-10-28T11:10 (X17.2) are: (a) 11:06:20-11:08:20~UT, (b) 11:08:20-11:10:20~UT, (c) 11:10:20-11:16:20~UT, (d) 11:16:20-11:18:20~UT,
and (e) 11:18:20-11:30:20~UT. 
The SOL2003-11-02T17:25 (X8.3) measurement was made between
17:16 and 17:26~UT, when the line was broad. 
The data point when the line was narrow is consistent with point~(e). 
The SOL2002-07-23T00:35 (X4.8) measurement was
integrated over the entire flare. 
The temperature scale and curves showing the
calculated 3$\gamma$/2$\gamma$ ratio vs. 511 keV line width for different densities are for a
fully ionized medium \citep[from][]{2004ApJ...615L.169S}.}
\index{flare (individual)!SOL2002-07-23T00:35 (X4.8)!illustration}
\index{flare (individual)!SOL2003-11-02T17:25 (X8.3)!illustration}
\index{flare (individual)!SOL2003-10-28T11:10 (X17.2)!illustration}
\index{flare (individual)!SOL2003-10-28T11:10 (X17.2)!broad 511~keV line}
\index{gamma-rays!511 keV line!illustration}
\index{gamma-rays!511 keV line!line width}
\label{share04}
\end{figure}

Positron production by flare fast ions has been studied in detail by \cite{1987ApJ...316..801K}, \cite{1987ApJS...63..721M} and \cite{2004ApJ...604..892K}\index{positrons}\index{gamma-rays!deuterium formation}.
Positrons may be produced via the decay of radioactive daughter nuclei, beta decay of excited states of target nuclei, or via $\pi^+$ production, e.g., in p(p,n$\pi^+$X)p$'$\index{ions!nuclear reactions!p(p,n$\pi^+$X)p$'$}.
In the first two cases cross-sections have thresholds mostly in the 1-10~MeV energy range, while $\pi^+$ production starts at 200-300~MeV/nucleon.\index{cross-sections!thresholds}\index{reactions!nuclear!thresholds}
Thus positron production, slowing down and annihilation will occur over a wide range of heights and the resulting $\gamma$-rays potentially carry information on deep layers of the atmosphere\index{positrons!effective ion energies!}.

The very broad energy range of ions contributing to positron production\index{positrons!ion energies for production} is illustrated in 
Figure~\ref{44_posi} from \citet{2007ApJS..168..167M}. 
Effective accelerated ion energies are shown for the production of positrons for accelerated $^3$He/$^4$He = 0 (white boxes) and 1 (grey boxes) and for accelerated $\alpha$/proton = 0.1 (panel a) and 0.5 (panel b). 
For very hard spectra dominated by protons, the effective ion energies range from 0.1~to~$>$1~GeV/nucleon because the positrons result mainly from very high-energy reactions producing charged pions.\index{reactions!nuclear!pion-producing}
At the opposite extreme, positron production from soft energy distributions rich in $\alpha$s and $^3$He is dominated by ions of 1-15~MeV/nucleon and even includes a significant contribution from $^3$He above $\sim$1~MeV/nucleon. 
Note that the effective energies shown in Figure~\ref{44_posi} are for the production of {\it{positrons}}. 
The escaping 0.511~MeV annihilation photons can be significantly attenuated if the positrons are produced deep in the solar atmosphere. 
Pion production can be very deep since the high energy ions responsible have very long ranges. As a result, even for very hard spectra, the annihilation photons 
that escape can be mainly from decay of radioactive nuclei rather than pions. 
Such nuclei are produced by ions with energies that are much lower than those indicated in the Figure.

Positron annihilation and consequent radiation are dealt with in great detail by \cite{2005ApJS..161..495M}\index{positrons!thermalization}\index{positrons!positronium formation}.
Positrons must slow down 
\index{gamma-rays!511 keV line}
and thermalize to contribute to 0.511~MeV radiation. 
They may annihilate directly on free electrons, or via positronium formation\index{positronium}\index{positronium!ortho- and para-}.
Positronium formation can involve both free or atomic electrons, in the latter case via
charge exchange reactions.\index{reactions!charge-exchange!and positronium}
Positronium in turn may exist as para-positronium (net spin~0) or ortho-positronium (net spin~1). 
Para-positronium annihilates to give two photons in the 0.511~keV line (2$\gamma$ mode)\index{positronium!2$\gamma$ and 3$\gamma$ decay}.
To conserve spin, ortho-positronium must annihilate producing three photons, at energies
$\le$~0.511~MeV (3$\gamma$ mode).
\index{positrons!three-photon continuum}\index{continuum!three-photon}
The lifetimes of the two spin states are different, so the proportions that annihilate (i.e., the ratio of the two-photon (line) and three-photon (continuum) spectral
components) depend on the rate of collisional destruction and thus on ambient density. 

Positrons in an ionized medium of temperature $T$~(K) produce a line of FWHM (full width at half maximum) width
$\sigma$ (keV) \citep{2005ApJS..161..495M}
\begin{equation}
\sigma \, = \, 0.011 T^{1/2}.
\label{epluswidth}
\end{equation}
For low temperatures an additional, broader component may be discriminated, from positronium formed via charge exchange in flight\index{positronium!formation by charge exchange}.
Equation~(\ref{epluswidth}) is valid for 0.511~MeV line emission formed via both direct annihilation and para-positronium formation on free electrons, but not for positronium formation via charge exchange with atoms or ions. 
Thus the ionization state of the medium also influences the shape and width of the line.\index{ionization state!gamma@$\gamma$-ray line shape}

In summary, then, the shape of the line and the relative intensities of the 2$\gamma$ and 3$\gamma$ components of the annihilation spectrum include information on the temperature, density and ionization state of the positron-annihilation region\index{spectrum!511 keV line}\index{gamma-rays!511 keV line!formation conditions}\index{gamma-rays!511 keV line!factors determining shape}

\textit{RHESSI} has observed 0.511~MeV line emission from the flares SOL2002-07-23T00:35 (X4.8), SOL2003-10-28T11:10 (X17.2), SOL2003-11-02T17:25 (X8.3) and SOL2005-01-20T07:01 (X7.1)
\citep{2003ApJ...595L..85S,2004ApJ...615L.169S,2005ApJS..161..495M}.
\index{flare (individual)!SOL2002-07-23T00:35 (X4.8)!511~keV line} 
\index{flare (individual)!SOL2003-10-28T11:10 (X17.2)!511~keV line}
\index{flare (individual)!SOL2003-11-02T17:25 (X8.3)!511~keV line}
\index{flare (individual)!SOL2005-01-20T07:01 (X7.1)!511~keV line}
During the greater part of these flares, until quite late times, the width of the line was determined to be 4-8~keV (see Figure~\ref{share04}). 
\index{gamma-rays!511 keV line!line width}

\begin{figure}
\begin{center}
\hspace*{-5mm} \includegraphics[scale=1.2]{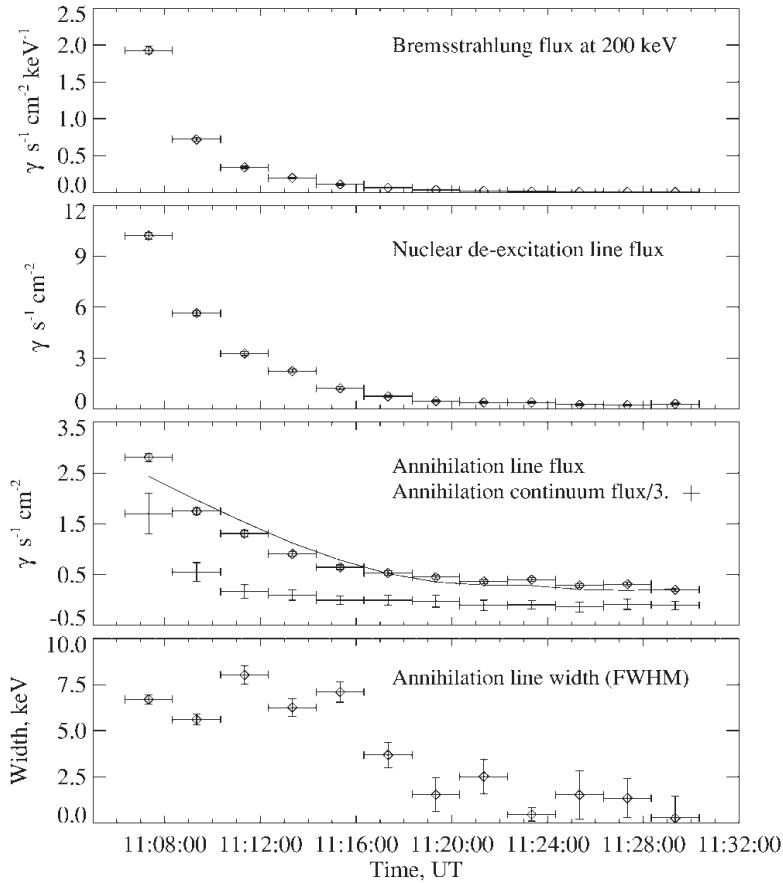}
\end{center}
\caption{Time histories of the bremsstrahlung, total nuclear de-excitation line, annihilation line and positronium continuum fluxes, and the width of the annihilation line during SOL2003-10-28T11:10 (X17.2) \citep[from][]{2004ApJ...615L.169S}.}
\index{flare (individual)!SOL2003-10-28T11:10 (X17.2)!illustration}
\label{share04a}
\end{figure}

The SOL2003-10-28T11:10 flare provided the best-observed 0.511~MeV line\index{positrons!511-keV line width}. 
For the first 10 minutes of the flare its width is more or less unchanged at $\sim$8~keV. Equation~(\ref{epluswidth}) then implies an ambient temperature of $\sim$$5 \times 10^5$ K (see Figure~\ref{share04a}).
\index{transition region!511-keV line width}
The measurements of the positronium continuum to line ratio (3$\gamma$/2$\gamma$ ratio) implies an ambient density in the range  $\sim 10^{13-15}$ cm$^{-3}$ (see above) \citep{2005ApJS..161..495M}. 
In this initial phase the positron annihilation region appears characterized by an unexpected combination of chromospheric density and transition-region temperature \citep{2004ApJ...615L.169S,2005ApJS..161..495M}. 
\index{chromospheric density model!and 511~keV line width}
The similarly broad line found in SOL2003-10-28T11:10 (X17.2) carries similar implications \citep{2004ApJ...615L.169S}.
The line then narrows in $\sim$2~min, settling down to a width of $\sim$1~keV~for the remaining 12~min of the flare. 
The best explanation of the late-time narrow line is in terms of a medium at a temperature of $\sim$5000~K, but with a relatively high degree of ionization ($\sim$20\%), to suppress the broader feature that results from positronium formation in flight (Figure \ref{share04}). 

The line in SOL2003-11-02T17:25 (X8.3) is more consistent with a ``normal'' solar atmosphere\index{atmospheric models!and 511-keV line shape}, specifically with the conditions in the layer of the \cite{1981ApJS...45..635V} model~C atmosphere at which the temperature is 5000~K\index{flare (individual)!SOL2003-11-02T17:25 (X8.3)!511~keV line width}.
So far the 0.511~MeV line shapes have been calculated for single regions with homogeneous conditions. 
As we see, these yield significant insight into at least average annihilation region conditions. 
There remains a need for such calculations to take account of the 
broad range of depths over which positrons will be produced, and the consequent wide range of conditions in which they will annihilate.

The conditions found for SOL2003-10-28T11:10 (X17.2) are particularly surprising, unlike those found at any layer of existing models for quiet-Sun or flaring atmospheres. 
\cite{2007ApJ...659..750R} study lines of O~{\sc vi}~UV lines and X-ray emission from SOL2002-07-23T00:35 (X4.8).
The density and temperature implied by the positron annihilation line would have to apply only in a very narrow layer, $\sim$100~m, to be consistent with the measured O~{\sc vi} line strengths. 
Even if such a narrow layer could be produced, bearing in mind the range of heights over which fast electrons would deposit energy, we would also have to find some mechanism which would localize
most positron annihilations there. 
A multi-wavelength study of this flare \citep{2006ApJ...650.1184S} further underlines 
the difficulty of accounting for the positron annihilation line width\index{positrons!511-keV line width!difficulty of interpretation}.

\subsubsection{Inclination of magnetic field lines}
\index{magnetic structures!tilted loops!gamma@$\gamma$-ray evidence for}\index{loops!tilted}

\begin{figure}
\begin{center}
%\hspace*{-5mm} \includegraphics[scale=0.20]{smith04.eps}
%\hspace*{-5mm} 
\includegraphics[scale=0.7]{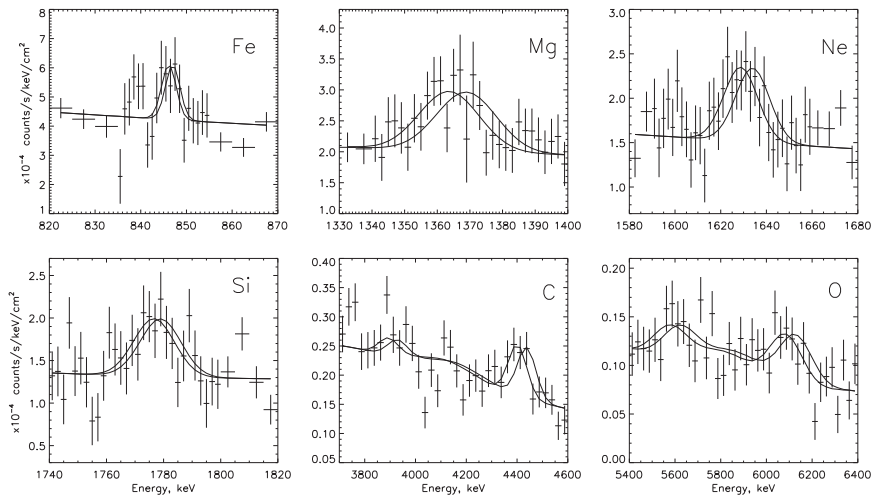}
\end{center}
\caption{\textit{RHESSI} background-subtracted count spectra for SOL2002-07-23T00:35 (X4.8). 
Each panel is labeled with the element primarily responsible for the line shown. 
The carbon and oxygen lines also show the secondary peak from the escape of a 511~keV positron-annihilation photon, which also contains information on the line shape. 
The thick curve shown in each panel is the Gaussian fit with a redshift of respectively 0.11\% for the Fe~line, 0.40\% for the Mg~line, 0.32\% for the Ne~line, 0.12\% for the Si~line, 0.79\% for the C~line and 0.58\% for the O~line plus the underlying bremsstrahlung continuum
and broad lines, convolved with the instrument response. 
The thinner line is the same fit forced to zero redshift for comparison. The error bars are
one $\sigma$ from Poisson statistics \citep[from][]{2003ApJ...595L..81S}.
}
\index{flare (individual)!SOL2002-07-23T00:35 (X4.8)!illustration}
\index{flare (individual)!SOL2002-07-23T00:35 (X4.8)!gamma@$\gamma$-rays}
\label{smith04}
\end{figure}

With the unprecedented spectral resolution in the $\gamma$-ray domain provided by \textit{RHESSI}, \cite{2003ApJ...595L..81S} were able to resolve for the first time the shapes of several of the strongest flare $\gamma$-ray lines in SOL2002-07-23T00:35 (X4.8), determining line 
widths and Doppler shifts (Figure~\ref{smith04}) (for a complete discussion of the interpretation of line shapes, see Section~\ref{ls}).
\index{flare (individual)!SOL2002-07-23T00:35 (X4.8)!line shapes} 
All lines were found to be redshifted by amounts that decrease with increasing target species mass, as expected for kinematic recoil.\index{gamma-rays!inelastic scattering!recoil redshifts detected}\index{scattering!pitch-angle!ion recoil}
The absolute values of these redshifts were larger than expected for a flare at S13E72, however, on the assumption of field lines normal to the solar surface. \cite{2003ApJ...595L..81S} thus suggested that field lines might be inclined  to the vertical, in this flare at least by $\sim$40$^{\circ}$.\index{magnetic structures! inclination determined via $\gamma$-rays}\index{gamma-rays!tilted fields} 
The alternative explanation, that fast ions are highly beamed, appears to be inconsistent with deductions from the $\alpha$-$\alpha$ line complex in the 0.4-0.5~MeV range \citep{2003ApJ...595L..89S}, as well as potentially raising difficult questions of velocity space stability \citep{1989ApJ...342..576T}\index{ions!nuclear reactions!alpha@$\alpha$/$\alpha$}.
Systematic field line tilts to the vertical have been suspected in the past, on the basis of Zeeman split lines\index{Zeeman effect} formed at different heights \citep{1978A&A....69..279W}. 
\cite{1990A&A...232..544M} had also suggested such a systematic tilt because of an apparent
asymmetry in the azimuthal distribution of events visible at photon energies $>$10~MeV. 
Field lines will fan out with altitude as a result of pressure balance \citep{1976RSPTA.281..339G} but this would produce only a broadening, not a systematic red or blue shift.

\begin{figure}
\begin{center}
\hspace*{-5mm} \includegraphics[scale=0.6]{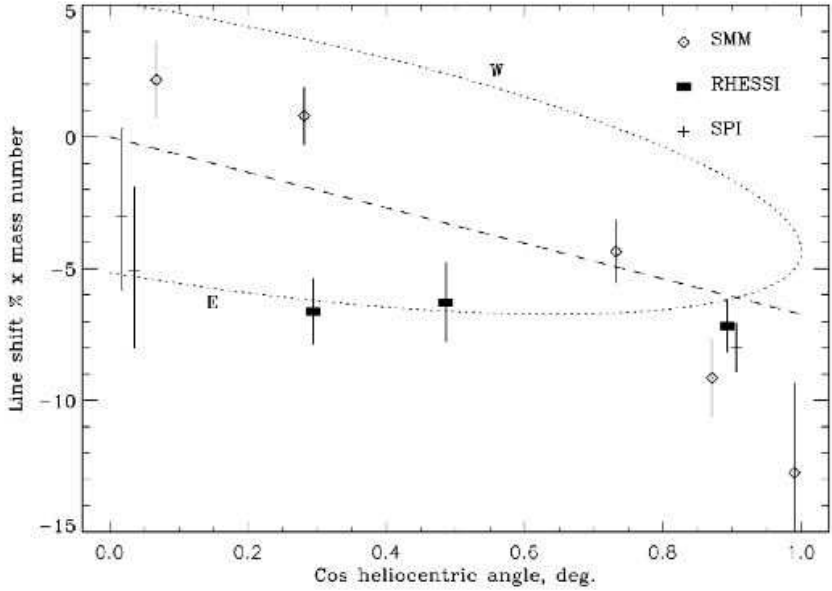}
\end{center}
\caption{Doppler shifts of $^{12}$C and $^{16}$O~line energies from rest, multiplied by the atomic mass of  the recoiling nucleus. 
Dashed line: expected Doppler shift from a simple model of 10 MeV protons incident on ambient solar atmosphere nuclei in a downward isotropic beam at 90$^{\circ}$. 
Dotted lines: expected shifts if the protons are incident along magnetic field lines tilted at 50$^{\circ}$ to the horizontal. 
If this is the explanation for the \textit{RHESSI} point at cos $\theta$  = 0.29 towards the east limb, then the Doppler shifts should follow the branches labelled ``E'' and ``W'' to the east and west of the Sun's center respectively. 
Sources of data points are from \textit{SMM} (19 measurements in bins of heliocentric angle; Share et al., 2002), \textit{RHESSI} measurements for the flares SOL2002-07-23T00:35, SOL2003-10-28 and SOL2003-11-02 \citep{2006GMS...165..177S}, and \textit{INTEGRAL}/SPI measurements for the flares SOL2003-10-28 \citep{2006A&A...445..725K}, SOL2003-11-04T19:53  and SOL2005-09-07T11:10 \citep{2007A&A...461..723H}; from \citep{2007A&A...461..723H}.
}
\index{flare (individual)!SOL2002-07-23T00:35 (X4.8)!illustration}
\index{flare (individual)!SOL2003-10-28T11:10 (X17.2)!illustration}
\index{flare (individual)!SOL2003-11-02T17:25 (X8.3)!illustration}
\index{flare (individual)!SOL2003-11-04T19:53 (X17.4)!illustration}
\index{flare (individual)!SOL2005-09-07T09:52 (X17.0)!illustration}
\label{harris}
\end{figure}

\cite{2007A&A...461..723H} studied \textit{INTEGRAL} spectra of SOL2003-11-04T19:53 (X17.4) and SOL2005-09-07T09:52 (X17.0)\index{flare (individual)!SOL2003-11-04T19:53 (X17.4)!gamma@$\gamma$-rays}\index{flare (individual)!SOL2005-09-07T09:52 (X17.0)!gamma@$\gamma$-rays}.
These took place on the west and east limbs respectively. 
Systematically tilted loops would produce a systematic difference in Doppler shifts: the 40$^{\circ}$ tilt suggested by \cite{2003ApJ...595L..81S} would imply blueshifted lines from the west limb\index{gamma-rays!tilted fields!and Doppler shifts}. 
The two flares studied both display (statistically weak) evidence for redshifts, but none for blueshifts. 
Thus it appears that both flares plus SOL2002-07-23T00:35 (X4.8) involve field lines oriented predominantly
away from the observer, but that there is no evidence so far for a systematic tilt in any preferred sense (Figure \ref{harris}). 
The small number of flares renders this study preliminary. 
At the heart of the interpretation is a picture of unidirectional field lines, either normal to the solar surface or all tilted at a single angle to the vertical, instructive but very simple. 
Major $\gamma$-ray flares originate from complex active regions and will undoubtedly involve a broad distribution of magnetic field directions. 

\nocite{2002ApJ...573..464S} 

\subsection {Narrow lines and ion energy distribution}
\label{dists}
\index{accelerated particles!ion primary spectrum}
\index{ions!energy distributions}
\index{gamma-rays!energy dependence of excitation}

Since the first detection of solar $\gamma$-ray lines in 1972, many solar 
$\gamma$-ray line flares have been observed with detectors aboard \textit{SMM}, 
\textit{Hinotori}, \textit{GRANAT}, \textit{CGRO} and \textit{Yohkoh}\index{satellites!SMM@\textit{SMM}}\index{satellites!Hinotori@\textit{Hinotori}}\index{satellites!GRANAT@\textit{GRANAT}}\index{satellites!CGRO@\textit{CGRO}}\index{satellites!Yohkoh@\textit{Yohkoh}}.
The analysis of these observations have led to quantitative analysis for more than 20~$\gamma$-ray line events. 
Apart from information on the solar atmospheric elemental abundances which have been deduced from the analysis of narrow $\gamma$-ray lines (see previous sections), energy spectra of protons have been deduced for 19~\textit{SMM}/GRS flares \citep{1995ApJ...452..933S,1995ApJ...455L.193R} and for one \textit{CGRO}/OSSE flare \citep{1997ApJ...490..883M}.  
Each line is characterized by energy-dependent cross-sections for excitation
by protons and by $\alpha$~particles.\index{gamma-rays!inelastic scattering!ion numbers}\index{scattering!inelastic}
Thus the measured fluence in each line provides a sum of energy-weighted measures of 
the numbers of protons and $\alpha$s \citep[e.g.,][]{1986psun....2..291R}, above the lesser of the threshold energies for proton and $\alpha$ particle excitation. 
The measured fluences in more than one line contain information on the energy distributions of protons and $\alpha$s.
\index{cross-sections!Ne inelastic scattering}
The 1.63~MeV cross-section of $^{20}$Ne has a lower cross-section than the
other de-excitation lines \citep{1995ApJ...452..933S}, so the ratio of the fluences in the lines at 1.63~and 6.13~MeV, for instance, gives a measure of fast ion spectral hardness, representing a variable range of ion energies of around 1-20~MeV (see Section~\ref{eff}). 
Higher ion energies, on average, are needed to liberate neutrons in nuclear collisions so the ratio of the 2.223~MeV line to one or more of the de-excitation lines similarly gives a measure of spectral shape. 
The authors listed above assumed the power-law form for ion energy distributions and deduced the values of energy power-law spectral index from either or both of these measures, for the 20 flares observed by \textit{SMM} or \textit{CGRO}. 
The measured ratios imply that the accelerated ion proton spectra should extend as unbroken power laws down to at least about 2~MeV/nucleon if a reasonable ambient Ne/O abundance ratio is used, i.e., in agreement with measurements of the Ne/O abundance ratio in the corona \citep{1995ApJ...455L.193R}\index{abundances!Ne/O ratio}.
The spectrum deduced is thereafter used to estimate the energy contained in accelerated protons above 1~MeV. 
This last determination is, however, largely dependent on the value of the ambient Ne/O abundance 
ratio and also on the composition of accelerated particles, in particular the ratio of energetic 
He nuclei to energetic protons ($\alpha$/p) \citep[see][]{1995ApJ...455L.193R}.

\begin{figure}
\begin{center}
\hspace*{-5mm} \includegraphics[scale=1]{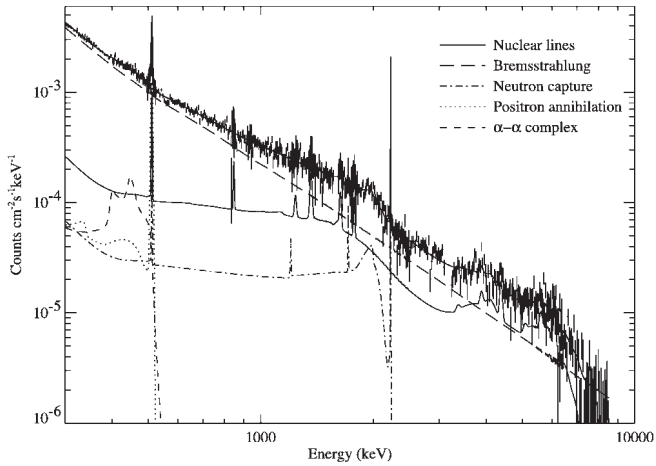}
\end{center}
\caption{\textit{RHESSI} $\gamma$-ray count spectrum for SOL2002-07-23T00:35 (X4.8). 
The lines show the different components of the model used to fit the spectrum \citep[from][]{2003ApJ...595L..69L}.}
\label{lin_2003}
\index{flare (individual)!SOL2002-07-23T00:35 (X4.8)!illustration}
\index{flare (individual)!SOL2002-07-23T00:35 (X4.8)!gamma@$\gamma$-ray spectrum}
\index{spectrum!gamma@gamma@$\gamma$-ray!illustration}

\end{figure}

Four $\gamma$-ray line events so far observed with \textit{RHESSI} provide new opportunities for such analyses: 
SOL2002-07-23T00:35 (X4.8) (see Figure \ref{lin_2003}) \citep[see][]{2003ApJ...595L..69L,2003ApJ...595L..81S,2003ApJ...595L..85S}, 
SOL2003-10-28T11:10 (X17.2) \citep{2004ApJ...615L.169S}; 
SOL2003-11-02T17:25 (X8.3) and 
SOL2005-01-20T07:01 (X7.1) \citep[see][for a review]{2006GMS...165..177S}.\index{flare (individual)!SOL2002-07-23T00:35 (X4.8)!ion primary spectrum}\index{flare (individual)!SOL2003-10-28T11:10 (X17.2)!ion primary spectrum}\index{flare (individual)!SOL2005-01-20T07:01 (X7.1)!ion primary spectrum}\index{flare (individual)!SOL2003-11-02T17:25 (X8.3)!ion primary spectrum}\index{gamma-rays!2.223~MeV line!intrinsic width}
\textit{RHESSI} has for the first time the energy resolution necessary to resolve all the $\gamma$-ray 
lines, except the intrinsically narrow 2.223~MeV line, and to determine the detailed line shapes 
expected from Doppler shifts, thus allowing deduction of velocity distributions of the interacting 
energetic ions \citep{2003ApJ...595L..81S}.  
The comparison of fluxes in the $^{20}$Ne line at 1.63~MeV 
and of the $^{12}$C and $^{16}$O lines at 4.44~and 6.13~MeV was performed for these four 
flares to provide information on the proton spectrum. 
Apart from SOL2002-07-23T00:35, for which the 
proton spectral slope was found to be $\sim$3.5 \citep{2003ApJ...595L..69L}, the other events have been reported to have much harder 
slopes ($\sim$2.5) \citep{2006GMS...165..177S}, 
harder than the average ($\sim$4.3) measured previously for the 19~\textit{SMM} flares. 
New analyses of the most recent events are currently being undertaken using the improved nuclear continuum modeling developed in \citet{2009ApJS..183..142M}, which could lead to softer proton spectra than was estimated in \citet{2006GMS...165..177S}.

There are also well-resolved \textit{INTEGRAL} spectra for three flares: SOL2003-10-28T11:10 (X17.2 \citep{2006A&A...445..725K}, SOL2003-11-04T19:53 (X17.4)
and SOL2005-09-07T09:52 (X17.0) \citep{2007A&A...461..723H}.
\index{flare (individual)!SOL2005-09-07T09:52 (X17.0)!gamma@$\gamma$-ray spectrum}
\index{flare (individual)!SOL2003-10-28T11:10 (X17.2)!gamma@$\gamma$-ray spectrum}
\index{flare (individual)!SOL2003-11-04T19:53 (X17.4)!gamma@$\gamma$-ray spectrum}
The proton spectral index was deduced for these events from the ratio of the 2.2~MeV 
line to the sum of the 4.44~and 6.13~MeV lines. 
Values of the power-law index in the range [3.5, -4] were found.
For SOL2003-10-28T11:10 (X17.2), observed both by \textit{RHESSI} and \textit{INTEGRAL}, the difference in the deduced power-law indices may result from the different methods used to infer the spectral index from observations. 
New analysis using improved nuclear line modeling \citep{2009ApJS..183..142M} could help to 
understand the apparent existing discrepancy.   

\subsection {Alpha particles}
\index{ions!alpha@$\alpha$ particles}

Both protons and $\alpha$s excite most of the strong de-excitation lines\index{gamma-rays!de-excitation lines}\index{ions!alpha@$\alpha$ particles}\index{ions!alpha@$\alpha/p$ ratio}.
Definitive information on the relative numbers of protons and $\alpha$s must be obtained by determining fluences of lines to which only one or the other species contribute\index{satellites!SMM@\textit{SMM}}\index{satellites!CGRO@\textit{CGRO}}.
Information on the $\alpha/p$ ratio had been deduced for five flares observed with \textit{SMM} \citep{1998ApJ...508..876S,1999ApJ...518..918M} from the fluence ratio of two lines: the prompt Fe~line at 0.847~MeV which is produced by the interaction of accelerated protons and $\alpha$s 
on ambient iron, and a pure $\alpha$ line at 0.339~MeV which only results from the interaction of energetic $\alpha$ particles on iron, producing an excited state of nickel. 
It was found by \citet{1999ApJ...518..918M} that for these five flares 
$\alpha/p$ exceeds the standard value of 0.1 and can even reach 0.5. 
Nuclear interactions of accelerated $^{3}$He with ambient $^{16}$O result in three $\gamma$-ray lines at 0.937, 1.04 and 1.08~MeV. 
While the fluence of the line at 0.937~MeV could be determined for a few flares observed by \textit{SMM}/GRS and \textit{CGRO}/OSSE, the other two lines cannot be separated from $\alpha$ lines at 1.05 and 1.00~MeV \citep{1998ApJ...508..876S}. 
\index{gamma-rays!unresolved feature at 1.02~MeV}
These lines combine to yield to an unresolved feature centered at 1.02~MeV. 
Using the information from this unresolved feature and from the line at 0.937~MeV, 
an estimate could be obtained of the ratio of accelerated $^{3}$He with respect to accelerated $^{4}$He. 
In 7 flares this led to an enhancement of this ratio (0.1 to 1) with respect to coronal values \citep{1999ApJ...518..918M}. 
In these analyses it was assumed that all accelerated species have the same energy distributions. 
As will be discussed in  Section \ref{content}, \cite{2004SoPh..223..155T} attempted to show how line fluences might be analyzed without this assumption. 
Finally, observations with good energy resolution help to reach stronger conclusions on these issues, and the  additional information from $\gamma$-ray line shapes also contributes, as discussed in the next section, to the determination of the $\alpha/p$ ratio. 

\subsection {Angular distributions of ions and $\alpha/p$ ratio as deduced from narrow line shapes}
\label{ls}
\index{ions!alpha@$\alpha/p$ ratio}

\begin{figure}
\begin{center}
\includegraphics[width=0.75\textwidth]{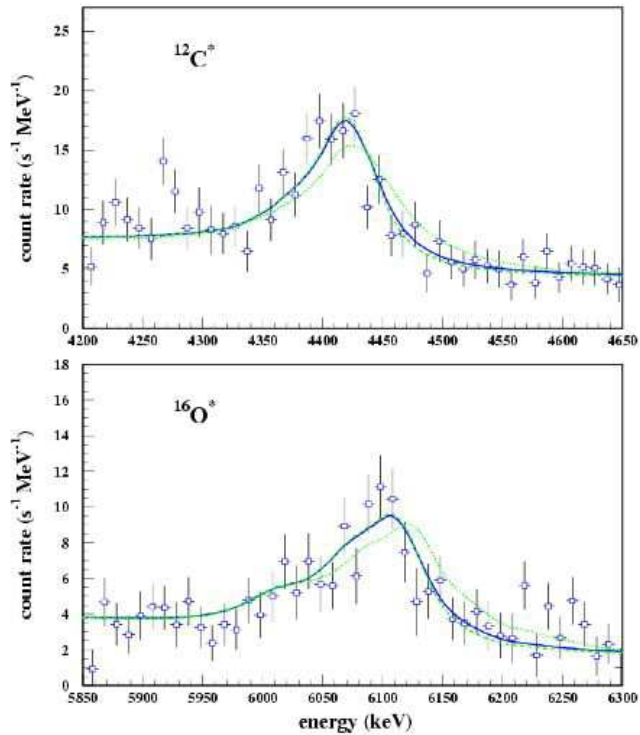}
\end{center}
\caption[]{Observed and calculated line shapes of the 4.44 and 6.13~MeV ambient $^{12}$C and 
$^{16}$O de-excitation lines observed by \textit{INTEGRAL}/SPI for SOL2003-10-28T11:10 (X17.2). The dashed line represents the calculated line shape for a downward isotropic distribution. The full and dotted lines represent the calculated line shapes for the distribution computed with pitch-angle scattering as in \cite{2007ApJS..168..167M} for $\lambda$=30 (full line), $\lambda$=300 (dotted line) \citep[from][]{2006A&A...445..725K}.}
\index{flare (individual)!SOL2003-10-28T11:10 (X17.2)!illustration}
\label{kiener}
\end{figure}

\begin{figure}
\begin{center}
\includegraphics[width=0.80\textwidth]{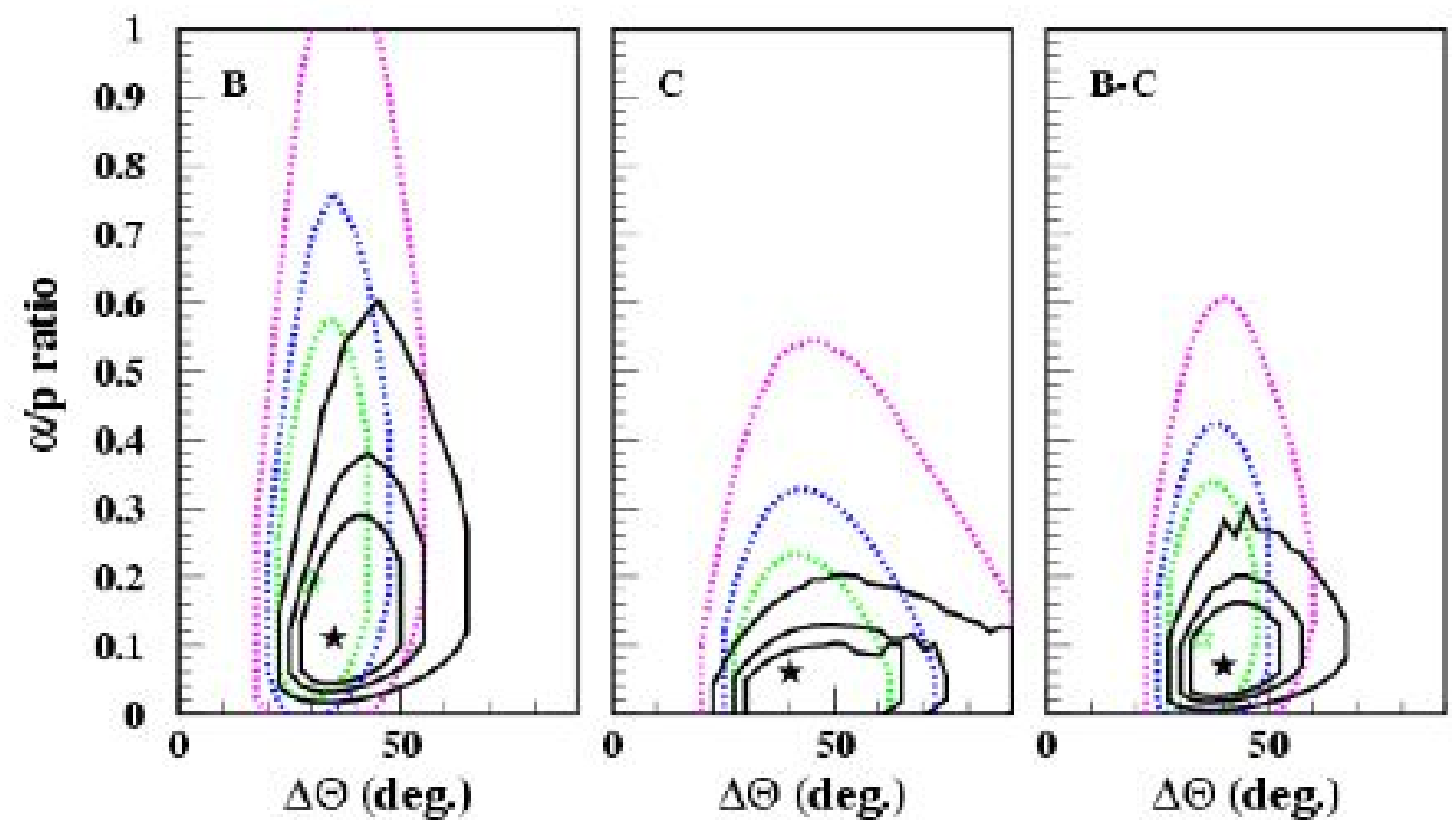}
\end{center}
\caption[]{Contour lines in the $\alpha/p$ -$\Delta$$\Theta$ (width of the injected particle distribution) parameter space of downward directed particle distributions for simultaneous fits to the 4.4 and 6.1 MeV lines for different phases of SOL2003-10-28T11:10 (X17.2). 
Confidence levels at 50\%, 70\%, 90\% are plotted for fits using an abundance of~C reduced by 50\% with respect to O as compared to the chromospheric abundance and for power-law particle spectra of $-3.0$ (dotted lines) and $-4.0$ (continuous lines). The best values of $\alpha/p$ and $\Delta$$\Theta$ for the two spectral indices are indicated by open ($s = -3.0$) and filled symbols ($s = -4.0$) \citep[from][]{2006A&A...445..725K}.
}
\label{kiener2}
\index{flare (individual)!SOL2003-10-28T11:10 (X17.2)!illustration}
\end{figure}

Profiles of narrow $\gamma$-ray lines can be used to derive the angular distributions of the emitting ions (in the interaction site) and of the accelerated ions, if the emission mechanisms are coupled with transport models\index{transport!ions}.
The nuclear excited states produced by collisions between accelerated and ambient particles have indeed very short lifetimes so that the decay $\gamma$-ray is emitted with a Doppler shift due to the recoil from the original collision. 
Furthermore, the collisions with the heavier $\alpha$ particles will produce greater recoil and should produce a broad ``red'' tail to the line shape. 
Fitting this tail, i.e., looking at the line shape, can constrain the $\alpha/p$ ratio independently of line fluxes and ratios. 
Accelerated nuclei heavier than He~also produce flare $\gamma$-ray lines 
but they are largely broadened and merge into a quasi-continuum $\gamma$-radiation see Section \ref{broad}). 
\index{continuum!gamma@$\gamma$-ray quasi-continuum}
The shape of the narrow $\gamma$-ray lines ultimately depends on the angular distribution and spectrum in the interaction site, the assumed $\alpha/p$ ratio, and the viewing angle of the observer 
\citep[see, e.g.,][]{1976ApJ...203..766R,1979ApJS...40..487R,1988ApJ...331.1029M,1990ApJS...73..349W,1991AIPC..232..445L,2001PhRvC..64b5803K,2007ApJS..168..167M}. 

Before the launch of \textit{RHESSI} and \textit{INTEGRAL}, line redshifts and widths had been studied for 19 flares using data from \textit{SMM} \citep{1995ApJ...452..933S,2002ApJ...573..464S}. 
Measurements of energies and widths of the strong flare $\gamma$-ray lines of C,~O~and Ne~were performed as a function of heliocentric angle using data with moderate spectral resolution provided by \textit{SMM}. 
Flares were grouped in five bins in heliocentric angle, with analysis performed on these sums of flares located in the same heliocentric interval (see Figure \ref{harris}). 
Redshifts of the Ne, C~and O~lines\index{gamma-rays!redshifts} of the order of~1\% are found 
for flares close to the disk center but no redshifts are found for flares close to the limb. 
The mean widths of the de-excitation lines were of the order of~3\% and did not exhibit any significant variation with heliocentric longitude\index{gamma-rays!de-excitation lines!mean widths}. 
It should be noted that the line shapes result from the $\gamma$-ray spectral analysis and are thus not completely independent of the way the analysis is performed. 
The observed redshifts are compared with predictions of line shapes in the case of angular distributions in the interacting sites being represented by a downward beam, a fan beam, and a downward or upward isotropic distribution.\index{ions!pitch-angle distributions}
Assuming a mean 
spectral index for ions of $\sim$4.2, $\alpha$/p=0.3 and Ne/O=0.25, \citet{2002ApJ...573..464S} concluded that the measured redshifts as a function of heliocentric angle are not consistent with beams in the interacting site but are rather consistent with a broad angular distribution in the downward direction.\index{beams!broad angular distribution}
These angular distributions are of course descriptive of the interaction sites and should be related to the angular distributions in the acceleration region itself by developing transport models \citep[see][]{2007ApJS..168..167M}.\index{transport!ions}\index{acceleration region!distinguished from energy-loss region}
 
Line widths were clearly observed for the first time in the flare SOL2002-07-23T00:35 (X4.8)
observed with \textit{RHESSI} \citep{2003ApJ...595L..81S,2003ApJ...595L..85S} (see Figure \ref{smith04})\index{gamma-rays!line shifts}.
\index{flare (individual)!SOL2002-07-23T00:35 (X4.8)!gamma@$\gamma$-ray line shifts}
\index{flare (individual)!SOL2003-10-28T11:10 (X17.2)!gamma@$\gamma$-ray line shifts}
\index{flare (individual)!SOL2003-11-03T09:55 (X3.9)!gamma@$\gamma$-ray line shifts}
\index{flare (individual)!SOL2003-11-02T17:25 (X8.3)!gamma@$\gamma$-ray line shifts}
It was found that redshifts varied as expected with the mass of the target species, but were larger than expected for line production by vertically precipitating ions in a flare at this location. 
The large shifts obtained rather suggest that the magnetic loops along which ions 
propagate are tilted with respect to the vertical (see Section~\ref{atmos}).
\index{magnetic structures!tilted loops!gamma@$\gamma$-ray evidence for}
Two other $\gamma$-ray line flares have been observed with \textit{RHESSI} for which line shapes have been analyzed (SOL2003-10-28T11:10 (X17.2) and SOL2003-11-02T17:25 (X8.3) \citep[see, e.g.,][]{2005AdSpR..35.1833G,2006GMS...165..177S}. 
The SOL2003-10-28T11:10 (X17.2) event was also observed at high spectral resolution with the \textit{INTEGRAL}/SPI spectrometer \citep{2004ESASP.552..669G,2006A&A...445..725K},
\index{flare (individual)!SOL2003-10-28T11:10 (X17.2)!INTEGRAL@\textit{INTEGRAL} observations}\index{satellites!INTEGRAL@\textit{INTEGRAL}} 
which provided high-resolution data for the 4.44~MeV $^{12}$C and 6.13~MeV $^{16}$O lines (see Figure \ref{kiener}). 
The comparison of the results of detailed calculations of line shapes to the observations places strong constraints on the $\alpha/p$ ratio \citep{2006A&A...445..725K}.  
In particular, the observed line width is inconsistent with an interpretation where too many $\alpha$ particles relative to the number of protons would be produced or otherwise the lines would be too broad when compared with the observations (Figure \ref{kiener})\index{inverse problem!for $\gamma$-ray observations}.
However, the line shapes depend on many parameters: the angular distribution of the interacting ions, the spectral index of the energetic ions, and the $\alpha/p$ ratio; these cannot be determined independently so the analysis results in extended regions of allowed parameter space. 

Using a spectral index between 3~and~4 deduced from the ratio of the 2.2~MeV to the fluence 
at 4.4~and 6.1~MeV \citep{2005astro.ph..1121T}, simultaneous fits of the C~and O~line shapes and line fluences have led \citet{2006A&A...445..725K} to favor the production of the $\gamma$-ray line emissions in the SOL2003-10-28T11:10 (X17.2) flare, by downward directed ion beams with a relatively low 
$\alpha/p$ ratio, around~0.1 (Figure \ref{kiener2}).\index{beams!downgoing}
\index{ions!alpha@$\alpha/p$ ratio}
High-resolution $\gamma$-ray line spectroscopy of two other weaker flares located at the limb was  also performed by \citet{2007A&A...461..723H} 
using \textit{INTEGRAL} data (SOL2003-11-04T19:53 and SOL2005-09-07T09:52; see also Section \ref{atmos}).
\index{flare (individual)!SOL2003-11-04T19:53 (X17.4)!INTEGRAL@\textit{INTEGRAL} observations}\index{flare (individual)!SOL2005-09-07T09:52 (X17.0)!INTEGRAL@\textit{INTEGRAL} observations}

In addition to the line width of the narrow flare $\gamma$-ray lines discussed above, the profile of $\alpha$/$^4$He  fusion lines around  0.452~MeV ($\alpha$/$\alpha$ lines) may also provide a potential diagnostic of the angular distributions of energetic particles in the interaction site.\index{ions!nuclear reactions!alpha@$\alpha$/$\alpha$}\index{ions!nuclear reactions!fusion}\index{reactions!nuclear!$\alpha$/$\alpha$}
From \textit{SMM} observations of two flares, it had been shown that the line shape 
of the $\alpha$/$^4$He fusion line was inconsistent with its production by a downward beam and was rather consistent with an isotropic or a fan-beam distribution (i.e., as obtained at a 
magnetic mirror point) in the interacting site \citep{1995ApJ...452..933S,1998ApJ...508..876S}. This is consistent with what was found by \citet{2002ApJ...573..464S} by analyzing line shapes for \textit{SMM} flares.\index{ions!pitch-angle distributions}

\begin{figure}
\begin{center}
\includegraphics[width=0.75\textwidth]{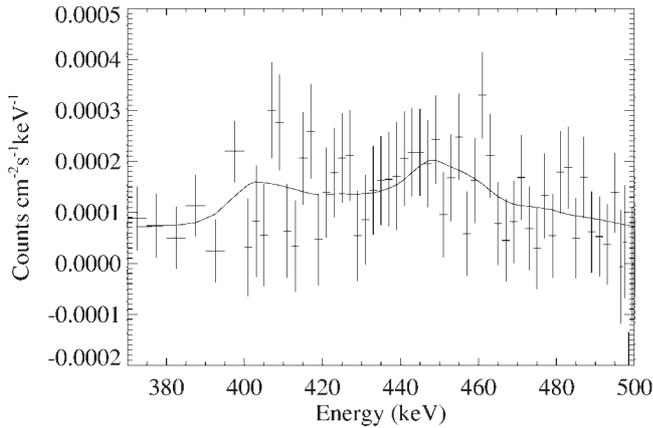}
\end{center}
\caption[]{Count spectrum of the $\alpha$-$^{4}$He line measured by \textit{RHESSI} for the SOL2002-07-23T00:35 (X4.8) event after subtraction of the other spectral components. 
The curve shows the \textit{RHESSI} response to the line shape in the case of an $\alpha$ particle 
distribution with saturated pitch-angle scattering (see Section \ref{shmu}) propagating in magnetic loops perpendicular to the solar surface at a heliocentric angle of 73$^{\circ}$ \citep[from][]{2003ApJ...595L..89S}.
}
\label{share_alpha}
\index{spectrum!alpha@$\alpha$-$^4$He line!illustration}
\index{flare (individual)!SOL2002-07-23T00:35 (X4.8)!illustration}
\end{figure}

Since the launch of \textit{RHESSI}, high-resolution spectra of the $\alpha$/$\alpha$ lines have been obtained for the SOL2002-07-23T00:35 (X4.8) (Figure \ref{share_alpha}) and SOL2003-10-28T11:10 (X17.2) events \citep{2003ApJ...595L..85S,2004ApJ...615L.169S}.
 For these two flares the shape of the line complex is also found to be consistent with a downward isotropic distribution. 
\index{flare (individual)!SOL2002-07-23T00:35 (X4.8)!and ion pitch-angle distribution}
\index{flare (individual)!SOL2003-10-28T11:10 (X17.2)!and ion pitch-angle distribution}

\subsection {Energy content in ions}
\index{accelerated particles!energy content of}
\label{content}

De-excitation $\gamma$-ray lines provide information about ions above $\sim$1~MeV/nucleon. 
\index{gamma-rays!de-excitation lines}
\index{ions!total energy}
The energetic ion distribution below this energy is essentially unknown but is of particular interest for the total ion energy content (see Section \ref{lep}). 
Since the first detection of 
$\gamma$-ray lines in 1972, quantitative results from $\gamma$-ray spectroscopy have been obtained for more than 20~events, yielding information on the energy content in ions. 
Several ratios of pairs of nuclear lines have been used to infer the ion energy spectrum and then the ion energy content: $\Phi_{2.223}$/$\Phi_{4-7}$ and $\Phi_{1.63}$/$\Phi_{6.13}$ \citep{1996AIPC..374..172R}. 
As an example, the threshold energies for excitation of Ne~and O~lines by fast protons are significantly different, $>$2 and $>$8~MeV, respectively, which allows the use of the ratio of their measured fluxes to infer the ion energy distribution 
\citep{1995ApJ...452..933S}. 
This determination is not independent of the assumed target abundances but combining results from
several such methods reduces some of this uncertainty.  
The comparison of the steepness of the ion spectra deduced from the different ratios, $\Phi_{2.223}$/$\Phi_{4-7}$ and $\Phi_{1.63}$/$\Phi_{6.13}$ for different flares, shows indeed some consistency only when an enhanced Ne abundance is assumed, $Ne/O = 0.25$ instead of the more standard value of 0.14 \citep{1996AIPC..374..172R} (see also Section \ref{abundances}). 
As most of the narrow lines can be produced by energetic protons and $\alpha$s, the usual hypothesis is  that the energy distributions of accelerated 
protons and $\alpha$s expressed in energy/nucleon are identical. 
The relaxation of this assumption will be discussed below.  
The determination of the  proton spectrum from the ratio of the excitation lines from Ne~and O~depends on the assumption made on the ratio of Ne/O in the flaring region and on the ratio $\alpha/p$ of fast $\alpha$ particles to energetic protons. It is indeed found in \citet{1995ApJ...455L.193R} that steeper ion spectra are needed to reproduce the same ratio of line fluences for smaller values of Ne/O and larger values of the 
$\alpha$/p\index{ions!alpha@$\alpha/p$ ratio} ratio.
The analysis based on the 19 events observed by \textit{SMM}/GRS led to the conclusion that the energy contained in $>$1~MeV ions lies in the range 
10$^{29-32}$~erg for $\gamma$-ray line flares (see Ramaty et al. 1995, 1996, and the summary by Miller et al. 1997).
\nocite{1995ApJ...455L.193R,1996AIPC..374..172R,1997JGR...10214631}
These estimations were done under the following assumptions: the slope of the ion power-law spectrum is deduced from the ratio of the Ne/O line assuming thick-target emission and values of Ne/O=0.25, $\alpha/p$ =0.1 and ``impulsive'' composition for other species (i.e. increase of Ne/O, Mg/O, Si/O, and S/O by a factor of $\sim$3 and by a factor of $\sim$10 for Fe/O with respect to coronal composition for the accelerated particles)\index{thick-target model!gamma@$\gamma$-rays}\index{abundances!assumptions in $\gamma$-ray analysis}.
The power-law spectrum is extended down to 1~MeV with a flat extension to lower energies. 
For the giant flares SOL1991-06-01T16:14 (X12) and SOL1991-06-04T03:37 (X12.0), observed either by \textit{GRANAT}/PHEBUS and \textit{CGRO}/OSSE, the energy contained in 1~MeV/nucleon ions has also been deduced and is found to be in the 10$^{32}$-10$^{33}$ ergs range \citep{1997ApJ...479..458R,1997ApJ...490..883M}\index{flare (individual)!SOL1991-06-04T03:37 (X12.0)!total ion energy}\index{flare (individual)!SOL1991-06-01T16:14 (X12)!total ion energy}.
For SOL1991-06-04T03:37, the total energy contained in accelerated protons is 
$1.7 \times 10^{32}$~erg  and the total energy contained in accelerated ions is $1.0 \times 10^{33}$~erg under the following assumptions: impulsive flare abundances for accelerated particles, identical energy dependence for all ion spectra, and $\alpha$/p~=~0.5.

The relaxation of the assumption of similar proton and $\alpha$ energy spectra in MeV/nucleon has been studied by \cite{2004SoPh..223..155T} and applied to the estimates of the ion energy content in flares\index{ions!nuclear reactions!alpha@$\alpha$/$\alpha$}\index{ions!total energy}.
As some of the narrow lines (in particular the $\alpha$-$\alpha$ lines at 0.429 and 0.478~MeV) result only from the interaction of energetic $\alpha$s with the solar atmosphere, the measured fluences of these lines can be used to infer the number of 
$\alpha$ particles $>$1~MeV/nucleon once the steepness ($\delta_\alpha$) of the $\alpha$ spectrum has been chosen.
\index{spectrum!alpha@$\alpha$ particles}
This allows a computation of the flux of the narrow lines at 1.63~MeV and 6.13~MeV produced by the 
$\alpha$s. 
The remaining fluences in the narrow lines should be produced by fast protons. 
The ratio of the remaining fluences is then used to determine the 
steepness ($\delta_p$)~of the proton spectrum and then the number of fast protons required to produce the remaining fluences. 
Note that some values of the steepness of the $\alpha$ spectra (excessively hard ones) 
can be excluded directly by the fact that the fluences of the produced narrow lines at 1.63~and 6.13~MeV would exceed the observed ones. 
The reexamination of the spectra and numbers of particles deduced from \textit{SMM}/GRS 
spectra of four flares chosen in the list of \citet{1995ApJ...452..933S} shows that harder spectra for 
$\alpha$ particles than for protons are generally deduced, that most of the energy still goes to protons (by a factor of 100~to 1000) and that as a whole less energy goes to ions. 
As an example, for the well~studied SOL1981-04-27T09:45 (X5.5) event, a total ion energy of 1.6~$\times$~10$^{31}$~erg is found with $\delta_\alpha$ =2.5, $\delta_p$=4.4, $\alpha/p$=0.004 from the method of \cite{2004SoPh..223..155T}, while the usual analysis (using the ratio of the fluences of the 1.63~MeV and 6.13~MeV lines) would lead, for the usually chosen value of $\alpha/p$= 0.5, to a total ion energy of 5.5~$\times$~10$^{31}$~erg  with an $\alpha$~and proton spectral index of~5.2\index{flare (individual)!SOL1981-04-27T09:45 (X5.5)!total ion energy}.
This discussion clearly shows the present limits and the uncertainties in deducing ion energy content from $\gamma$-ray line spectroscopy\index{inverse problem!for $\gamma$-ray problems!limitations}\index{caveats!inverse theory for $\gamma$-rays}.
It must finally be noted that in the analysis developed in \citet{2004SoPh..223..155T} the meaning of $\alpha/p$ is different from the standard one when both ion spectra are similar. 
In the case of different ion spectra, the value of $\alpha/p$ which is indicated above is the total number of $\alpha$ particles $\geq$1~MeV/nucleon to the total number of protons above 1~MeV. 

Other effects can also alter the amount of energy estimated for accelerated ions as 
deduced from $\gamma$-ray line spectroscopy. 
As stated above, the production of the $\gamma$-ray line emission is 
usually estimated under the assumption of thick-target production\index{flare (individual)!SOL2003-10-28T11:10 (X17.2)!illustration}.
This means that the energy loss rate of the radiating ions is also a parameter (even if well hidden) of the $\gamma$-ray line production. 
Most of the models of $\gamma$-ray line production in flares assume that the target in which the fast ions interact and produce lines is a ``cold'' neutral target (i.e., energetic particle speeds much larger than the target particle speeds). 
This may no longer be true if ions with energies close to the threshold energies for $\gamma$-ray line production interact in a hot flaring plasma (temperature of a few 10$^7$ K), i.e., if the $\gamma$-ray line emission site is partly coronal (see Section~ \ref{img} for some discussion on possible coronal sites  of $\gamma$-ray line emission). 
\citet{1997ApJ...485..430E}, and more recently \citet{2003A&A...409..745M}, have examined the effects of introducing a ``warm target'' for the production of $\gamma$-ray lines\index{thick-target model!warm target}.
In such a warm medium, fast ions with energies just above the threshold for nuclear line production will lose less energy to collisions than in a cold medium. 
This has the effect of increasing the efficiency of line production by ions just above the threshold. 
This increase is counterbalanced by the fact that the energy losses in an ionized (because hot) medium are larger than in a neutral medium.
\index{ions!interaction in warm medium}
\citet{2003A&A...409..745M} estimated the strength of the Ne~1.63~MeV line relative to the strength of the O~6.13~MeV line, used to estimate ion spectra, as a function of the temperature of the target and as as function of the $\alpha/p$ ratio, i.e. assuming as in most works a similar spectral shape for energetic protons and $\alpha$s in MeV/nucleon. 
They show that in a warm target, there is an 
increase of the relative production of the Ne~1.63~MeV line to the O~6.13~MeV line. 
This increase is more important for steeper spectra and for lower values of $\alpha/p$ (this simply illustrates the fact that the Ne line cross-sections have much lower thresholds for proton production than for $\alpha$ production). 
Flatter ion spectra and thus lower ion energy contents are then required to reproduce the observed line fluences ratio than usually deduced (a reduction by more than a factor of $\sim$10 of the total energy contained in protons and $\alpha$s can be achieved if the emission site has a temperature of $6 \times 10^7$~K and $\alpha/p$ is~0.1; see MacKinnon \& Toner, 2003).
 \nocite{2003A&A...409..745M}
 \index{accelerated particles!energy content of!reduction in warm medium}
Taking into account the effects of a warm target could finally reduce the discrepancies which were 
found in the ion spectral indices deduced from the different line ratio $\Phi_{2.223}$/$\Phi_{4-7}$ and $\Phi_{1.63}$/$\Phi_{6.13}$ for a standard value of Ne/O\index{thick-target model!warm target}\index{abundances!Ne/O ratio!and warm target}.
Indeed, even with a standard value of the Ne/0 abundance ratio, the inclusion of a warmer emission site leads to harder ion spectra to produce the same  $\Phi_{1.63}$/$\Phi_{6.13}$ \citep{2003A&A...409..745M}.

Observations performed by \textit{RHESSI} brought information on the ion energy content for three more $\gamma$-ray line events: 
SOL2002-07-23T00:35 (X4.8) \citep{2003ApJ...595L..69L,2004JGRA..10910104E,2007A&A...468..289D}, 
SOL2003-10-28T11:10 (X17.2) \citep{2005AdSpR..35.1833G,2006GMS...165..177S}, and
SOL2003-11-02T17:25 (X8.3)  \citep{2005AdSpR..35.1833G,2006GMS...165..177S}.
\index{flare (individual)!SOL2002-07-23T00:35 (X4.8)!ion energy content}
\index{flare (individual)!SOL2003-10-28T11:10 (X17.2)!ion energy content}
\index{flare (individual)!SOL2003-11-02T17:25 (X8.3)!ion energy content}
\index{ions!energy content}
Using a procedure similar to the one which has been used previously for \textit{SMM}/GRS and \textit{CGRO} events by \citet{1997ApJ...490..883M} and \citet{1997ApJ...479..458R}, \citet{2003ApJ...595L..69L} found a minimum total energy in accelerated protons above 2.5~MeV of 1.4~$\times$~10$^{30}$ ergs (10$^{31}$~erg for all ions) for SOL2002-07-23T00:35 (X4.8)  (assuming $\alpha/p$=0.5). 
Consistent values of the total energy in protons and $\alpha$~particles above 2.5 MeV/nucleon were found by \citet{2007A&A...468..289D} from the modeling of the $\gamma$-ray line time profiles. 
Values of the proton energy content derived for SOL2003-10-28T11:10 (X17.2) are consistently found to be around $1.5 \times 10^{31}$ ergs by \citet{2005AdSpR..35.1833G} and by \citet{2006GMS...165..177S}. 
For SOL2003-11-02T17:25 (X8.3) the proton energy content ranges from 
0.5~$\times$~10$^{31}$ ergs to 4.2~$\times$~10$^{31}$~erg depending on the value of the proton spectral index used ($3-3.6$).
This clearly shows the still-large uncertainties in the determination of these quantities. 

\subsection {Ion and electron acceleration in flares and variability}

Using the accelerated ion and electron spectra derived from $\gamma$-ray and X-ray spectroscopy, 
it is possible to compare the energies contained in flare-accelerated ions and electrons\index{accelerated particles!energy content, ions vs. electrons}. 
Such a study performed by \citet{2000IAUS..195..123R} for the~19 $\gamma$-ray line flares observed by \textit{SMM} showed that 
the energy contained in $>$1~MeV ions may be comparable to the energy contained in subrelativistic electrons ($>$20 keV)\index{electrons!relativistic!energy content}.
\citet{2004JGRA..10910104E} obtained a similar result for the first $\gamma$-ray line flare observed by \textit{RHESSI}. At present, such comparisons of subrelativistic electron and ion energy contents have not been performed for more flares. The published comparisons show that there is a large dispersion of the relative electron ($>$20 keV) and ion ($>$1~MeV/nucleon) energy contents from one flare to the other, the energy contained in the ions sometimes exceeding 
the energy in the electrons. 
However, it must be recalled that all these estimations strongly depend on the low energy cut-off of 
both energetic electrons and ions which is for electrons not always easy to determine
\citep[see the discussion in][]{Chapter3}\index{accelerated particles!energy content of!caveat}\index{caveats!particle energy content}.

\begin{figure}
\begin{center}
\includegraphics[width=0.75\textwidth]{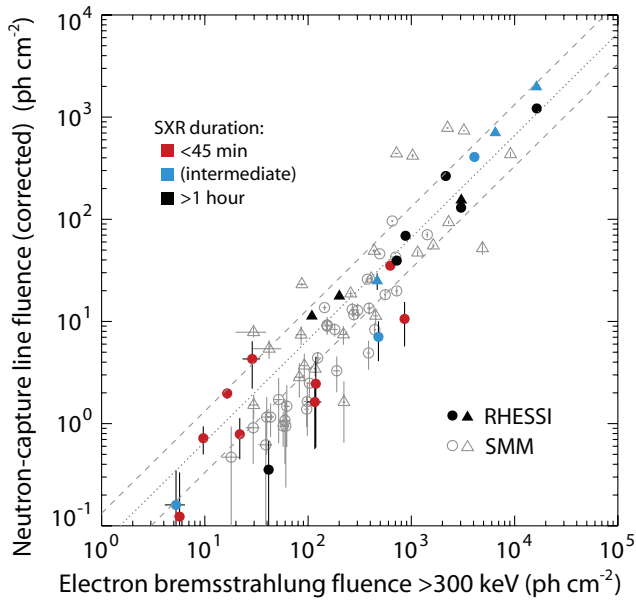}
\end{center}
\caption[]{Neutron-capture line fluence as a function of the electron bremsstrahlung fluence above 300~keV for flares with heliocentric angle $<$80$^{\circ}$  (\textit{RHESSI} in solid symbols, \textit{SMM} in open symbols). 
Circles (triangles) represent flares with complete (incomplete) coverage. 
Colors represent the \textit{GOES} soft X-ray (SXR) duration in three bins. 
The dotted line indicates the best fit in linear space that passes through the origin with a ratio of 0.066, and the dashed lines have slopes that differ by factors of $\sim$2 from the best-fit line \citep[from][]{2009ApJ...698L.152S}.
}
\label{shih}
\end{figure}

Another way of dealing with the question of the relative ion and electron acceleration in solar flares is to investigate the 
relationship between ion acceleration diagnosed by $\gamma$-ray line fluences and relativistic electron acceleration diagnosed by the $>$300 keV bremsstrahlung radiation. 
\index{accelerated particles!electron-to-proton ratio}
A first comparison was carried out on \textit{SMM}/GRS data between the $>$300~keV bremsstrahlung radiation and the excess rate in the 4-8~MeV range (i.e., the rate above the bremsstrahlung continuum which is attributed to the production of $\gamma$-ray lines by ions (see Figure~\ref{fig:Share00})) showing that there is a good correlation between ion production (normalized above 30~MeV/nucleon) and electron production above 300~keV \citep[see e.g.][]{1983AIPC..101....3F,1984ARA&A..22..359C}. 
The relationship between energetic ion and relativistic electron acceleration can also be studied by comparing the 2.223~MeV line fluence to the $>$300~keV X-ray emission (see, e.g., Murphy et al., 1993).
\index{gamma-rays!2.223~MeV line!and $>$300~keV fluences}
Since the launch of \textit{RHESSI}, 29 flares have been observed with significant $>300$~keV continuum fluence. 
The fluences of the 2.223~MeV lines have been estimated for these flares from spectral analysis 
and corrected for attenuation for flares close to the limb\index{gamma-rays!2.223~MeV line!limb darkening}\index{electrons!relativistic!correlation with 2.223~MeV $\gamma$-rays}.
The neutron-capture line fluences are found to be highly correlated to the $>$300~keV bremsstrahlung fluences for all the flares located at heliocentric angles less than 80$^{\circ}$ and over a range of more than three orders of magnitude in fluence.  
A similar correlation was found for the flares previously observed by \textit{SMM}. 
The ion energy range which produces the 2.223~MeV line depends in fact on the ion spectral index and on the accelerated and ambient abundances (see Section \ref{eff} and Murphy et al., 2007).
 \nocite{2007ApJS..168..167M}
As a first approximation, this line can be considered as a good indicator of 
the number of $>$30~MeV protons and its fluence is taken as being proportional to the number of protons above this energy\index{bremsstrahlung!$>$300~keV, correlation with $\gamma$-rays}.
When both \textit{SMM} and \textit{RHESSI} data sets are taken into account, the strong correlation obtained between the neutron-capture line fluence (corrected for attenuation) and the bremsstrahlung fluence ($>$300~keV) (see Figure~\ref{shih}) clearly shows a link between the total energy content 
in protons above 30~MeV and the total energy content in electrons above 300 keV \citep{2009ApJ...698L.152S}. 
This suggests that high-energy electrons and ions are directly linked by acceleration processes.
This may be a more direct link than for subrelativistic electrons and ions, given the larger variation between electron energy contents above 20~keV and ion energy contents above 1~MeV/nucleon discussed above. 
This is a constraint which should be borne in mind when discussing acceleration mechanisms in solar flares.

\index{accelerated particles!correlation of ions and relativistic electrons!illustration}

The correlation found by \citet{2009ApJ...698L.152S} between ion and relativistic electron acceleration seems to indicate that energetic ions are produced as soon as there is a significant production of energetic electrons above, e.g., 300~keV.
\index{flare types!``electron-dominated''} 
\index{accelerated particles!correlation of ions and relativistic electrons}
From this correlation, there does not appear to be a distinct class of electron-dominated flares
that produce much more high-energy bremsstrahlung radiation relative to nuclear emission \citep[e.g.,][]{1998SoPh..183..123R} although some episodes of the flares reported by \citet{2009ApJ...698L.152S} may exhibit higher relative numbers of electrons with respect to ions than the mean value over several flares as well as over the whole event. 
Extreme cases of the variability of the electron bremsstrahlung component to the nuclear line component have been reported in the \textit{SMM} era during short duration (a few seconds to a few tens of seconds) transient bursts observed above 10~MeV at any time during a flare. They are referred to as electron-dominated events (Rieger, 1994) 
and are characterized by weak or no detectable $\gamma$-ray line emission and by hard $\geq$1~MeV electron bremsstrahlung spectra\index{gamma-rays!variability relative to relativistic bremsstrahlung}\index{bremsstrahlung!independence of $\gamma$-ray line emission}\index{satellites!GAMMA-1@\textit{GAMMA-1}}.
They were first reported from \textit{SMM}/GRS observations \citep[e.g.,][]{1991max..conf...68R}, and were afterwards observed by \textit{GAMMA-1} \citep{1992SvAL...18...69A,1993A&AS...97..345L}, PHEBUS and SIGMA experiments aboard \textit{GRANAT} \citep{1992SoPh..140..121P,1994LNP...432..197V},
\textit{CGRO} \citep[e.g.,][]{1994AIPC..294..177D} and \textit{Yohkoh} \citep{1999ICRC....6....1Y}.
The spectral analysis of 12 electron-dominated events observed by \textit{SMM}/GRS \citep{1998SoPh..183..123R} confirmed the hardness of the bremsstrahlung spectra above 1~MeV (mean value of the power-slope around $-1.84$). 
The mean value of the spectral index between 0.3 and 1~MeV ($\sim$2.7) does not differ significantly from that of other flares. 
However, the apparent lack of $\gamma$-ray line emission does not rule out a simultaneous production of relativistic electrons and ions \citep{1998A&A...334.1099T,1999A&A...342..575V} 
for the electron-dominated events\index{flare types!``electron-dominated''} observed \textit{GRANAT}/PHEBUS. 
Indeed, if one assumes that the energy content in ions above 1~MeV is similar to that contained 
in electrons (i.e., around a few 10$^{29}$ ergs for the cases studied by \cite{1998A&A...334.1099T} and \cite{1999A&A...342..575V}, as appears to be the case for $\gamma$-ray line flares, one finds that no detectable $\gamma$-ray line fluence could be observed, given the hardness of the electron bremsstrahlung component and the limited spectral resolution (and thus line sensitivity) of the 
PHEBUS experiment. 
The only remaining question would then be to understand why the spectrum of accelerated electrons is so much harder in these events than at most other times.\index{satellites!GAMMA-1@\textit{GAMMA-1}}\index{satellites!SMM@\textit{SMM}}\index{satellites!GRANAT@\textit{GRANAT}}\index{satellites!CGRO@\textit{CGRO}}\index{satellites!Yohkoh@\textit{Yohkoh}}

Variability of the accelerated electron-to-proton ratio not only occurs from flare to flare as discussed above, but also on time scales of tens of seconds within individual events. \citet {1993A&A...275..602C} reported, e.g., for an \textit{SMM} event a variation of the relative number of electrons above 500 keV and of ions above 30~MeV on time scales of tens of seconds. \cite{2007A&A...468..289D} also reported variation of the relative number of electrons above 150~keV and of the number of energetic protons and $\alpha$ particles above 1~MeV/nucleon in the evolution of SOL2002-07-23T00:35 (X4.8) observed by \textit{RHESSI}\index{accelerated particles!electron-to-proton ratio}\index{flare (individual)!SOL2002-07-23T00:35 (X4.8)!electron-to-proton ratio}.
These variations were deduced from the analysis and modeling of time delays between $\gamma$-ray line emissions and hard X-rays\index{hard X-rays!timing relative to $\gamma$-rays}.
A question of obvious interest would be to determine whether these changes are connected to spatial source evolution. Such variations of the ratio of accelerated electrons to accelerated protons should also be taken into account when discussing acceleration processes.  

Another interesting result follows from the statistical analysis of \citet{2009ApJ...698L.152S}.
For eight of the flares with the largest 2.223~MeV line fluence, a correlation is found between the 2.223~MeV line fluence (i.e., the number of $>$30 MeV protons) and the fluence above 50~keV as well as with the \textit{GOES} class\index{satellites!GOES@\textit{GOES}}.
However, this correlation is found only when a threshold in the production of ions is reached. 
A similar relationship had been discussed at the time of \textit{SMM} by \citet{1994ApJ...426..767C}. 
This ultimately suggests that while the acceleration of protons above 30~MeV is closely related to the acceleration of relativistic electrons, the acceleration of subrelativistic electrons is only proportional to the acceleration of relativistic electrons and ions when a given threshold of high energy particles is reached. 
This may suggest two acceleration processes, one producing proportional quantities of relativistic electrons and ions and the other one producing mostly subrelativistic electrons. 

\subsection{Long-lived radioactivity from solar flares}
\index{radioactivity}

Interactions of flare-accelerated ions with the solar atmosphere can 
synthesize radioactive nuclei, whose decay can produce observable, delayed 
$\gamma$-ray lines in the aftermath of large flares. 
The detection of these delayed lines would provide new insights into the spectrum and composition of the flare fast ions. 
Many of these radioisotopes would be produced predominantly by
interactions of fast heavy ions with ambient hydrogen and helium, especially since
accelerated heavy nuclei are believed to be significantly enhanced
compared to the ambient medium composition \citep[e.g.,][]{1991ApJ...371..793M}.
\index{radioactivity!created by flares}\index{gamma-rays!long-lived radioisotopes}

One of the most promising of such lines is at 
0.847~MeV\index{gamma-rays!0.847 MeV $^{56}$Fe line} resulting from the first excited state of $^{56}$Fe which in turn has been produced via the decay of $^{56}$Co (half-life $T_{1/2}$=77.2~days)
\citep{2000IAUS..195..123R,2002ApJS..141..523K,2006ApJS..165..606T}.
Another line produced in this decay is at 1.238~MeV\index{gamma-rays!1.238 MeV $^{56}$Fe line}
\citep{2007AAS...210.6804S}.  
For a few days after a large flare however, the most intense delayed line is 
predicted to be the 511~keV
positron-electron annihilation line resulting from 
the decay of several long-lived $\beta^+$ radioisotopes. 
\index{gamma-rays!511 keV line!as radioactive decay product} 
\citet{2006ApJS..165..606T} have pointed out other delayed $\gamma$-ray lines that appear to be promising for detection, e.g., at 1.434~MeV from the radioactivity of both the isomer $^{52}$Mn$^m$ ($T_{1/2}$=21.1~min) and the ground state $^{52}$Mn$^g$ ($T_{1/2}$=5.59~days), 1.332 and 1.792~MeV from $^{60}$Cu ($T_{1/2}$=23.7~min), 1.369 and 2.754~MeV from $^{24}$Na 
 ($T_{1/2}$=15.0~hours), and 0.931~MeV from $^{55}$Co ($T_{1/2}$= 17.5~hours). 
The delayed lines should be very narrow, because the radioactive nuclei are 
stopped by energy losses in the solar atmosphere before they decay. 
Unless the flare is very close to the solar limb, the $\gamma$-ray line attenuation in the 
solar atmosphere should not be significant \citep[see][]{1989ApJ...341..516H}, 
as long as the radioactive nuclei do not plunge deep into the solar convection zone. 
Searches for upper limits of the radioactivity line at 1.238~MeV were performed with 
\textit{RHESSI} \citep {2007AAS...210.6804S} for periods after the large flares of July 2002, October-November 2003 and January 2005. 
No strong conclusion has been drawn so far.

In addition to $\gamma$-ray lines emitted from de-excitation of daughter nuclei, 
radioactive X-ray lines can be produced from the decay of proton-rich 
isotopes by orbital electron capture or the decay of isomeric nuclear levels 
by emission of a conversion electron\index{gamma-rays!de-excitation lines}.
The strongest delayed X-ray line is predicted to be the Co~K$\alpha$ at 6.92~keV \citep{2006ApJS..165..606T}, which is produced from both the decay of the isomer $^{58}$Co$^m$ 
($T_{1/2}$=9.04~hours) by the conversion of a K-shell electron and the decay of 
$^{57}$Ni ($T_{1/2}$=35.6~hours) by orbital electron capture\index{gamma-rays!delayed emission!6.92~keV Co~K$\alpha$}.
The attenuation of this line by photoelectric absorption in the solar atmosphere should be 
$<$10\% for flares occuring at low heliocentric angles, as long as the 
radioisotopes produced in the chromosphere and upper photosphere are not 
transported to greater depths.\index{transport!radioisotopes}\index{absorption!solar atmosphere!Co~K$\alpha$}
Distinguishing this atomic line from the thermal X-ray emission can be challenging until the flare plasma has significantly cooled down. 
However, a few hours after the flare end time the thermal emission will be gone, or significantly reduced, and the delayed Co~K$\alpha$ line will be more easily detected. 

\begin{figure}
\begin{center}
\includegraphics[width=0.75\textwidth]{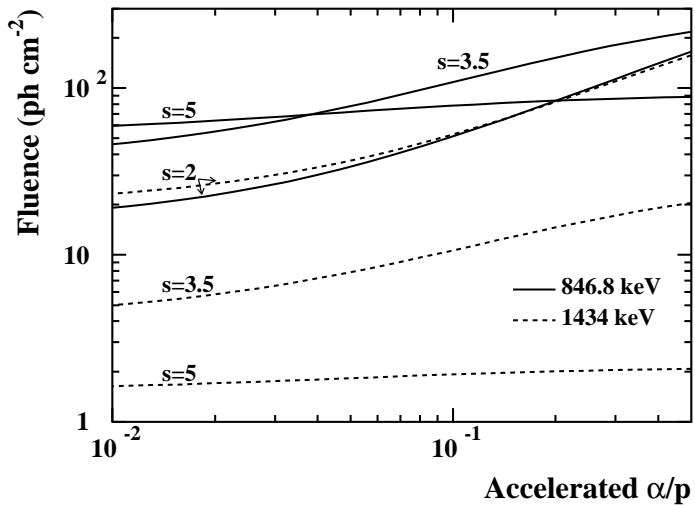}
\end{center}
\caption[]{Total fluences of the delayed lines at 0.847~MeV 
({\it solid curves}) and 1.434~MeV ({\it dashed curves}) as a function of the 
accelerated $\alpha/p$ abundance ratio (see text), for spectral indices of the
accelerated ion power-law spectrum $s$=2, 3.5 and 5. 
The calculations are normalized to a total fluence of 300 photons~cm$^{-2}$ emitted during the 
$\gamma$-ray flare in the sum of the 4.44 and 6.13~MeV ambient $^{12}$C and 
$^{16}$O de-excitation lines \citep[from][]{2006ApJS..165..606T}.
}
\label{fig:tatischeff_fig1}
\end{figure}

To illustrate the potential of these lines for revealing details of the heavy 
ion energy distribution, we show in Figure~\ref{fig:tatischeff_fig1} 
calculated fluences of the 0.847~MeV and 1.434~MeV lines as a function of the 
accelerated $\alpha/p$ abundance ratio. 
The calculations were normalized to a total fluence of the summed 4.44 and 6.13~MeV prompt narrow lines of 300 photons~cm$^{-2}$, which is the approximate fluence observed in the 
SOL2003-10-28T11:10 (X17.2) flare\index{flare (individual)!SOL2003-10-28T11:10 (X17.2)!gamma@$\gamma$-ray fluences}.
We assumed for the abundances of fast~C and heavier elements 
relative to $\alpha$-particles the average composition of solar energetic 
particles measured in impulsive events \citep{1999SSRv...90..413R}. 
Thus, the predicted fluence variations with accelerated $\alpha/p$ actually show the 
relative contributions of reactions induced by fast protons. 
The fluences decrease for decreasing $\alpha/p$ ratio (i.e., increasing proton abundance), 
because, for $\alpha/p$ $>$ 0.05, the radioisotopes are predominantly 
produced by spallation of accelerated Fe, whereas the ambient 
$^{12}$C and $^{16}$O lines largely result from fast proton interactions\index{ions!nuclear reactions!spallation}.
This effect is less pronounced for $s$=5, because for this very soft spectrum, the 
contribution of $\alpha$-particle reactions to the prompt line emission is more 
important. 
Thus, a concomitant detection of the 0.847~MeV and 1.434~MeV lines would 
allow measurement of not only the abundance of accelerated ions, but also 
of their energy spectrum. 

As well as serving as diagnostics of the flare fast ion population, the 
radioisotopes can serve as tracers to study solar atmospheric mixing 
\citep{2000IAUS..195..123R}. 
A measurement of the decay profile with time of several delayed lines would place very useful constraints on the extent and timescale of mixing processes in the outer convection zone.
A future observation of the size and development of the radioactive patch on 
the solar surface would furthermore provide unique information on both the 
transport of flare-accelerated particles and dynamics of solar active regions. 
Note that solar radioactivity can be the only way to study flares that had 
recently occurred over the east limb\index{radioactivity!and convection}.

\subsection {Low-energy protons}
\label{lep}
\index{accelerated particles!low-energy protons}
\index{protons!low energies}

The total energy content of a power-law particle energy distribution, $F(E) \propto E^{-s}$, is dominated by the lowest energy $E$ for which this form holds (as long as $s > 2$)\index{ions!energy content}. 
The low-energy form of the distribution is thus important for assessing the ions' overall importance in flare energetics and energy transport. 
Power-law ion distributions measured in space appear unbroken to 0.02~MeV/nucleon \citep{1997ApJ...483..515R}. 
If this were to be true also at the flare site, extrapolation to these low energies of the ion distributions deduced above $\sim$2~MeV \citep{1997ApJ...490..883M} would yield estimates of ion energy content well in excess of any other measure of flare total energy. 
Unfortunately, there are few diagnostics bearing on ion distributions below 1~MeV. 

Proton-capture cross-sections typically have resonances in the 0.1-1~MeV energy range\index{cross-sections!proton capture below 1~MeV}.
Since these resonances result from formation of the compound nucleus in excited states, they also give rise to $\gamma$-ray lines as these states decay to lower-lying energy levels.
The cross-sections for these lines are small compared to those for de-excitation, but 
useful constraints on total ion energy could result from upper limits at the $10^{-5}$~cm$^{-2}$
level \citep{1989A&A...226..284M}\index{gamma-rays!de-excitation lines}.
The two strongest lines are at 2.37~MeV, from $^{12}\mathrm{C(p},\gamma)^{13}$N,
and at 8.07~MeV, from $^{13}$C(p,$\gamma$)$^{14}$N, constraining the proton energy distribution above 0.46 and 0.555~MeV, respectively. 
Attempts to constrain the fluxes in these lines observationally have proven inconclusive, however, 
whether in flares \citep{2001SoPh..201..191S} or the quiet Sun \citep{1997ICRC...1..13M}. 
The line at 8.07~MeV would suffer less competition from the strong de-excitation lines, but attempts to constrain it would still require fine energy resolution and/or low instrumental background.

Other possible diagnostics for lower energy ions include Doppler-shifted Lyman-$\alpha$
line emission \citep{1985ApJ...295..275C} and line impact polarization of H$\alpha$, H$\beta$ and other lines \citep{1990ApJS...73..303H,1996SoPh..164..345V}\index{Lyman-$\alpha$!Doppler shift}.
Observations of H$\alpha$ and H$\beta$  linear polarization in flares seem increasingly well-established \index{polarization!collisional impact}
\citep{2005ApJ...631..618X,2006ApJ...650.1193X,2008SoPh..249...53F} although there remains an ambiguity in its interpretation between ion beams and streaming electron
distributions (e.g., return currents).\index{return current}\index{beams!return current}
While there is a claimed observation of Doppler-shifted Lyman-$\alpha$\index{Lyman-$\alpha$!Doppler shift}
emission in a stellar flare \citep{1992ApJ...397L..95W}, \cite{2001ApJ...555..435B}
finds no evidence for $\alpha$-particle beams in He~{\sc ii} Lyman-$\alpha$ from a C3.8 flare\index{He~{\sc ii} 304\AA!Doppler shift}.\index{beams!alpha@$\alpha$-particle}

Their importance for the flare energy budget notwithstanding, flare-site ions 
below 1~MeV/ nucleon remain more or less unconstrained by observations. 
A promising new technique could be to detect suprathermal ions through charge exchange with the ambient hydrogen (see the first detection of energetic neutral hydrogen atoms in a solar flare by 
Mewaldt et al. 2009).
\nocite{2009ApJ...693L..11M}
\index{energetic neutral atoms}

\subsection {Heavy energetic ions as diagnosed from broad lines }
\label{broad}
\index{gamma-rays!broad-line spectrum}
\index{spectrum!broad $\gamma$-ray lines}
\index{accelerated particles!heavy ions}

The relative abundances of accelerated ions heavier than He~can be determined from the relative fluences of the broad de-excitation lines\index{gamma-rays!de-excitation lines}.
These lines are produced when heavy energetic ions interact with ambient H~and He. 
Because the heavy ions lose little of their kinetic energy in the reactions, the excited nucleus produced has a large velocity and the de-excitation $\gamma$-ray line can be strongly Doppler-broadened by up to $\sim$20\% FWHM. 
Because the lines are so broad, many of them overlap, making the heavy-ion abundance determination much more difficult than the corresponding determination of ambient abundances using the narrow de-excitation lines. In addition, such broad lines are difficult to distinguish from the underlying continua formed by the electron bremsstrahlung and the nuclear continuum. 
The nuclear continuum is composed of hundreds of relatively weak, closely spaced de-excitation lines that also result from the nuclear interactions along with the strong lines clearly visible in flare spectra. These lines are so numerous and closely spaced that they merge and appear as a continuum to $\gamma$-ray detectors.\index{gamma-rays!unresolved continuum}\index{spectrum!gamma@$\gamma$-ray pseudo-continuum} \index{continuum!gamma@$\gamma$-ray pseudo-continuum}
This continuum, however, does have structure, on the order of several hundred keV, and knowledge of this structure is critical to reliably separate it from the broad lines. 
The early attempts to derive heavy-ion abundances from flare $\gamma$-ray data 
that we discuss here used relatively crude modeling of this nuclear continuum. 
On-going analyses using improved nuclear-continuum modeling recently obtained from modern nuclear reaction codes will improve the reliability of such determinations \citep[see][]{2009ApJS..183..142M}.\index{reactions!nuclear!codes}

\begin{figure}[tbp]
\begin{center}
\hspace*{-5mm} \includegraphics[scale=0.60]{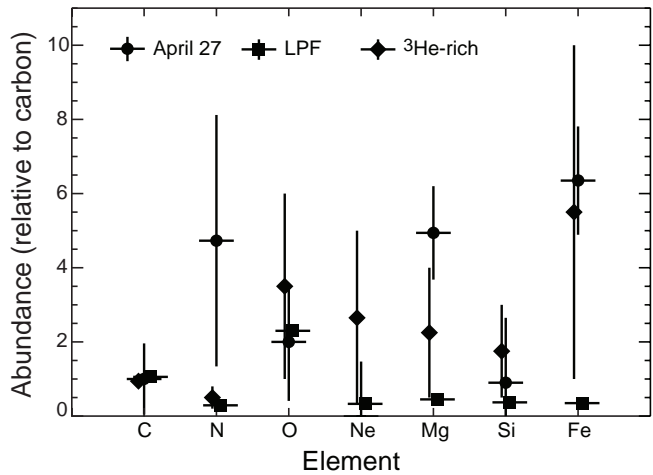}
\end{center}
\caption{Derived accelerated-ion abundances relative to accelerated carbon (circles) for SOL1981-04-27T09:45 (X5.5) \citep[adapted from][]{1991ApJ...371..793M}. 
Also shown are large proton flare (LPF, squares) and impulsive $^3$He-rich event (diamonds) 
abundances (Murphy, private communication). }
\label{apr27_acc_abund}
\end{figure}

\citet{1991ApJ...371..793M} used the $\gamma$-ray de-excitation line code from \citet{1979ApJS...40..487R}  \citep[see the revised version in][]{2002ApJS..141..523K}
to calculate both narrow and broad line spectra from the most abundant elements in the solar atmosphere, and the most abundant accelerated ions\index{satellites!SMM@\textit{SMM}}.
The abundances were varied to obtain the best fit to the \textit{SMM} data for SOL1981-04-27T09:45 (X5.5).
\index{flare (individual)!SOL1981-04-27T09:45 (X5.5)!abundances}\index{accelerated particles!abundances}
The resulting accelerated-ion abundances relative to accelerated carbon are shown in Figure \ref{apr27_acc_abund}. 
Also shown for comparison are similar element abundances measured in space from large proton flares (LPF) and impulsive, $^3$He-rich events. 
The uncertainties for most of the elements are quite large, but Mg and Fe are seen to be significantly ($>$$3-4\ \sigma$) enhanced relative to~C and~O, similar to the enhancements seen in the impulsive $^3$He-rich events\index{solar energetic particles (SEPs)!$^3$He-rich events}.

\index{flare (individual)!SOL1981-04-27T09:45 (X5.5)!illustrationt}

Observations of the behind-the-limb flare SOL1991-06-01T16:14 (X12) with \textit{GRANAT}/ PHEBUS \citep{1994ApJ...425L.109B} have moreover shown that the enhancements in heavy ions may increase with time in the course of the flare, reaching towards its end the highest values observed for solar energetic particles in space \citep{1996AIPC..374..153T,1997ApJ...479..458R}\index{occulted sources!gamma@$\gamma$-rays}\index{flare (individual)!SOL1991-06-01T16:14 (X12)!solar energetic particles}.
As a thin-target production of the emission\index{thin target!ions} is required to account for the very high observed ratio of 1.1-1.8~MeV flux to 4.1-7.6~MeV flux, the temporal evolution of the abundances of accelerated ions is to be related to the evolution of the accelerated particles themselves.
\index{thin target!gamma@$\gamma$-rays, SOL1991-06-01T16:14}
\index{coronal sources!gamma-rays@$\gamma$-rays}
It must finally be noticed that, although a behind-the-limb flare, this event is associated with one of the largest $\gamma$-ray line fluences observed so far. 
This flare is also surprisingly associated with the observation of a strong flux of neutrons by the OSSE experiment \citep{1999ApJ...510.1011M}.\index{flare (individual)!SOL1991-06-01T16:14 (X12)!neutrons}\index{satellites!CGRO@\textit{CGRO}}As the neutrons are also expected to be produced as a thin target, a hard ion spectral index (power-law index around $-2$) is deduced by \citet{1999ApJ...510.1011M} up to at least 50~MeV.\index{thin target!neutron production} 

\section {Pion-decay radiation in solar flares}
\index{gamma-rays!$\Pi^0$-decay}

Observations of pion-decay radiation from solar flare $\gamma$-rays and of neutrons (see next section) combined with observations of flare $\gamma$-ray lines give a complete picture of the accelerated ion distribution above a few MeV/nucleon to several GeV/nucleon
\citep[see, e.g.,][for reviews]{1986psun....2..291R,1984ARA&A..22..359C,1996AIPC..374....3C,2003LNP...612..127V,2009RAA.....9...11C}. 
Studies of high energy emissions from the flare site started with \textit{SMM} \citep{1987ApJS...63..721M}, continued with events observed with \textit{GAMMA-1} \citep{1992SvAL...18...69A},
\textit{CGRO} \citep[e.g.,][]{1992ApJ...389..739M,1993A&AS...97..349K,1999SoPh..187...45D}, and now with \textit{CORONAS-F} \citep {2003ICRC....6.3183K}.
Pion-decay radiation together with observations of neutrons provide information on the highest ion energies produced during flares.
Charged pions decay to yield electrons and positrons, which in turn produce $\gamma$-rays by bremsstrahlung\index{bremsstrahlung!from charged 
pion decay}\index{continuum!pi@$\pi^0$-decay}.
Positrons also contribute to the continuum by annihilating in flight. 
\index{continuum!pi@$\pi^0$-decay}
Neutral-pion ($\pi^0$) decay results in two photons, with one emitted at a high energy. 
This results in a very flat, broad ``bump'' feature which has a maximum at 67~MeV\index{spectrum!neutral-pion decay}\index{gamma-rays!67-MeV ``bump'' from $\pi^0$ decay}.
Synchrotron losses of electrons and positrons\index{synchrotron energy losses} may be important and shorten their lifetimes, therefore reducing their contribution with respect to the radiation from neutral-pion decay \citep{1987ApJS...63..721M}\index{synchrotron emission!gamma@$\gamma$-ray continuum}.
Energetic electrons above 10~MeV are also produced in solar flares and produce bremsstrahlung continuum, potentially masking pion-decay radiation\index{bremsstrahlung!competition with $\pi^0$ decay}.

\begin{figure}
\begin{center}
\hspace*{-5mm} \includegraphics[scale=0.55,angle=90]{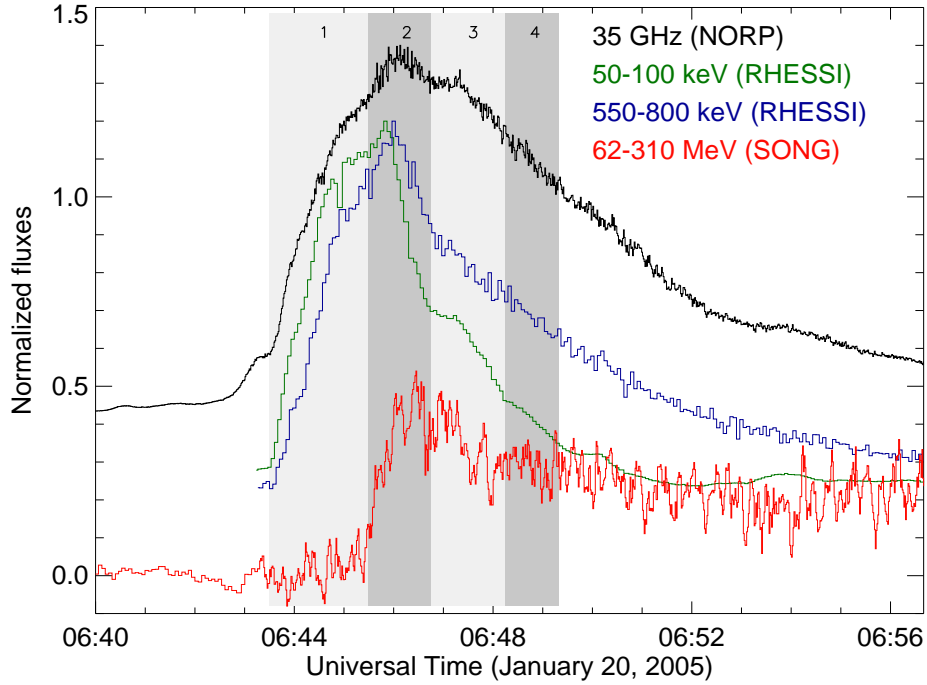}
\end{center}
\caption{Time profiles of the normalized flux density at 35~GHz (top) and normalized count rates of hard X-rays and $\gamma$-rays (\textit{RHESSI}, \textit{CORONAS-F}/SONG) at different energies. The high energy $\gamma$-rays observed by SONG are pion-decay photons from primary protons at energies above 300~MeV \citep[from][]{2009SoPh..257..305M}.
}
\label{masson}
\index{flare (individual)!SOL2005-01-20T07:01 (X7.1)!illustration}
\end{figure}

\begin{figure}
\begin{center}
\hspace*{-5mm} \includegraphics[scale=0.50]{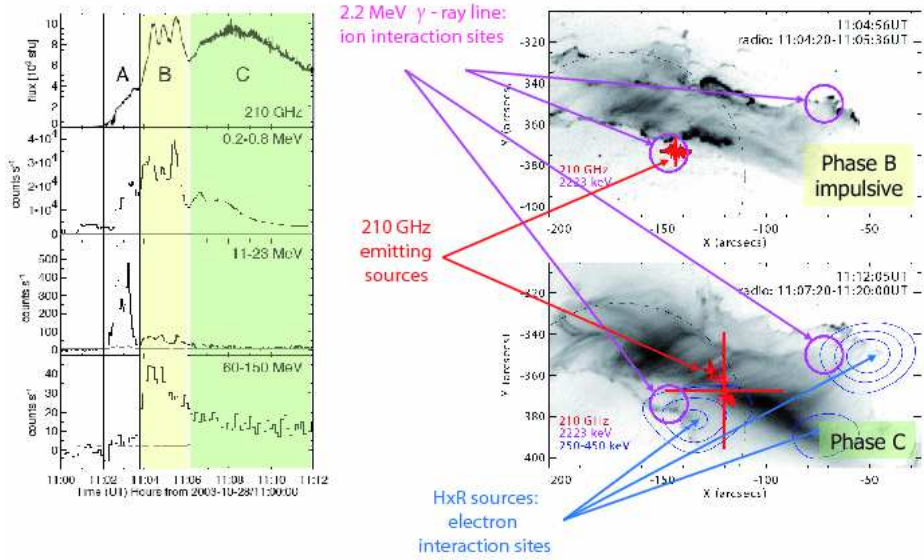}
\end{center}
\caption{\textit{Left:} 210~GHz submillimeter radio flux from KOSMA
(K{\" o}lner Observatorium f{\" u}r SubMillimeter Astronomie) 
and X-ray/$\gamma$-ray fluxes from \textit{CORONAS-F}/SONG; \textit{right:} imaging in radio at 210~GHz (red crosses), UV 
(background image from \textit{TRACE}), X-rays and $\gamma$-rays from \textit{RHESSI} (blue and pink contours from Hurford et al. 2006) in phase B~and C showing, respectively, electron and ion interaction sites \citep[adapted from][]{2008ApJ...678..509T}.
}
\label{trottet_pion}
\end{figure}
\index{pions!and submillimeter radiation!illustration}
\index{submillimeter emission!illustration}
\index{radio emission!submillimeter!illustration}
\index{flare (individual)!SOL2003-10-28T11:10 (X17.2)!illustration}

Pion decay radiation was first observed with the \textit{SMM}/GRS instrument for the flare SOL1980-06-21T02:00 (X2.6) (upper limit) and then for the flare SOL1982-06-03T13:26 (X8.0) \citep{1985ICRC....4..146F}.
\index{flare (individual)!SOL1980-06-21T02:00 (X2.6)!$\pi^0$ continuum}
\index{flare (individual)!SOL1982-06-03T13:26 (X8.0)!$\pi^0$ continuum}
The latter event provided the first convincing observations of $\pi^0$-decay radiation in a solar flare. 
This event showed that the production of pions can occur early in the impulsive phase defined by the production of hard X-rays near 100~keV\index{hard X-rays!and pion production}.
However, a significant portion of the high-energy emission was also observed well after the impulsive phase in an extended phase mostly observed at high energies. 
High-energy neutrons were also detected for that event (see next section).    
Together with the development of models for ion acceleration and $\gamma$-ray production, these observations led to the first determination of high-energy ion distributions in flares \citep{1987ApJS...63..721M}.
Given the assumed spectral shape of energetic protons (either a Bessel function or a power-law spectrum), numbers and spectra of energetic protons were estimated for both the impulsive and extended phases from the ratio of the differential pion decay radiation at 100 MeV to the 4.1-6.4 MeV nuclear emission. 
It was found that the first phase was characterized by a significantly steeper proton spectrum than the extended phase. 
Pion-decay radiation was subsequently reported for other flares observed with \textit{SMM}/GRS 
(Dunphy \& Chupp 1991, 1992; see also Chupp \& Dunphy 2000, for a review).
\nocite{1991ICRC....3...65D,1992AIPC..264..253D,2000ASPC..206..400C}
\index{satellites!SMM@\textit{SMM}}\index{satellites!CORONAS-F@\textit{CORONAS-F}}\index{satellites!CGRO@\textit{CGRO}}\index{satellites!GRANAT@\textit{GRANAT}}

After the end of the \textit{SMM} mission in November 1989 and before the launch of \textit{CORONAS-F} at the end of July 2001, there was no specifically solar-dedicated mission for high-energy radiation, but several satellites still provided $\gamma$-ray data (\textit{GRANAT}/PHEBUS, \textit{GAMMA-1}. and \textit{CGRO}). 
 
Until now, around 20 events have been observed with significant pion production \citep[see][]{1997SoPh..173..151L,2007AdSpR..40.1929M,2009RAA.....9...11C}.
\index{gamma-rays!pi@$\pi^0$-decay}
Some of the most recent events have been observed in a wide energy range by both \textit{RHESSI} and \textit{CORONAS-F} \citep[e.g.,][]{2008ICRC....1..121K,2008ApJ...678..509T,2009SoPh..257..305M} (see, e.g., Figure~\ref{masson}).
\index{flare (individual)!SOL2005-01-20T07:01 (X7.1)!pi@$\pi^0$ continuum}
For some of the events, high-energy emission has been observed for hours after the impulsive phase of the flare, revealing that high energy ions must be continuously accelerated on long timescales in some flares \citep[e.g.,][]{1993A&AS...97..349K,1994AIPC..294...89R,2000SSRv...93..581R,2001A&A...378.1046R}.\index{acceleration!extended flare}
Quantitative analysis of a few of the events with significant pion production has been performed providing information on the ion energy spectrum at energies greater than 300~MeV and allowing a comparison of this spectrum with the one deduced at lower energies from $\gamma$-ray line spectroscopy \citep[see e.g.,][]{1994SoPh..151..147A,1999SoPh..187...45D,1994SoPh..150..267K,1998A&A...340..257K,2003A&A...412..865V}. 
These comparisons generally show that the ion energy distribution does not have a simple power-law form from the $\gamma$-ray line emitting energy domain (1-10~MeV) to the pion-producing energy domain ($>$300~MeV).  

In recent years, some of the events with significant pion-decay radiation have been found to be associated with submillimeter (above 200~GHz) emissions \citep{2004ApJ...603L.121K,2004A&A...415.1123L}.\index{submillimeter radiation}\index{radio emission!submillimeter}
\index{pions!and submillimeter radiation}
These observations have revealed the existence of a new emission component with intensities well above the extrapolation of the radio synchrotron spectrum seen at lower frequencies. 
Moreover, the radio spectrum increases with increasing frequency contrary 
to what is expected from an optically thin synchrotron spectrum. 
The origin of this new component is still under discussion. 
One possibility could be that this emission be produced by synchrotron emission from pion-decay positrons. 
This last process was described a long time ago \citep{1967P&SS...15.1303L} and could be considered as a good candidate for the high frequency observations, given the association between events exhibiting pion-decay radiation and events showing a spectral increase above 200~GHz \citep{2004cosp...35.1511M,2007SoPh..245..311S}.
\index{gamma-rays!pi@$\pi^0$-decay!and submillimeter emission}
This is illustrated in Figure~\ref{trottet_pion} from \citet {2008ApJ...678..509T} which shows the observations of SOL2003-10-28T11:10 (X17.2) with KOSMA\footnote{K{\" o}lner Observatorium f{\" u}r SubMillimeter Astronomie.} at 210~GHz and with the SONG experiment on \textit{CORONAS-F} at energies above 0.2~MeV.\index{observatories!KOSMA}\index{satellites!CORONAS-F@\textit{CORONAS-F}}\index{flare (individual)!SOL2003-10-28T11:10 (X17.2)!submillimeter emission}
The onset of the impulsive component of the submillimeter component (phase~B on the Figure) is 
clearly associated with the start of the radiation above 60~MeV, i.e., the start of 
$>$300~MeV/nucleon ion acceleration. 
The 210~GHz source size is compact ($<$10$''$) in this phase and its location is co-spatial with the site of interacting ions (revealed by the 2.2~MeV line emission site) but not to the site of interacting electrons\index{radio emission!submillimeter}.
In phase~C, when no impulsive submillimeter emission and no strong emission above 60~MeV are observed, the submillimeter sources are quite different. 
The close correlation in time and space of the impulsive submillimeter emission and  of the strong 
production of neutral pions thus suggests that synchrotron emission from charged pion-decay positrons could be responsible for the submillimeter emission. 
However the submillimeter flux predicted from the number of positrons 
derived from charged pion decay seems inadequate to account for observed fluxes.  
An alternative interpretation was given by \citet{2007SoPh..245..311S} for SOL2003-11-02T09:55 (X3.9).
\index{flare (individual)!SOL2003-11-03T09:55 (X3.9)!submillimeter emission}
It was concluded there that the most likely source of the submillimeter emission is gyrosynchrotron emission from non-thermal electrons but this requires high local magnetic fields, very large densities of accelerated electrons and very small source sizes. 
Although many interpretations have been proposed in the literature 
\citep[see above and][]{2006PhPl...13g0701K,2007A&A...474L..33S,2010ApJ...709L.127F}, the origin of this submillimeter component remains a challenging topic of discussion. 

Another challenging topic linked to the observations of pion decay radiation in a few flares is related to the origin of coronal hard X-ray sources reported in, particularly, SOL2005-01-20T07:01 (X7.1). 
\citet{2010A&A...510A..29M} have indeed shown that some coronal hard X-ray sources might be interpreted as photospheric, optical photons inverse Compton scattered to deka-keV energies by electrons, or indeed positrons in the 10-100~MeV energy range.\index{Compton scattering!and coronal hard X-ray sources}\index{coronal sources!inverse Compton radiation}\index{flare (individual)!SOL2005-01-20T07:01 (X7.1)!inverse Compton emission}\index{scattering!pitch-angle}
In particular, the coronal hard X-ray source reported in SOL2005-01-20T07:01
\citep{2008ApJ...678L..63K} could be explained in this way if just a few percent of the secondary positrons implied by estimated fast ion numbers \citep{2009SoPh..257..305M} were present in the corona. 
Coronal hard X-ray sources in locations that seem too tenuous for a bremsstrahlung explanation \citep[see][for a review]{2008A&ARv..16..155K} may thus offer a novel window on $\sim$GeV-energy ions.

\section{Neutron observations in space and on Earth}
\label{neu}
\index{neutrons}

Many reactions between energetic flare ions and ambient nuclei produce energetic neutrons as secondaries. 
These secondary neutrons undergo multiple, elastic scattering and either thermalize in the solar atmosphere or escape into interplanetary space.\index{scattering!neutrons}
Measurements of the escaping neutrons augment knowledge of flare ion distributions gained from 
$\gamma$-ray lines. In particular they fill a ``diagnostic gap'' between the de-excitation lines, 
dominated by ions in the 1-100~MeV energy range, and the pion-decay radiation which needs ions of $\ge$300~MeV \citep{1997SoPh..173..151L}. 
Neutrons in space give valuable information on the most energetic flare ions, especially when measurements 
of pion decay $\gamma$-rays are not available. 
They may also play a role in constraining the numbers of heavier fast ions, prolific producers of free neutrons via evaporation reactions.\index{reactions!nuclear!evaporation}\index{ions!nuclear reactions!evaporation}

Having no electric charge, they arrive at the detector undeviated by active region magnetic fields or interplanetary medium plasma turbulence.
\index{turbulence!interplanetary medium}\index{plasma turbulence}
Free neutrons are unstable to $\beta$-decay, with a rest-frame mean lifetime of 886~s, 
so many decay before they can be detected at Earth.
\index{neutrons!half-life}

\subsection{Neutron detections}
\index{neutrons}

Energetic neutrons from flares have been detected both in space and on the ground. 
They give a signal in scintillators that may be distinguished by various means from photons or charged particles (pulse shape, combination of signals in multiple elements, etc.) 
\citep[e.g.,][]{1987ApJ...318..913C,1999ApJ...510.1011M}. 
On the ground they have been detected by neutron monitors, and by 
specially designed neutron  telescopes \citep[e.g.,][]{2006ApJ...636.1135W}\index{neutron monitors!direct detection of solar neutrons}\index{neutrons!advantages of observing}.

The global neutron-monitor network\index{neutron monitors!global network} has great value for studying extremes of solar particle acceleration, offering detectors of a stopping power and effective area that would be impractical in space.
They are located inside Earth's magnetic field in a way that aids the discrimination of neutrons
from other particle signals \citep{1997AnGeo..15..375U, 1999SSRv...88..483L}.
\index{solar energetic particles (SEPs)!and neutron monitors}
Neutron-monitor count-rate enhancements following solar flares are generally due to solar energetic particles interacting with Earth's atmosphere \citep[e.g.,][]{2006GMS...165..283L} but neutrons genuinely of solar origin may be distinguished on grounds of timing and geographical distribution. 
Neutron telescopes provide neutron spectral and angular resolution in the $>$50~MeV energy range and can discriminate neutrons from charged particles \citep{1992ApJ...400L..75M,2001NIMPA.463..183T}\index{neutron telescopes}.
They have been installed in seven high-altitude locations around the world. 

A neutron of 150~MeV travels a distance of 1~AU in a neutron lifetime \citep{2007A&A...462..763M}.
Lesser energies are more likely to decay \emph{en route} from the Sun while time dilatation extends the lifetime of neutrons at energies above 1~GeV. 
The neutron survival probability to Earth thus plummets below $\sim$100~MeV. 
Nonetheless, solar flare neutrons have arrived at Earth in detectable numbers in the energy ranges 10-100~MeV
\citep{1993AdSpR..13..255R} and 50-360~MeV \citep{1992ApJ...400L..75M}. 
Minimum  neutron energies for detection on the ground are 50~MeV \citep[for neutron telescopes, see e.g.,][]{1992ApJ...400L..75M} or 200-500~MeV \citep[neutron monitors; see][]{1989NIMPA.278..573D,1994JGR....99.6651S}.
\index{neutron monitors!neutron energy range}
\index{neutron telescopes!neutron energy range}

The first detection of flare neutrons was by the \textit{SMM}/GRS instrument, from the flare SOL1980-06-21T02:00 (X2.6) \citep{1982ApJ...263L..95C}.
\index{flare (individual)!SOL1980-06-21T02:00 (X2.6)!neutrons}
\index{flare (individual)!SOL1982-06-03T13:26 (X8.0)!neutrons}
The flare SOL1982-06-03T13:26 (X8.0) produced energetic neutrons that were detected in space by \textit{SMM}/GRS and on the ground using the Jungfraujoch and other neutron monitors
\citep{1987ApJ...318..913C}. 
\index{satellites!SMM@\textit{SMM}}
\index{observatories!Jungfraujoch}
Neutron-decay protons were also detected in space from this flare \citep{1983ApJ...274..875E}.
\index{protons!neutron-decay}

Ground-based detections of solar neutrons have been recently reviewed by \citet{2005ICRC....1...37W} and \citet{2009AdSpR..43..565V}\index{neutrons!ground-level detection}. 
Since SOL1982-06-0313:26 (X8.0), ten more solar neutron events have been clearly recognized in neutron monitor data: SOL1990-05-24T21:45 (X9.3) \citep{1991GeoRL..18.1655S, 1995ICRC....4..171S, 1993ApJ...409..822D, 1997ApJ...479..997D}, SOL1991-03-22T22:45 (X9.4) \citep{1991ICRC....3...53P}, SOL1991-06-04T03:37 (X12.0) and SOL1991-06-06T03:37 (X12.0) \citep{1992ApJ...400L..75M, 1994ApJ...429..400S} and several occasions during Cycle 23 \citep{2003ApJ...592..590W, 2005GeoRL..3203S02B, 2006ApJ...636.1135W, 2006ApJ...651L..69S}.
The second such event was observed by the IGY-type neutron monitor located at Climax and several other stations in North America: SOL1990-05-24T21:45 (X9.3)
\citep{1991GeoRL..18.1655S, 1995ICRC....4..171S, 1993ApJ...409..822D, 1997ApJ...479..997D}.
\index{flare (individual)!SOL1990-05-24T21:45 (X9.3)!neutrons}
In the SOL1991-06-04T03:37 (X12.0) event, solar neutron signals were recorded by three different detectors (neutron monitor, solar neutron telescope, and muon telescope\index{muon telescope}) located at Mt.~Norikura.
\index{observatories!Mt. Norikura}
\index{muon telescope}
\index{flare (individual)!SOL1982-06-03T13:26 (X8.0)!neutron detection}
\index{flare (individual)!SOL1990-05-24T21:45 (X9.3)!neutron detection}
\index{flare (individual)!SOL1991-03-22T22:45 (X9.4)!neutron detection}
\index{flare (individual)!SOL1991-06-04T03:37 (X12.0)!neutron detection}
\index{flare (individual)!SOL1991-06-06T01:12 (X12.0)!neutron detection}

In space, apart from the \textit{SMM}/GRS instrument, energetic flare neutrons have been detected with the OSSE and COMPTEL instruments on the
\textit{CGRO} mission \citep{1999ApJ...510.1011M,1999SoPh..187...45D} as well as with the PHEBUS instrument aboard \textit{GRANAT} \citep{1997ApJ...479..997D} \citep[see, e.g.,][for a review]{2009RAA.....9...11C}. 
Pulse-shape analysis in OSSE's NaI scintillators discriminated neutrons from photons \citep{1999ApJ...510.1011M}.
In COMPTEL, signals from two arrays of detectors were combined to deduce arrival directions and energies for individual neutrons, via  elastic scattering kinematics.\index{neutrons!elastic-scattering detection}\index{scattering!neutrons}
Such imaging spectroscopy of 10-100~MeV neutrons was possible for several flares in June 1991 \citep[e.g.,][]{1993AdSpR..13..255R}.\index{imaging spectroscopy!neutrons}

\subsection{Key features of neutron observations}

Solar neutron events are too few in number for statistical studies. 
Nonetheless detections so far share some common features summarized 
in \citet{2005ICRC....1...37W}. 
Detectable neutrons have come from large flares, SOL2000-11-24T15:13 (X2.3) being the
smallest of these \citep{2003ApJ...592..590W}\index{flare (individual)!SOL2000-11-24T15:13 (X2.3)!neutron detection}.
Of course, this could be observational bias rather than a physical feature of flares.
No correlation is evident between the flare soft X-ray flux and the neutron emissivity.\index{soft X-rays!and neutron emission}\index{neutron emission}
Detectable neutrons at Earth do not appear to come preferentially from limb or disk centre flares\index{neutrons!lack of limb dependence}.

Using, e.g., the $\gamma$-ray de-excitation line flux time profile as a proxy for the rate of ion production at the Sun, neutron energy distributions have been deduced from neutron-monitor count-rate time profiles\index{neutron monitors}.
Neutrons at the Sun are generally found to have $E^{-\delta}$ distributions, with $\delta$ in the range 3-4 \citep{2003ApJ...592..590W,2006ApJ...636.1135W}\index{spectrum!neutrons}.
Via model calculations using the code of \citet{2002ApJS..140..563H}, the ion energy distribution at the Sun is then found to be roughly one power of~$E$ softer. 
Total emissivity at the Sun is in the range 4~$\times$~10$^{28}$-8~$\times$~10$^{30}$~sr$^{-1}$ for neutrons in the range 0.05-1.5~GeV.

\citet{1997SoPh..173..151L} and \citet{1997ApJ...479..997D} emphasized that the de-excitation line flux does not always give the best proxy for the time development of neutron production, urging use instead of the pion-decay flux when this is measured. 
Neutron and pion production by energetic ions  ($>$300~MeV) are closely related (see Section~\ref{nprod}) so this is not surprising.
\index{flare (individual)!SOL2005-09-07T09:52 (X17.0)!neutron detection}
The flare SOL2005-09-07T09:52 (X17.0), for instance,
was detected on the ground by both neutron telescopes and neutron monitors and found to produce neutrons over a more extended period than implied by the flare impulsive phase, as defined by hard X-ray emission. 
The class of extended $\gamma$-ray flares \citep{2000SSRv...93..581R} 
often show pion-decay $\gamma$-radiation over tens of minutes. 
Similarly time-extended neutron emission would be a corollary, as was found using COMPTEL for SOL1991-06-09T04:24 (X10.0) and SOL1991-06-15T11:17 (X12.0) \citep{1993AdSpR..13..255R}.
\index{flare (individual)!SOL1991-06-09T04:24 (X10.0) (X10.0)!neutron detection}\index{flare (individual)!SOL1991-06-15T11:17 (X12.0)!neutron detection}\index{neutron emission!time-extended}

\subsection{Production of solar neutrons}
\label{nprod}
Ions are believed to be accelerated in the corona, losing their energy mostly in the chromosphere and photosphere and producing the bulk of their secondaries (photons, neutrons) there. 
The eventual escaping neutron velocity distribution is governed by the initial velocity distribution of neutrons, which in turn depends on the primary ion distribution throughout the 
atmosphere, and the consequences of mutiple neutron elastic scattering.\index{scattering!neutrons} 
Thus a complete modeling of neutron production needs to include ion transport,
neutron production and scattering. 
Such a code was constructed by \citet{1987SoPh..107..351H} and improved and updated by 
\citet{2002ApJS..140..563H}, and the consequences of varying loop and acceleration parameters was described in detail by \citet{2007ApJS..168..167M}. 
The code of \citet{2002ApJS..140..563H} is freely available and may be used to calculate 2.223~MeV line flux and neutron distributions expected at 1~AU 
under a wide variety of assumptions for both physical and acceleration parameters.

Neutron-producing reactions of fast ions have been enumerated by \cite{1987SoPh..107..351H,1987ApJS...63..721M,2002ApJS..140..563H}. 
The reaction p(p,n$\pi^+$)p$'$ has a threshold of 270~MeV\index{ions!neutron production!(p,p) reaction}\index{neutrons!production in (p,p) reactions}. 
It is a major contributor to total neutron yields for harder ion distributions 
\citep[e.g., energy power law $E^{-s}$ with $s >3$;][]{2007ApJS..168..167M}. 
Thus neutron emission from flares will often be accompanied by pion-decay $\gamma$-rays.\index{neutron emission}

In contrast are the cross-sections for collisions involving heavier nuclei\index{cross-sections!heavy nuclei}\index{neutrons!production by evaporation process}.
Neutrons in these cases can be produced via evaporation processes, 
without the creation of pions, so cross-sections often have much lower threshold energies, 
below 1 MeV/nucleon in some cases. Neutron measurements thus have some potential for constraining heavy ion acceleration in flares\index{cross-sections!neutron evaporation}.
Especially with softer assumed ion energy distributions, neutron yields will depend sensitively on assumptions about fast ion abundances
\citep{2007ApJS..168..167M}.

\begin{figure}
\begin{center}
\hspace*{-5mm} \includegraphics[scale=0.20]{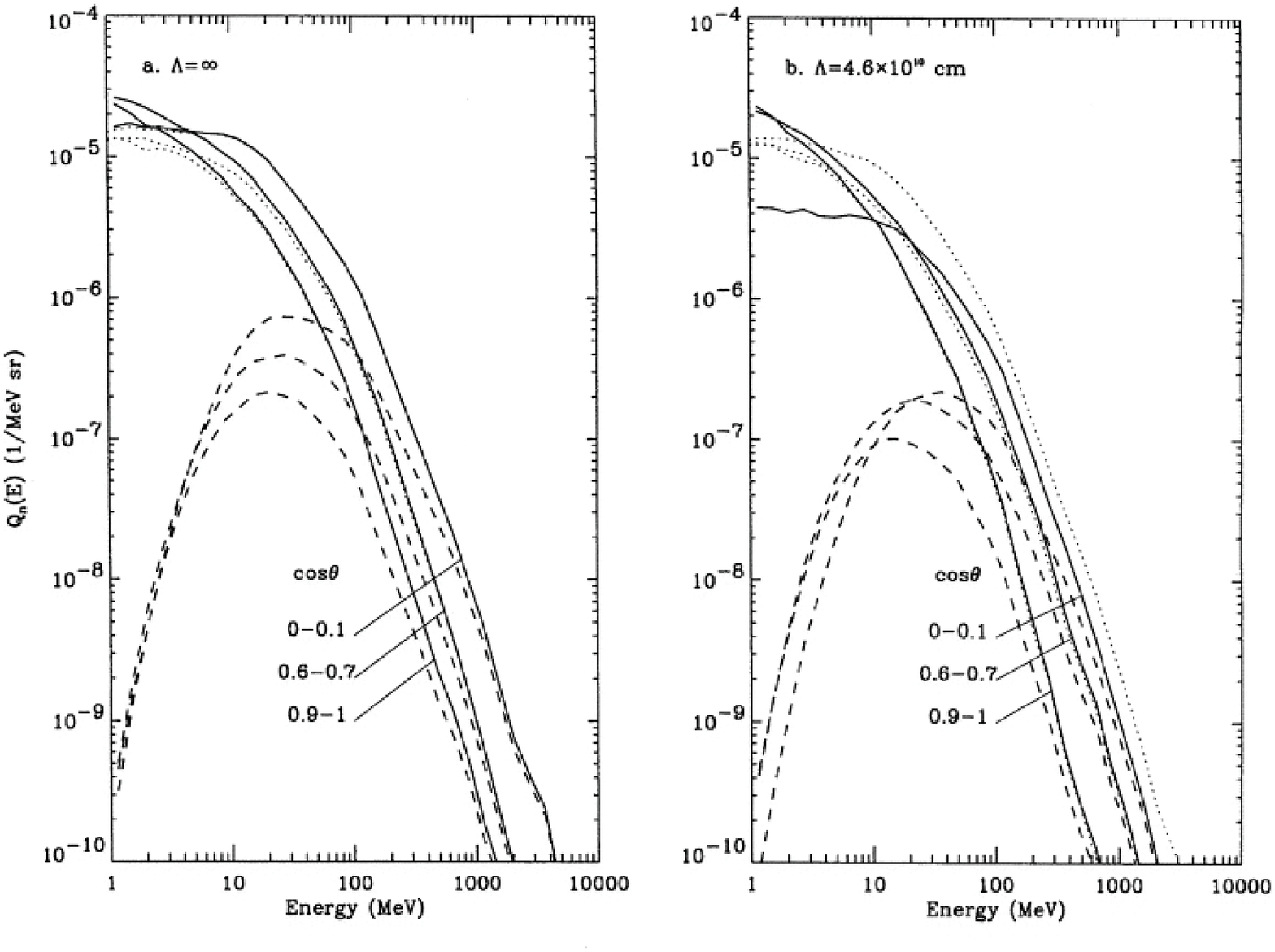}
\end{center}
\caption{Angular-dependent energy spectra of the escaping neutrons at the Sun (solid curves) and of those surviving at 1 AU (dashed curves) at different cos $\theta$ 
(zenith angle) compared to their production spectra at the Sun (dotted curves) in the case with no pitch-angle scattering of the ions (left) and of nearly saturated pitch-angle scattering 
(right).
See Section~\ref{kp} for a discussion of scattering. \citep[From][] {2002ApJS..140..563H}.}
\label{hua}
\end{figure}

Figure~\ref{hua} shows neutron distributions produced by a population of fast ions, calculated using the code of \citet{2002ApJS..140..563H}. 
A power-law ion energy distribution has been assumed, $\propto$E$^{-s}$ with $s = 2$, and fast ion abundances representative of ``gradual'' particle events in space. 
No magnetic field convergence or MHD pitch-angle scattering have been assumed in the
corona so ions precipitate freely to the dense atmosphere. 
The two figures show the energy distribution of neutrons produced in the solar atmosphere, and the energy distribution of neutrons that actually escape from the Sun. 
Above $\sim$100~MeV the neutron energy distribution reflects the primary ion energy distribution; for assumed power laws the neutron distribution is about one power of 
energy harder \citep{2005ICRC....1...37W}. 
At lower energies the neutron energy distribution is dominated by the relevant nuclear physics, particularly since many of the reactions are near threshold, and by neutron scattering in the atmosphere.\index{scattering!neutron}
Any factor that increases neutron production at greater depth (e.g., pitch-angle scattering in the strong limit which maximizes 
precipitation -- Section~\ref{kp}) decreases the escaping neutron flux \citep{2007ApJS..168..167M}.
\index{pitch-angle scattering!and neutron production}\index{precipitation}\index{scattering!pitch-angle}

\subsection{Neutron decay products}
Energetic protons from decaying flare neutrons were detected in space from SOL1980-06-21T02:00 (X2.6) \citep{1990ApJS...73..273E}, SOL1982-06-03T13:26 (X8.0)
\citep{1983ApJ...274..875E} and SOL1984-04-25T01:40 (X13) \citep{1990ApJS...73..273E}.
\index{flare (individual)!SOL1980-06-21T02:00 (X2.6)!neutrons}
\index{flare (individual)!SOL1982-06-03T13:26 (X8.0)!neutrons}
\index{flare (individual)!SOL1984-04-25T01:40 (X13)!neutrons}
\index{neutrons!detection in space} 
When the detector is not magnetically well connected to the flare site, neutron decay products may be clearly distinguished from flare-associated fast particles.
\index{flare (individual)!SOL1990-05-24T21:45 (X9.3)!neutrons}
An energetic proton ``precursor'' in SOL1990-05-24T21:45 (X9.3),
arriving at the \textit{GOES} spacecraft before the bulk of the flare-associated fast ions, has also been interpreted as a signature of decaying neutrons 
\citep{1996SoPh..169..181K}\index{neutrons!detection via decay products}.\index{precursor!neutron-decay}
Energetic neutron-decay electrons were observed from SOL1980-06-21T02:00 (X2.6)
\citep{1996ApJ...464L..87D}.
\index{flare (individual)!SOL1980-06-21T02:00 (X2.6)!neutron-decay electrons}

Neutron-decay protons carry off most of the energy of their parent particles and are straightforwardly detected, given an appropriately placed spacecraft
\citep{1990ApJS...73..273E}. 
These advantages have to be weighed against the complication of accounting correctly for proton propagation in the turbulent interplanetary medium \citep{1991ApJ...382..688R}. 
Clearly they could offer a useful window on flare ion acceleration but we are
aware of only one such claimed detection in Cycles~22 and~23 \citep{1996SoPh..169..181K}, and the argument in this case rested on fast proton arrival time rather than a lack of magnetic connection to the flare location\index{neutrons!advantages of observing}.

\subsection {Low-energy neutrons}
\index{neutrons!low energy}
 
Neutrons $<$100~MeV mostly decay long before they reach 1~AU. 
A neutron detector placed inside the orbit of Mercury would detect four to five 
orders of magnitude more neutrons in the 1-20~MeV energy range than a detector at 1~AU (in contrast to a detector of photons, which would gain roughly a single order of magnitude). 
There are several good scientific reasons to attempt neutron measurements in the inner heliosphere 
\citep[e.g.,][]{2001ESASP.493..405V}. 
Much greater neutron fluxes will be measured from large flares by detectors on future inner-heliosphere space missions. 
As we mentioned above, heavy-ion populations may be usefully constrained,
even though these do not produce narrow de-excitation lines. 
Constraints on ion acceleration in small flares and even the quiet Sun will be greatly strengthened\index{accelerated particles!in quiet Sun}.
It should be emphasized that very recently, the \textit{MESSENGER\footnote{MErcury Surface, Space ENvironment, GEochemistry, and Ranging.}} neutron spectrometer made the first claimed detection of 1-8~MeV solar neutrons continuously 
produced for several hours, in good consistency with the previously observed extended production of $\gamma$-ray line emissions or pion-decay radiation 
(see previous sections) \citep{2010JGRA..11501102F}. 
However, local neutron production by solar energetic protons and 
$\alpha$ particles cannot be completely excluded in these observations.
\index{satellites!MESSENGER@\textit{MESSENGER}}
\index{neutrons!MESSENGER@\textit{MESSENGER} detection}
These considerations motivate ongoing development of multiple scatter neutron detectors
\index{neutrons!multiple scatter detectors} \citep{2009NIMPA.603..406P}, whose angular resolution in particular 
is crucial to discriminate genuine solar neutrons from those produced locally by interactions of solar energetic particles in the spacecraft\index{solar energetic particles (SEPs)!local neutron production}.

\section{Imaging solar $\gamma$-rays}
\label{img}
\index{gamma-rays!imaging}

\subsection{Introduction}

The spatial distribution of $\gamma$-ray emission provides a channel of information on the acceleration and transport of energetic ions that is independent of spectroscopic analyses.  
As with electron bremsstrahlung, nuclear $\gamma$-rays disclose not where particle acceleration occurs, but rather where accelerated particles interact with the ambient atmosphere.  
Nevertheless it provides the only currently available method for placing accelerated ions in a solar magnetic context.  

Prior to the launch of \textit{RHESSI}, the only information on the spatial distribution of solar flare
 $\gamma$-ray emission originated from non-imaging observations of $\gamma$-ray
line emissions from flares believed to be behind the limb. \citet {1993ApJ...409L..69V}, for example, reported a strong neutron-capture line emission from a behind-the-limb flare.
\index{occulted sources!gamma@$\gamma$-rays}  
Such an observation would imply that in addition to a compact source predicated in many models \citep[e.g.,][]{1987SoPh..113..229H}, the $\gamma$-ray line emission also originated from a large scale source, probably diffuse, that was far from the flare itself\index{coronal sources!gamma-rays@$\gamma$-rays}.
Another event observed with strong prompt $\gamma$-ray line emission 
by \citet {1994ApJ...425L.109B} was also believed to be associated with a behind-the-limb flare.  
In this case, the absence of neutron capture line emission as well as the smooth time profile of the prompt $\gamma$-ray line emission led to the conclusion that 
this emission was produced in the low corona at densities below $1-5 \times 10^{11}$~cm$^{-3}$.  
On the other hand, measurements of $\gamma$-ray line emissions (time profiles with respect to HXR 
time profiles) have been interpreted as generally consistent with their production in a compact  region located in the chromosphere at densities 
$\geq$ 10$^{12}$ cm$^{-3}$  \citep[e.g.,][]{1996AIPC..374....3C}. 
  
\textit{RHESSI} has provided the first opportunity to locate the $\gamma$-ray emission from solar flares directly.
\textit{RHESSI} $\gamma$-ray imaging to date is based mainly on the narrow neutron-capture line at 2.223~MeV\index{gamma-rays!2.223~MeV line!imaging}.
This line is generated when fast neutrons, created by the interaction of accelerated protons and ions with the ambient atmosphere, are thermalized and subsequently captured by protons, producing deuterium and a narrow $\gamma$-ray line at 2.223~MeV\index{gamma-rays!2.223~MeV line!limb darkening}\index{gamma-rays!2.223~MeV line!absence of Doppler-broadening}.
Narrow because of the absence of Doppler broadening, the line is an optimal choice for imaging for four reasons: i) it is relatively intense; ii) since locally-generated fast neutrons cannot thermalize in the limited mass of the spacecraft, there is no locally-generated background line emission; iii) the line is narrow so that the continuum background is minimized; iv) the narrowness of the line also limits the contribution of the solar electron-bremsstrahlung continuum to the resulting images.  
The multiple elastic collisions required for neutron thermalization require an average of $\sim$100~s \cite[e.g.,][]{2007ApJS..168..167M}, which results in a characteristic time delay in this line compared to the prompt lines.  
Nevertheless, the corresponding average spatial offset between the primary interaction site and the capture site is only $\sim$500~km (R. Murphy, private communication).  
As a result, despite its two-step formation mechanism, a neutron-capture line image can faithfully indicate the site of the original nuclear interaction to within an arcsecond. 

\textit{RHESSI} imaging of nuclear $\gamma$-ray emission is based on time- and energy-tagged counts from the thick rear-segments of \textit{RHESSI} detectors 6~and~9 (35$''$ and 183$''$ resolution, respectively).  
These are the only detectors for which the grids are sufficiently thick (2~and 3~cm, respectively) to modulate the high energy photon flux.  
Rear segments of the detectors are shielded from the intense flux of low energy X-rays, and so potential concerns about instrumental live time or pulse pileup effects (\citet {2002SoPh..210...33S} can be discounted\index{pulse pileup}.
On the other hand, since only two of the nine collimators (RMCs) are relevant to imaging and their thick grids block some of the flux, the $\gamma$-ray throughput for such imaging is only about 14\% of that for spectroscopy, and so the sensitivity for imaging is necessarily lower than that for spectroscopy.  The count data provide the starting point for algorithms such as back-projection to reconstruct the image of the source \citep{2002SoPh..210...61H}.  
The count statistics and the relatively few spatial frequencies measured do not permit \textit{RHESSI} to reproduce complex source morphologies. 
 However, the data {\it are} well-suited to the quantitative characterization of simple sources and 
 $\gamma$-ray images for five flares have been obtained.  

\subsection{Neutron-capture line imaging}

\begin{table}
\caption{\textit{RHESSI} Events with Neutron Capture Line Imaging}
\begin{tabular}{llllll}
\hline\noalign{\smallskip}
Parameter&2002-07-23&2003-10-28&2003-10-29&2003-11-02&2005-01-20\\
\hline\noalign{\smallskip}
Time range&00:27:20-&11:06:20-&20:43:00-&17:16:00-&06:44:00-\\
 &00:43:20&11:29:40&21:00:00&17:29:00&06:56:00\\
\hline\noalign{\smallskip}
\textit{GOES} class&X4.8&X17&X10&X8&X7\\
\hline\noalign{\smallskip}
Energy range&2223$\pm$5&2223$\pm$5&2223$\pm$5&2223$\pm$5&2223$\pm$8\\
(keV)& & & & & \\
\hline\noalign{\smallskip}
Total counts &336,240&2280,1643&463,313&781,557&961,518\\
in RMC 6, 9& & & & & \\
\hline\noalign{\smallskip}
Relative visibility&1.61$\pm$0.45&0.91$\pm$0.20&...&0.76$\pm$0.26&0.98$\pm$0.27\\
 RMC 6 & & & & & \\
\hline\noalign{\smallskip}
Relative visibility&1.28$\pm$0.26&0.82$\pm$0.10&1.28$\pm$0.28&0.84$\pm$0.15&0.93$\pm$0.21\\
 RMC 9 & & & & & \\
\hline\noalign{\smallskip}
Source size&$<$22&$<$30&$<$94&$<$44&$<$20\\
FWHM (arcsec) & & & & & \\
(2 $\sigma$ upper limit)& & & & & \\
\hline
\end{tabular}
\label{tab:imaging}
\end{table}     

\begin{figure}
\begin{center}
\hspace*{-5mm} \includegraphics[scale=0.63]{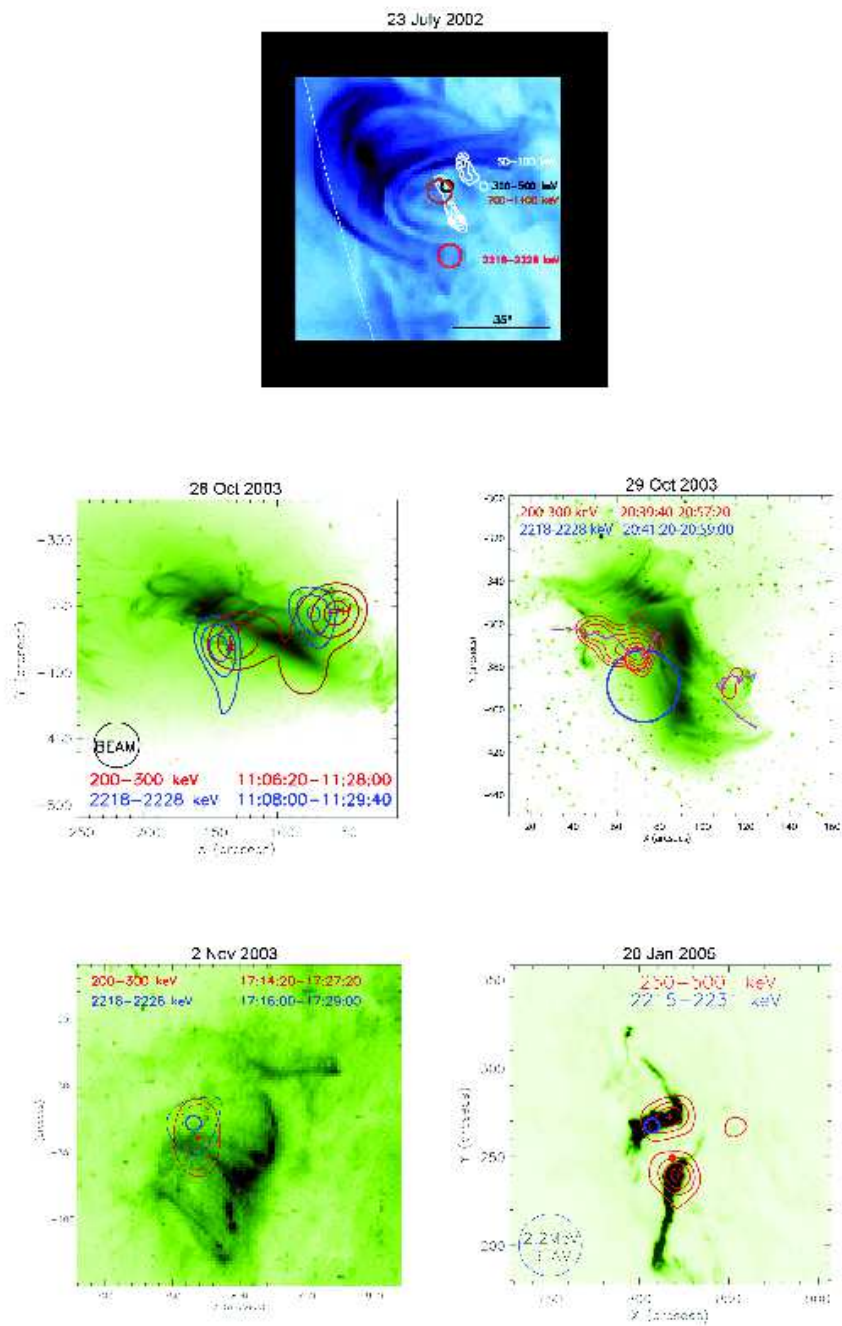}
\end{center}
\caption{Respective locations of $\gamma$-ray sources at 2.223~MeV and of hard X-ray sources observed with \textit{RHESSI} for different flares. 
The SOL2002-07-23T00:35 (X4.8) event; the circles represent the 1~$\sigma$ errors for the 300-500~keV (black), 700-1400 keV (red) and 2218-2228 keV (pink) maps made with identical parameters.  
The white contours show the high resolution 50-100~keV map with 3~arcsec resolution. The background is a \textit{TRACE} image showing the post-flare loops \citep[adapted from][]{2003ApJ...595L..77H}. 
The SOL2003-10-28T11:10 (X17.2) event; overlay of contours at 50\%, 70\%, and 90\% of 35$'' \gamma$-ray images made with detectors~6 and~9 on a \textit{TRACE} image. 
The red plus signs indicate 200-300~keV footpoint locations at different times. 
The SOL2003-10-29T20:49 (X10.0) event; fine red contours: (50\%-90\%) of the 200-300~keV high resolution maps superposed on a \textit{TRACE} image. 
The purple line indicates the motions of the 200-300~keV footpoints. 
Thick red and blue circles: centroid locations with 1~$\sigma$ error for detector~9 imaging at 200-300 and 2218-2228~keV, respectively. 
The SOL2003-11-02T17:25 (X8.3) event; overlay of 200-300~keV (red contours) and 2218-2228~keV (blue contours) sources on a postflare \textit{SOHO}/EIT image. 
The dashed and thin solid red contours (50\%) show the result of 200-300 keV mapping with 7$''$ and 35$''$ respectively. 
The thin blue contour (50\%) is the corresponding map at 2218-2228~keV made with detectors~6 and~9. 
The thicker 1$\sigma$ error circles show the detector~6 source centroid location \citep[adapted from][]{2006ApJ...644L..93H} for
the SOL2005-01-20T07:01 (X7.1) event; contours of the 250-500~keV emission overlaid on a \textit{TRACE} image and centroids (in red and blue) of the map performed in the 250-500~keV and 2215-2231~keV range.
}
\index{flare (individual)!SOL2002-07-23T00:35 (X4.8)!illustration}
\index{flare (individual)!SOL2003-10-28T11:10 (X17.2)!illustration}
\index{flare (individual)!SOL2003-10-29T20:49 (X10.0)!illustration}
\index{flare (individual)!SOL2003-11-02T17:25 (X8.3)!illustration}
\index{flare (individual)!SOL2005-01-20T07:01 (X7.1)!illustration}
\label{hurford}
\end{figure}

The complete set of images obtained to date is shown in Figure \ref{hurford}  with quantitative details summarized in Table~\ref{tab:imaging} (Hurford et al. 2003, 2006a, 2006b).
 \nocite{2003ApJ...595L..77H,2006ApJ...644L..93H,2006SPD....37.2804H}
Rather than deal with each flare in turn, we summarize these observations from three perspectives:  the number of distinct spatial components; their size; and their location\index{flare (individual)!SOL2002-07-23T00:35 (X4.8)!gamma@$\gamma$-ray imaging}\index{flare (individual)!SOL2003-10-28T11:10 (X17.2)!gamma@$\gamma$-ray imaging}\index{flare (individual)!SOL2003-10-29T20:49 (X10.0)!gamma@$\gamma$-ray imaging}\index{flare (individual)!SOL2003-11-02T17:25 (X8.3)!gamma@$\gamma$-ray imaging}\index{flare (individual)!SOL2005-01-20T07:01 (X7.1)!gamma@$\gamma$-ray imaging}.

Considering first the number of distinct components, four of the five events presented a single, unresolved source whereas the fifth event, SOL2003-10-28T11:10 (X17.2), showed a double source that straddled an arcade of loops.\index{loops!gamma@$\gamma$-ray image structure} 
This pattern does not necessarily imply, however, that double footpoint sources are an atypical feature of the $\gamma$-ray emission.  
Two conditions must be met to image a double footpoint source:  the separation of the footpoints must be sufficiently large that they can be resolved with the available 35~arcsec resolution; and there must be sufficient signal-to-noise that two sources (rather than one) can be confidently detected.  
Referring to the count statistics in Table~\ref{tab:imaging} and using the separation of the electron-bremsstrahlung components as a guide, SOL2003-10-28T11:10 (X17.2) was the only one of the five that satisfied both criteria and that event {\it did} display a double source.  
Had $\gamma$-ray emission from the other events also been in the form of double sources, this could not have been seen\index{footpoints!gamma@$\gamma$-rays}\index{gamma-rays!footpoint sources}.
Therefore, based on the number of observed components, there is no evidence to suggest that the true neutron-capture line sources are predominantly single sources, a result that would be in contrast to typical hard X-ray double sources\index{gamma-rays!image properties contrasted with HXR images}.
 
A second perspective is provided by the estimates of the size of the unresolved components.  
Such estimates can be obtained from comparisons of the total observed line flux to that imaged in the unresolved source.  
The ratio of imaged flux to total flux, called the relative visibility\index{gamma-rays!imaging!visibilities}, can be used to put upper limits on the size of an unresolved source.
\index{RHESSI@\textit{RHESSI}!visibilities}
\index{visibilities!relative} 
Ignoring the modulation, the total number of observed counts in the line provides a direct indication of the total, spatially-integrated flux in the narrow energy range associated with the line.  
Such counts are due to a combination of real solar neutron-capture line events, continuum solar emission and background.  
Since the number of such counts that are due to solar continuum or background continuum can be reliably estimated from energy ranges just above 2.223~MeV, an accurate determination of the spatially integrated solar neutron-capture line flux can be made, independent of background models.  The imaged flux does not contain any unmodulated background, but could include a contribution from the solar continuum.  Although this contribution could be estimated by imaging in a narrow energy range above 2.223~MeV, it is easy to demonstrate that, thanks to the narrow energy range required for imaging the neutron-capture line (10-16 keV), its contribution is negligible.  Note that since the relative visibilities are estimated by comparing the imaged to total counts observed with a given rotating modulation collimator, the measurement of relative visibilities is independent of potential errors in estimating the detector efficiency.\index{rotating modulation collimator}

The observed relative visibilities (Table~\ref{tab:imaging}) show that in all cases the flux in the unresolved neutron-capture line source(s) is consistent with the spatially-integrated flux.  
This implies that there is no evidence for a large-scale diffuse component.  
By assuming a Gaussian source model, upper limits to the FWHM source diameters are obtained that correspond to a few tens of arcseconds.

A third perspective on the neutron-capture line sources is provided by their location.   
An unanticipated result was that in some, if not all, cases, the neutron-capture line sources were not co-spatial with the corresponding electron-bremsstrahlung sources\index{bremsstrahlung!sources displaced from 2.223-MeV sources}\index{gamma-rays!sources displaced from bremsstrahlung sources}.
Since the character and statistical significance of the displacements differed from event to event, we first consider them individually.  
For SOL2003-10-28T11:10 (X17.2), double footpoint $\gamma$-ray sources straddled the same loop arcade as the electron sources.  
However, they were displaced along the loop arcade by 14$''$~and 17$''\pm$5$''$ respectively (Figure~\ref{hurford}).  
For the other events, the location of the single unresolved sources are expressed as a centroid location with appropriate statistical errors.  
For SOL2003-07-23T00:35 (X4.8), the neutron-capture line source was displaced by 25$''\pm$5$''$ from the corresponding electron source.  
Although it was not located near any obvious EUV or H$\alpha$ line emission, its location was consistent with the footpoint of a post-flare loop.  
For SOL2005-01-20T07:01 (X7.1) the neutron source was clearly associated with one of the two electron-bremsstrahlung footpoints.  
For SOL2003-11-02T17:25 (X8.3) the location also appeared to be preferentially associated with one of the two electron-bremsstrahlung footpoints, although in this case the displacement of the 2.2~MeV emission compared to the centroid of the electron-bremsstrahlung source was significant only at the 2$\sigma$ level. 
 For SOL2003-10-29T20:49 (X10.0), the measured separation of the corresponding centroids was 13$''\pm$17$''$, so a separation, if any, could not be established.  
 In summary, statistically significant displacements were observed in three of the five events, marginally in a fourth but not seen in the fifth and statistically least significant event\index{gamma-rays!imaging!event summary}.  
 \index{flare (individual)!SOL2002-07-23T00:35 (X4.8)!$\gamma$-ray imaging}
\index{flare (individual)!SOL2003-10-28T11:10 (X17.2)!$\gamma$-ray imaging}
\index{flare (individual)!SOL2003-10-29T20:49 (X10.0)!$\gamma$-ray imaging}
\index{flare (individual)!SOL2003-11-02T17:25 (X8.3)!$\gamma$-ray imaging}
\index{flare (individual)!SOL2005-01-20T07:01 (X7.1)!$\gamma$-ray imaging}

\subsection{Other $\gamma$-ray imaging}
\index{gamma-rays!imaging!continuum}
\index{imaging!gamma@$\gamma$-ray}

In addition to the neutron-capture line, imaging of the $\gamma$-ray continuum in the 200-800 keV range dominated by electron bremsstrahlung has also been reported \citep{2005AGUFMSH23A0313H,2006SPD....37.2804H,2008ApJ...678L..63K}.
\index{gamma-rays!imaging!continuum}
At higher energies, where nuclear emission is more important, only \textit{RHESSI} detector~9 (183$''$) was able to modulate effectively with the events studied.  
The tentative explanation for the reduced efficiency of detector~6 modulation for imaging in a broad energy band at high energies is that photons passing through front grid slits into rear grid slats can scatter into the detector and be counted as part of the modulated flux.  
The modulation of such events, however, is 180$^{\circ}$ out of phase with that of the valid (front slit-to-rear slit) events and so tends to reduce the effective modulation efficiency.  
This issue is a factor for detector~6 (but not detector~9) since the detector~6 grids are not thick enough to suppress the scattered photons.  
This issue is not relevant for neutron-capture line imaging since photons that are scattered by the front grid scatter away from the narrow window used for spectral analysis.

Nevertheless, the centroid locations obtained for the SOL2005-01-20T07:01 (X7.1)  $\gamma$-ray continuum  with detector~9 in four bands between 600~keV and 8~MeV were of interest particularly because of their energy dependence.
\index{flare (individual)!SOL2005-01-20T07:01 (X7.1)!$\gamma$-ray imaging}  
At the lowest energies in this band, their centroid locations matched that of the centroid of the X-ray electron-bremsstrahlung source.  
However, at higher energies where the nuclear contribution became more important, the $\gamma$-ray continuum location trended toward the neutron-capture line centroid \citep[][]{2006SPD....37.2804H}.  
Such a pattern is consistent with the displacement between the sites of electron-bremsstrahlung collisions and nuclear interactions.

Imaging in a narrow energy range surrounding the 511~keV line was also attempted for SOL2005-01-20.
\index{gamma-rays!imaging!in 511 keV line}  
Images obtained independently from the front and rear segments of detector~9 were found to be self-consistent in terms of strength and intensity.  
However, modeling suggests that although a narrow 500-520~keV energy range was used, the imaged fluence was dominated by the electron-bremsstrahlung continuum.

\subsection{Discussion}

One well-established result of the neutron-capture line mapping was that the flux for all five events was consistent with coming from compact source component(s) in the flaring active region.  This would suggest that the flare-associated $\gamma$-ray emission was associated with flare-accelerated ions, and not with ions that are shock-accelerated at high altitudes.\index{shocks!particle acceleration!and $\gamma$-rays}\index{ions!shock acceleration}\index{acceleration!shock}
If that were the case, the ions might be expected to interact with ambient material in a more diffuse pattern, perhaps at some distance from the flare. This shows that the $\gamma$-ray emission does not result from the interaction of solar energetic particles with the solar wind as alternatively proposed by \citet{2008ApJ...675..846K}.

Given the resolution and sensitivity of the imaging and the limited sample of events, there is no evidence that the morphologies of nuclear $\gamma$-ray sources (e.g., double vs. single footpoints) differ systematically from that of hard X-ray sources\index{gamma-rays!limitations of imaging}\index{caveats!gamma@$\gamma$-ray imaging}.

The most striking result, of course, is the differences in the locations of the electron-bremsstrahlung and neutron-capture line sources.  
Hypotheses that attempt to account for such differences generally fall into three catagories:  i) the displacement is an instrumental artifact; ii) the displacement has a solar, but relatively mundane origin; or iii) the displacement reflects differences in acceleration and/or transport processes for flare-accelerated ions and electrons\index{transport!ions!and $\gamma$-ray imaging}.

We first consider the possibility that the displacements are an artifact of the imaging process.  
There are three independent arguments indicating that this is highly unlikely.  
First, the measurements of the locations of the sources at two different energy regimes were based on analyses that differed only in the energy of the counts upon which the imaging is based.  
The same time interval, detector grid optics, aspect solution, imaging algorithm and imaging parameters were used for both so that the potential for systematic co-location errors can be discounted.  
Second, if the relative displacements had a systematic instrumental origin they would likely be with respect to the only axis of symmetry associated with the imaging process, namely the line joining the direction of the rotation axis to that of the flare.  
However, the displacements were neither systematically parallel to or perpendicular to this direction\index{gamma-rays!imaging}.
Third, in all but SOL2002-07-23T00:35, the direction of the displacement, if any, was in a magnetically significant solar direction, as opposed to some randomly-chosen direction\index{magnetic structures!interpretation of $\gamma$-ray imaging}.
As result of these three considerations, it is considered highly unlikely that the displacements have an instrumental origin.  

A second class of potential explanations for the displacements is that they have a solar, but relatively mundane explanation.  
One such explanation might be based on the average distance between the locations of the original nuclear interaction and that of the neutron-capture process.  
However, as mentioned, such a displacement is expected to be less than $\sim$1$''$.  
Another possibility is that the displacement is related to the time-delay associated with thermalization.  
This scenario notes that the neutron-capture line image reflects the location of nuclear interactions that occurred $\sim$100 seconds earlier.  
To investigate this hypothesis, for each event the neutron-capture line images were compared both to cotemporaneous electron-bremsstrahlung images and to electron-bremsstrahlung images made with integration times shifted by 100~s.  
No significant differences were observed.

These considerations therefore support a solar-physics-based interpretation of the displacement, based on a difference in acceleration and/or propagation between accelerated ions and electrons.  Several such interpretations have been put forward.  
\citet {2004ApJ...602L..69E} discussed a stochastic acceleration\index{acceleration!stochastic} mechanism whereby ions are preferentially accelerated in larger loops than electrons.\index{loops!particle acceleration}
Imaging for SOL2002-07-23 \citep{2003ApJ...595L..77H} would support this interpretation\index{hard X-rays!timing relative to $\gamma$-rays}.
The interpretation of time delays between $\gamma$-ray lines and hard X-rays by trap-plus-precipitation models\index{flare models!trap-plus-precipitation} for the same flare \citep{2007A&A...468..289D} also confirm the fact that ions
are injected in longer loops than electrons.\index{models!trap-plus-precipitation!and $\gamma$-ray imaging}

However, such an explanation would not appear to be appropriate for the SOL2003-10-28T11:10 (X17.2) where the footpoint separations for the electrons and ions were similar.  
\index{flare (individual)!SOL2003-10-28T11:10 (X17.2)!$\gamma$-ray imaging}
Gradient and curvature drifts were also considered by \citet{2006ApJ...644L..93H}, but the predicted displacements between the ions and electrons were two orders of magnitude too small.  
Another possibility, raised by \citet{1993SoPh..146..127L}, was that electric-field acceleration would lead to separation of particles with opposite charge \citep{2004ApJ...604..884Z}.  
Its potential applicability in these cases has not yet been pursued.

In two of the five events (SOL2005-01-20T17:25 and SOL2003-11-02T17:25) the ions were preferentially associated with one of the two electron bremsstrahlung footpoints.  
This raises the possibility that at least for these two events, the different footpoint weighting between the two species might be attributed to differences between electron and ion propagation in asymmetrical fields\index{gamma-rays!imaging!and asymmetric footpoints}.
This would be a result of the significant differences in the scattering properties and ionization-loss stopping distances between the accelerated electrons and ions.  
Modeling to support such a hypothesis has not been carried out, however.
\index{flare (individual)!SOL2002-07-23T00:35 (X4.8)!$\gamma$-ray imaging}
\index{flare (individual)!SOL2005-01-20T07:01 (X7.1)!$\gamma$-ray imaging}

With only five imaged events, it is too early to establish the systematics of the displacements. 
 It is also apparent that \textit{RHESSI}'s 35$''$ resolution is insufficient to resolve the individual $\gamma$-ray sources.  
 Therefore, it is to be hoped that \textit{RHESSI} will observe additional events in the coming maximum so that their systematic properties can be further explored and that new instrumentation will become available with improved angular resolution and sensitivity in the $\gamma$-ray regime.

\section{Conclusions and future perspectives}

Two particular observational advances characterize the \textit{RHESSI} era: spectral resolution adequate to reveal line shapes and to better constrain the line fluences; and $\gamma$-ray imaging.
\index{eras!RHESSI@\textit{RHESSI}}
The former provides in principle previously inaccessible information on proton and $\alpha$-particle energy and angular distributions. 
The latter has revealed the sites of ion interaction and neutron production and most dramatically shown that they do not coincide with the HXR-identified locations of electron interaction. 
Also, it has not confirmed the \textit{SMM}-era
\index{eras!SMM@\textit{SMM}} 
suggestion that 2.223~MeV line
emission might be produced over an extended region \citep{1993ApJ...409L..69V}. 
The observations are also not consistent with a production of $\gamma$-rays 
through the interaction of solar energetic protons with the solar wind as examined by \citet{2008ApJ...675..846K}.
We may look forward to further advances, both observational \citep[e.g., with GRIPS;][]{2009SPD....40.1810S}, and theoretical, over the next few years.

At the same time as \textit{RHESSI} as well as \textit{INTEGRAL}/SPI observations have brought pioneering observations of $\gamma$-ray spectra at high resolution, it was also realized that the determination of all the parameters which characterize the $\gamma$-ray spectrum (target abundances and ionization states, energy and angular distributions of energetic ions, relative abundances of accelerated $\alpha$ particles and heavier ions, transport and energy losses of energetic ions between the acceleration region and interaction region if not co-spatial, etc.) is still very challenging.\index{acceleration region!distinguished from energy-loss region}
In particular, the different parameters cannot be determined independently from the assumptions made on others: both ambient abundances and parameters of energetic particles must be deduced together even in the ``simple" analysis of a $\gamma$-ray spectrum. 
Combining the information from narrow line shapes and line fluence ratios can, however, help to disentangle the effects of the different parameters. 
The use of lines purely produced by $\alpha$ particles (in particular the line at 0.339 MeV formed by the interaction of $\alpha$ particles with iron) would also considerably improve the analysis of 
$\gamma$-ray spectra by providing cleaner constraints on the distribution of $\alpha$ particles. 
A few attempts had been made at the time of \textit{SMM}, but no further studies have been published since then based on spectra obtained at higher resolution. 
This is also the case for the lines produced by energetic $^3$He on ambient $^{16}$O around 1~MeV.
 Line shapes and widths of the 511~keV annihilation line and positronium continuum have been measured at high spectral resolution leading for a few flares to unexpected combination of chromospheric densities and transition region temperatures for the annihilation region.
 \index{transition region!511-keV line width}
Taking this into account in the future, the production and annihilation of positrons over a broad range of heights and thus of conditions in the medium is probably the next step to undertake to better understand the observations of shapes and widths of the annihilation line and continuum. 

Heavier ($Z > 2$) accelerated ion species are harder to diagnose but they may nonetheless be significant in overall flare energetics 
\citep{1997ApJ...490..883M,2004JGRA..10910104E}. The detailed treatment now available of the unresolved component of nuclear $\gamma$-rays should bear significantly on this question \citep{2009ApJS..183..142M}. 
Heavier species are prolific producers of free neutrons via evaporation-type reactions so instruments planned to measure, on future inner-heliosphere missions, neutrons in the 1-10~MeV range not yet explored systematically  \citep[e.g.,][]{2009SPIE.7438E..19W} will yield new constraints.

Do all flares accelerate ions\index{ions!ubiquity?}? 
It seems unlikely that they do not, but all flares do not produce detectable $\gamma$-ray lines. 
\textit{RHESSI}'s eight years of operation have accumulated a data set bearing on this question \citep{2009ApJ...698L.152S}. 
A good correlation is found between the energy content of electrons above 300 keV and protons above 30~MeV suggesting a strong correlation between the acceleration of high-energy electrons and $\gamma$-ray line producing ions. 
A larger variation is found between the energy content of electrons above 20~keV and the ion energy content. 
Some episodes in flares or some specific flares exhibit indeed higher relative numbers of electrons with respect to ions. 
Based on the \textit{RHESSI} data set, it was also found that there is a strong link between the production of energetic ions and sub-relativistic electrons, but only when a threshold in the production of ions is attained. In conclusion, while the acceleration of protons above 30~MeV is related to the acceleration of relativistic electrons, the acceleration of sub-relativistic electrons is linked to the acceleration of these energetic protons and relativistic electrons, but only when a threshold of energetic ions and electrons is reached. 
These conclusions must be kept in mind when addressing acceleration mechanisms in flares. 
The content in low energy ions from flares is mostly unknown and no strong constraints have been derived on the production of low energy protons in solar flares. Attempts to constrain fluxes from low energy proton capture lines could be a way to address this issue. 
Another promising way could be to detect suprathermal ions through charge exchange with the ambient hydrogen \citep[see the first detection of energetic neutral hydrogen atoms in a solar flare by][]{2009ApJ...693L..11M}.\index{suprathermal populations!and charge-exchange reactions}\index{reactions!charge-exchange}

Apart from \textit{CORONAS-F}/SONG \citep[e.g.,][]{2006SoSyR..40..104K}, solar observations at the 
highest $\gamma$-ray energies have been lacking in recent years\index{satellites!CORONAS-F@\textit{CORONAS-F}}.
This is one of the only ways (with the detection of neutrons) to address the issue of the highest energetic particles accelerated in solar flares and to determine the shape of the energetic ion spectrum over several decades of energy range. 
The global network of neutron detectors thus plays at the present time a particularly important role, yielding otherwise unobtainable information on ion energy distributions in the GeV energy range \citep[e.g.,][]{2009SoPh..257..305M}. 
As the solar cycle picks up, \textit{Fermi} observations of the high-energy continuum should become critical to probing the highest ion energies attained in flares.
\index{satellites!Fermi@\textit{Fermi}}
Further observations of flares at submillimeter (and even shorter) wavelengths \citep[see, e.g.,][]{2006via..conf...49K} as well as the search for X-ray inverse Compton radiation in coronal sources 
\citep{2010A&A...510A..29M} could additionally provide information on electrons and positrons produced in the 10-100 MeV energy range in flares.
\index{coronal sources!inverse Compton radiation}\index{inverse Compton radiation}

Recent years have seen a re-examination of the thick-target interpretation of flare HXRs \citep{2008ApJ...675.1645F,2009A&A...508..993B}\index{thick-target model!hard X-rays}.
Since HXR-emitting electrons are believed to embody a large fraction of the total energy released in the flare, this is a question that goes
beyond interpretation of radiation signatures to the heart of the flare energy release process. 
The existing thick-target models assume that 
electrons radiate negligibly in the (coronal) acceleration region, and that they slow down collisionally in the (thick-target, chromospheric) interaction region.\index{acceleration region!distinguished from energy-loss region}
If acceleration and interaction regions were one and the same, fewer electrons would be needed to account for HXRs. 
A similar reconsideration of $\gamma$-ray source regions may prove necessary. 

\begin{acknowledgements}

This chapter is dedicated to Reuven Ramaty who was one of the initiators of the field of high energy physics in solar flares, inducting two of us (NV and ALM) into the field and inspiring many other high energy solar physicists. 
The authors are very grateful to the organizers of the series of \textit{RHESSI} workshops and to the team leaders of the working group on energetic ions and particle acceleration who stimulated the many discussions and collaborations that led to many recent results summarized in this chapter.  
We would like to express our special thanks to G. Share, R. Murphy, V. Tatischeff and K. Watanabe for useful suggestions for this chapter and discussions. 
Nicole Vilmer acknowledges support from the Centre National d'Etudes Spatiales (CNES) and from the French program on Solar-Terrestrial Physics (PNST) of INSU/CNRS for the participation to the \textit{RHESSI} project. 
Collaborative work on the chapter in Meudon was supported by a British Council Franco-British Alliance grant. 
Solar physics research in Glasgow is supported by an STFC Rolling Grant.

\end{acknowledgements}

\bibliographystyle{ssrv}
\bibliography{ch4,book_chapters}

\begin{thebibliography}{205}
\expandafter\ifx\csname natexlab\endcsname\relax\def\natexlab#1{#1}\fi
\expandafter\ifx\csname url\endcsname\relax
  \def\url#1{{\tt #1}}\fi
\expandafter\ifx\csname urlprefix\endcsname\relax\def\urlprefix{URL }\fi
\providecommand{\eprint}[2][]{\url{#2}}

\bibitem[{{Akimov} et~al.(1992){Akimov}, {Afanasev}, {Belousov}, {Blokhintsev},
  {Volzhenskaya}, {Kalinkin}, {Leikov}, {Nesterov}, {Galper}, {Voronov},
  {Zemskov}, {Kirillov-Ugryumov}, {Luchkov}, {Ozerov}, {Popov}, {Rudko},
  {Runtso}, {Chesnokov}, {Kurnosova}, {Rusakovich}, {Topchiev}, {Fradkin},
  {Chuikin}, {Tugaenko}, {Tian}, {Ishkov}, {Gros}, {Grenier}, {Barouch},
  {Wallin}, {Baser-Bachi}, {Lavigne}, {Olive}, \&
  {Juchniewicz}}]{1992SvAL...18...69A}
V.~V. {Akimov}, V.~G. {Afanasev}, A.~S. {Belousov}, I.~D. {Blokhintsev}, V.~A.
  {Volzhenskaya}, L.~F. {Kalinkin}, N.~G. {Leikov}, V.~E. {Nesterov}, A.~M.
  {Galper}, S.~A. {Voronov}, V.~M. {Zemskov}, V.~G. {Kirillov-Ugryumov}, B.~I.
  {Luchkov}, Y.~V. {Ozerov}, A.~V. {Popov}, V.~A. {Rudko}, M.~F. {Runtso},
  V.~Y. {Chesnokov}, L.~V. {Kurnosova}, M.~A. {Rusakovich}, N.~P. {Topchiev},
  M.~I. {Fradkin}, E.~I. {Chuikin}, V.~Y. {Tugaenko}, T.~N. {Tian}, V.~I.
  {Ishkov}, M.~{Gros}, I.~{Grenier}, E.~{Barouch}, P.~{Wallin}, A.~R.
  {Baser-Bachi}, J.~M. {Lavigne}, J.~F. {Olive}, J.~{Juchniewicz}, Soviet
  Astronomy Letters {\bf 18\/}, 69 (1992)

\bibitem[{{Alexander} et~al.(1994){Alexander}, {Dunphy}, \&
  {MacKinnon}}]{1994SoPh..151..147A}
D.~{Alexander}, P.~P. {Dunphy}, A.~L. {MacKinnon}, \solphys {\bf 151\/}, 147
  (1994), doi:10.1007/BF00654088

\bibitem[{{Asplund} et~al.(2005){Asplund}, {Grevesse}, \&
  {Sauval}}]{2005ASPC..336...25A}
M.~{Asplund}, N.~{Grevesse}, A.~J. {Sauval}, in {\em Cosmic Abundances as
  Records of Stellar Evolution and Nucleosynthesis\/}, ed. by T.~G. {Barnes},
  III, F.~N. {Bash} (2005), volume 336 of {\em Astronomical Society of the
  Pacific Conference Series\/}, pp. 25--+

\bibitem[{{Barat} et~al.(1994){Barat}, {Trottet}, {Vilmer}, {Dezalay}, {Talon},
  {Sunyaev}, {Terekhov}, \& {Kuznetsov}}]{1994ApJ...425L.109B}
C.~{Barat}, G.~{Trottet}, N.~{Vilmer}, J.~{Dezalay}, R.~{Talon}, R.~{Sunyaev},
  O.~{Terekhov}, A.~{Kuznetsov}, \apjl {\bf 425\/}, L109 (1994),
  doi:10.1086/187322

\bibitem[{{Bieber} et~al.(2005){Bieber}, {Clem}, {Evenson}, {Pyle}, {Ruffolo},
  \& {S{\'a}iz}}]{2005GeoRL..3203S02B}
J.~W. {Bieber}, J.~{Clem}, P.~{Evenson}, R.~{Pyle}, D.~{Ruffolo},
  A.~{S{\'a}iz}, \grl {\bf 32\/}, 3 (2005), doi:10.1029/2004GL021492

\bibitem[{{Biermann} et~al.(1951){Biermann}, {Haxel}, \&
  {Schl{\"u}ter}}]{1951ZNatA...6...47B}
L.~{Biermann}, O.~{Haxel}, A.~{Schl{\"u}ter}, Zeitschrift Naturforschung Teil A
  {\bf 6\/}, 47 (1951)

\bibitem[{{Brosius}(2001)}]{2001ApJ...555..435B}
J.~W. {Brosius}, \apj {\bf 555\/}, 435 (2001), doi:10.1086/321438

\bibitem[{{Brown} et~al.(2009){Brown}, {Turkmani}, {Kontar}, {MacKinnon}, \&
  {Vlahos}}]{2009A&A...508..993B}
J.~C. {Brown}, R.~{Turkmani}, E.~P. {Kontar}, A.~L. {MacKinnon}, L.~{Vlahos},
  \aap {\bf 508\/}, 993 (2009), \eprint{0909.4243},
  doi:10.1051/0004-6361/200913145

\bibitem[{{Canfield} \& {Chang}(1985)}]{1985ApJ...295..275C}
R.~C. {Canfield}, C.-R. {Chang}, \apj {\bf 295\/}, 275 (1985),
  doi:10.1086/163371

\bibitem[{{Chupp}(1984)}]{1984ARA&A..22..359C}
E.~L. {Chupp}, \araa {\bf 22\/}, 359 (1984),
  doi:10.1146/annurev.aa.22.090184.002043

\bibitem[{{Chupp}(1996)}]{1996AIPC..374....3C}
E.~L. {Chupp}, in {\em American Institute of Physics Conference Series\/}, ed.
  by {R.~Ramaty, N.~Mandzhavidze, \& X.-M.~Hua} (1996), volume 374 of {\em
  American Institute of Physics Conference Series\/}, pp. 3--34,
  doi:10.1063/1.50997

\bibitem[{{Chupp} et~al.(1987){Chupp}, {Debrunner}, {Flueckiger}, {Forrest},
  {Golliez}, {Kanbach}, {Vestrand}, {Cooper}, \& {Share}}]{1987ApJ...318..913C}
E.~L. {Chupp}, H.~{Debrunner}, E.~{Flueckiger}, D.~J. {Forrest}, F.~{Golliez},
  G.~{Kanbach}, W.~T. {Vestrand}, J.~{Cooper}, G.~{Share}, \apj {\bf 318\/},
  913 (1987), doi:10.1086/165423

\bibitem[{{Chupp} \& {Dunphy}(2000)}]{2000ASPC..206..400C}
E.~L. {Chupp}, P.~P. {Dunphy}, in {\em High Energy Solar Physics Workshop -
  Anticipating Hess!\/}, ed. by {R.~Ramaty \& N.~Mandzhavidze} (2000), volume
  206 of {\em Astronomical Society of the Pacific Conference Series\/}, pp.
  400--+

\bibitem[{{Chupp} et~al.(1973){Chupp}, {Forrest}, {Higbie}, {Suri}, {Tsai}, \&
  {Dunphy}}]{1973Natur.241..333C}
E.~L. {Chupp}, D.~J. {Forrest}, P.~R. {Higbie}, A.~N. {Suri}, C.~{Tsai}, P.~P.
  {Dunphy}, \nat {\bf 241\/}, 333 (1973), doi:10.1038/241333a0

\bibitem[{{Chupp} et~al.(1981){Chupp}, {Forrest}, {Ryan}, {Cherry}, {Reppin},
  {Kanbach}, {Rieger}, {Pinkau}, {Share}, \& {Kinzer}}]{1981ApJ...244L.171C}
E.~L. {Chupp}, D.~J. {Forrest}, J.~M. {Ryan}, M.~L. {Cherry}, C.~{Reppin},
  G.~{Kanbach}, E.~{Rieger}, K.~{Pinkau}, G.~H. {Share}, R.~L. {Kinzer}, \apjl
  {\bf 244\/}, L171 (1981), doi:10.1086/183505

\bibitem[{{Chupp} et~al.(1982){Chupp}, {Forrest}, {Ryan}, {Heslin}, {Reppin},
  {Pinkau}, {Kanbach}, {Rieger}, \& {Share}}]{1982ApJ...263L..95C}
E.~L. {Chupp}, D.~J. {Forrest}, J.~M. {Ryan}, J.~{Heslin}, C.~{Reppin},
  K.~{Pinkau}, G.~{Kanbach}, E.~{Rieger}, G.~H. {Share}, \apjl {\bf 263\/}, L95
  (1982), doi:10.1086/183931

\bibitem[{{Chupp} \& {Ryan}(2009)}]{2009RAA.....9...11C}
E.~L. {Chupp}, J.~M. {Ryan}, Research in Astronomy and Astrophysics {\bf 9\/},
  11 (2009), doi:10.1088/1674-4527/9/1/003

\bibitem[{{Chupp} et~al.(1993){Chupp}, {Trottet}, {Marschhauser}, {Pick},
  {Soru-Escaut}, {Rieger}, \& {Dunphy}}]{1993A&A...275..602C}
E.~L. {Chupp}, G.~{Trottet}, H.~{Marschhauser}, M.~{Pick}, L.~{Soru-Escaut},
  E.~{Rieger}, P.~P. {Dunphy}, \aap {\bf 275\/}, 602 (1993)

\bibitem[{{Cliver} et~al.(1994){Cliver}, {Crosby}, \&
  {Dennis}}]{1994ApJ...426..767C}
E.~W. {Cliver}, N.~B. {Crosby}, B.~R. {Dennis}, \apj {\bf 426\/}, 767 (1994),
  doi:10.1086/174113

\bibitem[{{Dauphin} \& {Vilmer}(2007)}]{2007A&A...468..289D}
C.~{Dauphin}, N.~{Vilmer}, \aap {\bf 468\/}, 289 (2007),
  doi:10.1051/0004-6361:20066247

\bibitem[{{Debrunner} et~al.(1989){Debrunner}, {Fl{\"u}ckiger}, \&
  {Stein}}]{1989NIMPA.278..573D}
H.~{Debrunner}, E.~O. {Fl{\"u}ckiger}, P.~{Stein}, Nuclear Instruments and
  Methods in Physics Research A {\bf 278\/}, 573 (1989),
  doi:10.1016/0168-9002(89)90881-4

\bibitem[{{Debrunner} et~al.(1997){Debrunner}, {Lockwood}, {Barat},
  {Buetikofer}, {Dezalay}, {Flueckiger}, {Kuznetsov}, {Ryan}, {Sunyaev},
  {Terekhov}, {Trottet}, \& {Vilmer}}]{1997ApJ...479..997D}
H.~{Debrunner}, J.~A. {Lockwood}, C.~{Barat}, R.~{Buetikofer}, J.~P. {Dezalay},
  E.~{Flueckiger}, A.~{Kuznetsov}, J.~M. {Ryan}, R.~{Sunyaev}, O.~V.
  {Terekhov}, G.~{Trottet}, N.~{Vilmer}, \apj {\bf 479\/}, 997 (1997),
  doi:10.1086/303895

\bibitem[{{Debrunner} et~al.(1993){Debrunner}, {Lockwood}, \&
  {Ryan}}]{1993ApJ...409..822D}
H.~{Debrunner}, J.~A. {Lockwood}, J.~M. {Ryan}, \apj {\bf 409\/}, 822 (1993),
  doi:10.1086/172712

\bibitem[{{Dingus} et~al.(1994){Dingus}, {Sreekumar}, {Bertsch}, {Fichtel},
  {Hartman}, {Hunter}, {Thompson}, {Schneid}, {Brazier}, {Kanbach}, {von
  Montigny}, {Mayer-Hasselwander}, {Lin}, {Michelson}, {Nolan}, {Kniffen}, \&
  {Mattox}}]{1994AIPC..294..177D}
B.~L. {Dingus}, P.~{Sreekumar}, D.~L. {Bertsch}, C.~E. {Fichtel}, R.~C.
  {Hartman}, S.~D. {Hunter}, D.~J. {Thompson}, E.~J. {Schneid}, K.~T.~S.
  {Brazier}, G.~{Kanbach}, C.~{von Montigny}, H.~A. {Mayer-Hasselwander}, Y.~C.
  {Lin}, P.~F. {Michelson}, P.~L. {Nolan}, D.~A. {Kniffen}, J.~R. {Mattox}, in
  {\em High-Energy Solar Phenomena - a New Era of Spacecraft Measurements\/},
  ed. by {J.~Ryan \& W.~T.~Vestrand} (1994), volume 294 of {\em American
  Institute of Physics Conference Series\/}, pp. 177--182, doi:10.1063/1.45187

\bibitem[{{Drake} \& {Testa}(2005)}]{2005Natur.436..525D}
J.~J. {Drake}, P.~{Testa}, \nat {\bf 436\/}, 525 (2005),
  \eprint{arXiv:astro-ph/0506182}, doi:10.1038/nature03803

\bibitem[{{Droege} et~al.(1996){Droege}, {Ruffolo}, \&
  {Klecker}}]{1996ApJ...464L..87D}
W.~{Droege}, D.~{Ruffolo}, B.~{Klecker}, \apjl {\bf 464\/}, L87+ (1996),
  \eprint{arXiv:astro-ph/9604019}, doi:10.1086/310093

\bibitem[{{Dunphy}(1999)}]{1999ICRC....6....9D}
P.~P. {Dunphy}, in {\em International Cosmic Ray Conference\/} (1999), volume~6
  of {\em International Cosmic Ray Conference\/}, pp. 9--+

\bibitem[{{Dunphy} \& {Chupp}(1991)}]{1991ICRC....3...65D}
P.~P. {Dunphy}, E.~L. {Chupp}, in {\em International Cosmic Ray Conference\/}
  (1991), volume~3 of {\em International Cosmic Ray Conference\/}, pp. 65--+

\bibitem[{{Dunphy} \& {Chupp}(1992)}]{1992AIPC..264..253D}
P.~P. {Dunphy}, E.~L. {Chupp}, in {\em Particle Acceleration in Cosmic
  Plasmas\/}, ed. by {G.~P.~Zank \& T.~K.~Gaisser} (1992), volume 264 of {\em
  American Institute of Physics Conference Series\/}, pp. 253--256,
  doi:10.1063/1.42736

\bibitem[{{Dunphy} et~al.(1999){Dunphy}, {Chupp}, {Bertsch}, {Schneid},
  {Gottesman}, \& {Kanbach}}]{1999SoPh..187...45D}
P.~P. {Dunphy}, E.~L. {Chupp}, D.~L. {Bertsch}, E.~J. {Schneid}, S.~R.
  {Gottesman}, G.~{Kanbach}, \solphys {\bf 187\/}, 45 (1999)

\bibitem[{{Emslie} et~al.(1997){Emslie}, {Brown}, \&
  {MacKinnon}}]{1997ApJ...485..430E}
A.~G. {Emslie}, J.~C. {Brown}, A.~L. {MacKinnon}, \apj {\bf 485\/}, 430 (1997),
  doi:10.1086/304423

\bibitem[{{Emslie} et~al.(2004{\natexlab{a}}){Emslie}, {Kucharek}, {Dennis},
  {Gopalswamy}, {Holman}, {Share}, {Vourlidas}, {Forbes}, {Gallagher}, {Mason},
  {Metcalf}, {Mewaldt}, {Murphy}, {Schwartz}, \&
  {Zurbuchen}}]{2004JGRA..10910104E}
A.~G. {Emslie}, H.~{Kucharek}, B.~R. {Dennis}, N.~{Gopalswamy}, G.~D. {Holman},
  G.~H. {Share}, A.~{Vourlidas}, T.~G. {Forbes}, P.~T. {Gallagher}, G.~M.
  {Mason}, T.~R. {Metcalf}, R.~A. {Mewaldt}, R.~J. {Murphy}, R.~A. {Schwartz},
  T.~H. {Zurbuchen}, Journal of Geophysical Research (Space Physics) {\bf
  109\/}, 10104 (2004{\natexlab{a}}), doi:10.1029/2004JA010571

\bibitem[{{Emslie} et~al.(2004{\natexlab{b}}){Emslie}, {Miller}, \&
  {Brown}}]{2004ApJ...602L..69E}
A.~G. {Emslie}, J.~A. {Miller}, J.~C. {Brown}, \apjl {\bf 602\/}, L69
  (2004{\natexlab{b}}), doi:10.1086/382350

\bibitem[{{Evenson} et~al.(1990){Evenson}, {Kroeger}, {Meyer}, \&
  {Reames}}]{1990ApJS...73..273E}
P.~{Evenson}, R.~{Kroeger}, P.~{Meyer}, D.~{Reames}, \apjs {\bf 73\/}, 273
  (1990), doi:10.1086/191462

\bibitem[{{Evenson} et~al.(1983){Evenson}, {Meyer}, \&
  {Pyle}}]{1983ApJ...274..875E}
P.~{Evenson}, P.~{Meyer}, K.~R. {Pyle}, \apj {\bf 274\/}, 875 (1983),
  doi:10.1086/161500

\bibitem[{{Feldman} et~al.(2010){Feldman}, {Lawrence}, {Goldsten}, {Gold},
  {Baker}, {Haggerty}, {Ho}, {Krucker}, {Lin}, {Mewaldt}, {Murphy}, {Nittler},
  {Rhodes}, {Slavin}, {Solomon}, {Starr}, {Vilas}, \&
  {Vourlidas}}]{2010JGRA..11501102F}
W.~C. {Feldman}, D.~J. {Lawrence}, J.~O. {Goldsten}, R.~E. {Gold}, D.~N.
  {Baker}, D.~K. {Haggerty}, G.~C. {Ho}, S.~{Krucker}, R.~P. {Lin}, R.~A.
  {Mewaldt}, R.~J. {Murphy}, L.~R. {Nittler}, E.~A. {Rhodes}, J.~A. {Slavin},
  S.~C. {Solomon}, R.~D. {Starr}, F.~{Vilas}, A.~{Vourlidas}, Journal of
  Geophysical Research (Space Physics) {\bf 115\/}, 1102 (2010),
  doi:10.1029/2009JA014535

\bibitem[{{Firstova} et~al.(2008){Firstova}, {Polyakov}, \&
  {Firstova}}]{2008SoPh..249...53F}
N.~M. {Firstova}, V.~I. {Polyakov}, A.~V. {Firstova}, \solphys {\bf 249\/}, 53
  (2008), doi:10.1007/s11207-008-9165-0

\bibitem[{{Fleishman} \& {Kontar}(2010)}]{2010ApJ...709L.127F}
G.~D. {Fleishman}, E.~P. {Kontar}, \apjl {\bf 709\/}, L127 (2010),
  \eprint{0911.5335}, doi:10.1088/2041-8205/709/2/L127

\bibitem[{{Fletcher} \& {Hudson}(2008)}]{2008ApJ...675.1645F}
L.~{Fletcher}, H.~S. {Hudson}, \apj {\bf 675\/}, 1645 (2008),
  \eprint{0712.3452}, doi:10.1086/527044

\bibitem[{{Forrest}(1983)}]{1983AIPC..101....3F}
D.~J. {Forrest}, in {\em Positron-Electron Pairs in Astrophysics\/}, ed. by
  {R.~Ramaty} (1983), volume 101 of {\em American Institute of Physics
  Conference Series\/}, pp. 3--14, doi:10.1063/1.34119

\bibitem[{{Forrest} et~al.(1985){Forrest}, {Vestrand}, {Chupp}, {Rieger},
  {Cooper}, \& {Share}}]{1985ICRC....4..146F}
D.~J. {Forrest}, W.~T. {Vestrand}, E.~L. {Chupp}, E.~{Rieger}, J.~F. {Cooper},
  G.~H. {Share}, in {\em International Cosmic Ray Conference\/}, ed. by
  {F.~C.~Jones} (1985), volume~4 of {\em International Cosmic Ray
  Conference\/}, pp. 146--149

\bibitem[{{Gabriel}(1976)}]{1976RSPTA.281..339G}
A.~H. {Gabriel}, Royal Society of London Philosophical Transactions Series A
  {\bf 281\/}, 339 (1976)

\bibitem[{{Gan}(2005)}]{2005AdSpR..35.1833G}
W.~Q. {Gan}, Advances in Space Research {\bf 35\/}, 1833 (2005),
  doi:10.1016/j.asr.2004.11.025

\bibitem[{{Gros} et~al.(2004){Gros}, {Tatischeff}, {Kiener}, {Cordier},
  {Chapuis}, {Weidenspointner}, {Vedrenne}, {von Kienlin}, {Diehl}, {Bykov}, \&
  {NM{\'e}ndez}}]{2004ESASP.552..669G}
M.~{Gros}, V.~{Tatischeff}, J.~{Kiener}, B.~{Cordier}, C.~{Chapuis},
  G.~{Weidenspointner}, G.~{Vedrenne}, A.~{von Kienlin}, R.~{Diehl},
  A.~{Bykov}, M.~{NM{\'e}ndez}, in {\em 5th INTEGRAL Workshop on the INTEGRAL
  Universe\/}, ed. by {V.~Schoenfelder, G.~Lichti, \& C.~Winkler} (2004),
  volume 552 of {\em ESA Special Publication\/}, pp. 669--+

\bibitem[{{Harris} et~al.(2007){Harris}, {Tatischeff}, {Kiener}, {Gros}, \&
  {Weidenspointner}}]{2007A&A...461..723H}
M.~J. {Harris}, V.~{Tatischeff}, J.~{Kiener}, M.~{Gros}, G.~{Weidenspointner},
  \aap {\bf 461\/}, 723 (2007), \eprint{arXiv:astro-ph/0610859},
  doi:10.1051/0004-6361:20066084

\bibitem[{{Henoux} et~al.(1990){Henoux}, {Chambe}, {Smith}, {Tamres},
  {Feautrier}, {Rovira}, \& {Sahal-Brechot}}]{1990ApJS...73..303H}
J.~C. {Henoux}, G.~{Chambe}, D.~{Smith}, D.~{Tamres}, N.~{Feautrier},
  M.~{Rovira}, S.~{Sahal-Brechot}, \apjs {\bf 73\/}, 303 (1990),
  doi:10.1086/191466

\bibitem[{{Holman et al.}(2011)}]{Chapter3}
G.~D. {Holman et al.}, \ssr pp. XXX--XXX (2011)

\bibitem[{{Hua} et~al.(2002){Hua}, {Kozlovsky}, {Lingenfelter}, {Ramaty}, \&
  {Stupp}}]{2002ApJS..140..563H}
X.~{Hua}, B.~{Kozlovsky}, R.~E. {Lingenfelter}, R.~{Ramaty}, A.~{Stupp},
  Astrophys.J.Supp. {\bf 140\/}, 563 (2002), doi:10.1086/339372

\bibitem[{{Hua} \& {Lingenfelter}(1987{\natexlab{a}})}]{1987SoPh..113..229H}
X.~{Hua}, R.~E. {Lingenfelter}, \solphys {\bf 113\/}, 229 (1987{\natexlab{a}}),
  doi:10.1007/BF00147702

\bibitem[{{Hua} et~al.(1989){Hua}, {Ramaty}, \&
  {Lingenfelter}}]{1989ApJ...341..516H}
X.~{Hua}, R.~{Ramaty}, R.~E. {Lingenfelter}, \apj {\bf 341\/}, 516 (1989),
  doi:10.1086/167513

\bibitem[{{Hua} \& {Lingenfelter}(1987{\natexlab{b}})}]{1987SoPh..107..351H}
X.-M. {Hua}, R.~E. {Lingenfelter}, \solphys {\bf 107\/}, 351
  (1987{\natexlab{b}}), doi:10.1007/BF00152031

\bibitem[{{Hudson} et~al.(1980){Hudson}, {Bai}, {Gruber}, {Matteson}, {Nolan},
  \& {Peterson}}]{1980ApJ...236L..91H}
H.~S. {Hudson}, T.~{Bai}, D.~E. {Gruber}, J.~L. {Matteson}, P.~L. {Nolan},
  L.~E. {Peterson}, \apjl {\bf 236\/}, L91 (1980), doi:10.1086/183205

\bibitem[{{Hulot} et~al.(1992){Hulot}, {Vilmer}, {Chupp}, {Dennis}, \&
  {Kane}}]{1992A&A...256..273H}
E.~{Hulot}, N.~{Vilmer}, E.~L. {Chupp}, B.~R. {Dennis}, S.~R. {Kane}, \aap {\bf
  256\/}, 273 (1992)

\bibitem[{{Hulot} et~al.(1989){Hulot}, {Vilmer}, \&
  {Trottet}}]{1989A&A...213..383H}
E.~{Hulot}, N.~{Vilmer}, G.~{Trottet}, \aap {\bf 213\/}, 383 (1989)

\bibitem[{{Hurford} et~al.(2006{\natexlab{a}}){Hurford}, {Krucker}, {Lin},
  {Schwartz}, {Share}, \& {Smith}}]{2006SPD....37.2804H}
G.~J. {Hurford}, S.~{Krucker}, R.~P. {Lin}, R.~A. {Schwartz}, G.~H. {Share},
  D.~M. {Smith}, in {\em Bulletin of the American Astronomical Society\/}
  (2006{\natexlab{a}}), volume~38 of {\em Bulletin of the American Astronomical
  Society\/}, pp. 255--+

\bibitem[{{Hurford} et~al.(2006{\natexlab{b}}){Hurford}, {Krucker}, {Lin},
  {Schwartz}, {Share}, \& {Smith}}]{2006ApJ...644L..93H}
G.~J. {Hurford}, S.~{Krucker}, R.~P. {Lin}, R.~A. {Schwartz}, G.~H. {Share},
  D.~M. {Smith}, \apjl {\bf 644\/}, L93 (2006{\natexlab{b}}),
  doi:10.1086/505329

\bibitem[{{Hurford} et~al.(2005){Hurford}, {Krucker}, {Lin}, {Schwartz}, \&
  {Smith}}]{2005AGUFMSH23A0313H}
G.~J. {Hurford}, S.~{Krucker}, R.~P. {Lin}, R.~A. {Schwartz}, D.~M. {Smith},
  AGU Fall Meeting Abstracts pp. A313+ (2005)

\bibitem[{{Hurford} et~al.(2002){Hurford}, {Schmahl}, {Schwartz}, {Conway},
  {Aschwanden}, {Csillaghy}, {Dennis}, {Johns-Krull}, {Krucker}, {Lin},
  {McTiernan}, {Metcalf}, {Sato}, \& {Smith}}]{2002SoPh..210...61H}
G.~J. {Hurford}, E.~J. {Schmahl}, R.~A. {Schwartz}, A.~J. {Conway}, M.~J.
  {Aschwanden}, A.~{Csillaghy}, B.~R. {Dennis}, C.~{Johns-Krull}, S.~{Krucker},
  R.~P. {Lin}, J.~{McTiernan}, T.~R. {Metcalf}, J.~{Sato}, D.~M. {Smith},
  \solphys {\bf 210\/}, 61 (2002), doi:10.1023/A:1022436213688

\bibitem[{{Hurford} et~al.(2003){Hurford}, {Schwartz}, {Krucker}, {Lin},
  {Smith}, \& {Vilmer}}]{2003ApJ...595L..77H}
G.~J. {Hurford}, R.~A. {Schwartz}, S.~{Krucker}, R.~P. {Lin}, D.~M. {Smith},
  N.~{Vilmer}, \apjl {\bf 595\/}, L77 (2003), doi:10.1086/378179

\bibitem[{{Kahler} \& {Ragot}(2008)}]{2008ApJ...675..846K}
S.~W. {Kahler}, B.~R. {Ragot}, \apj {\bf 675\/}, 846 (2008), doi:10.1086/526416

\bibitem[{{Kanbach} et~al.(1993){Kanbach}, {Bertsch}, {Fichtel}, {Hartman},
  {Hunter}, {Kniffen}, {Kwok}, {Lin}, {Mattox}, \&
  {Mayer-Hasselwander}}]{1993A&AS...97..349K}
G.~{Kanbach}, D.~L. {Bertsch}, C.~E. {Fichtel}, R.~C. {Hartman}, S.~D.
  {Hunter}, D.~A. {Kniffen}, P.~W. {Kwok}, Y.~C. {Lin}, J.~R. {Mattox}, H.~A.
  {Mayer-Hasselwander}, \aaps {\bf 97\/}, 349 (1993)

\bibitem[{{Kaufmann} \& {Raulin}(2006)}]{2006PhPl...13g0701K}
P.~{Kaufmann}, J.~{Raulin}, Physics of Plasmas {\bf 13\/}(7), 070701 (2006),
  doi:10.1063/1.2244526

\bibitem[{{Kaufmann} et~al.(2004){Kaufmann}, {Raulin}, {de Castro}, {Levato},
  {Gary}, {Costa}, {Marun}, {Pereyra}, {Silva}, \&
  {Correia}}]{2004ApJ...603L.121K}
P.~{Kaufmann}, J.~{Raulin}, C.~G.~G. {de Castro}, H.~{Levato}, D.~E. {Gary},
  J.~E.~R. {Costa}, A.~{Marun}, P.~{Pereyra}, A.~V.~R. {Silva}, E.~{Correia},
  \apjl {\bf 603\/}, L121 (2004), doi:10.1086/383186

\bibitem[{{Kennel} \& {Petschek}(1966)}]{1966JGR....71....1K}
C.~F. {Kennel}, H.~E. {Petschek}, \jgr {\bf 71\/}, 1 (1966)

\bibitem[{{Kiener} et~al.(2001){Kiener}, {de S{\'e}r{\'e}ville}, \&
  {Tatischeff}}]{2001PhRvC..64b5803K}
J.~{Kiener}, N.~{de S{\'e}r{\'e}ville}, V.~{Tatischeff}, \prc {\bf 64\/}(2),
  025803 (2001), \eprint{arXiv:astro-ph/0105277},
  doi:10.1103/PhysRevC.64.025803

\bibitem[{{Kiener} et~al.(2006){Kiener}, {Gros}, {Tatischeff}, \&
  {Weidenspointner}}]{2006A&A...445..725K}
J.~{Kiener}, M.~{Gros}, V.~{Tatischeff}, G.~{Weidenspointner}, \aap {\bf
  445\/}, 725 (2006), \eprint{arXiv:astro-ph/0511091},
  doi:10.1051/0004-6361:20053665

\bibitem[{{Klein} et~al.(2006){Klein}, {Trottet}, {Molodij}, \&
  {S{\'e}mery}}]{2006via..conf...49K}
K.~{Klein}, G.~{Trottet}, G.~{Molodij}, A.~{S{\'e}mery}, in {\em Visions for
  Infrared Astronomy, Instrumentation, Mesure, M{\'e}trologie\/}, ed. by
  {V.~Coud{\'e} du Foresto, D.~Rouan, \& G.~Rousset} (2006), pp. 49--54

\bibitem[{{Kocharov} et~al.(1998){Kocharov}, {Debrunner}, {Kovaltsov},
  {Lockwood}, {McConnell}, {Nieminen}, {Rank}, {Ryan}, \&
  {Schoenfelder}}]{1998A&A...340..257K}
L.~{Kocharov}, H.~{Debrunner}, G.~{Kovaltsov}, J.~{Lockwood}, M.~{McConnell},
  P.~{Nieminen}, G.~{Rank}, J.~{Ryan}, V.~{Schoenfelder}, \aap {\bf 340\/}, 257
  (1998)

\bibitem[{{Kocharov} et~al.(1994){Kocharov}, {Kovaltsov}, {Kocharov},
  {Chuikin}, {Usoskin}, {Shea}, {Smart}, {Melnikov}, {Podstrigach}, \&
  {Armstrong}}]{1994SoPh..150..267K}
L.~G. {Kocharov}, G.~A. {Kovaltsov}, G.~E. {Kocharov}, E.~I. {Chuikin}, I.~G.
  {Usoskin}, M.~A. {Shea}, D.~F. {Smart}, V.~F. {Melnikov}, T.~S.
  {Podstrigach}, T.~P. {Armstrong}, \solphys {\bf 150\/}, 267 (1994),
  doi:10.1007/BF00712889

\bibitem[{{Kocharov} et~al.(1996){Kocharov}, {Torsti}, {Vainio}, {Kovaltsov},
  \& {Usoskin}}]{1996SoPh..169..181K}
L.~G. {Kocharov}, J.~{Torsti}, R.~{Vainio}, G.~A. {Kovaltsov}, I.~G. {Usoskin},
  \solphys {\bf 169\/}, 181 (1996), doi:10.1007/BF00153840

\bibitem[{{Kozlovsky} et~al.(1987){Kozlovsky}, {Lingenfelter}, \&
  {Ramaty}}]{1987ApJ...316..801K}
B.~{Kozlovsky}, R.~E. {Lingenfelter}, R.~{Ramaty}, \apj {\bf 316\/}, 801
  (1987), doi:10.1086/165244

\bibitem[{{Kozlovsky} et~al.(2002){Kozlovsky}, {Murphy}, \&
  {Ramaty}}]{2002ApJS..141..523K}
B.~{Kozlovsky}, R.~J. {Murphy}, R.~{Ramaty}, \apjs {\bf 141\/}, 523 (2002),
  doi:10.1086/340545

\bibitem[{{Kozlovsky} et~al.(2004){Kozlovsky}, {Murphy}, \&
  {Share}}]{2004ApJ...604..892K}
B.~{Kozlovsky}, R.~J. {Murphy}, G.~H. {Share}, \apj {\bf 604\/}, 892 (2004),
  doi:10.1086/381969

\bibitem[{{Krucker} et~al.(2008{\natexlab{a}}){Krucker}, {Battaglia},
  {Cargill}, {Fletcher}, {Hudson}, {MacKinnon}, {Masuda}, {Sui}, {Tomczak},
  {Veronig}, {Vlahos}, \& {White}}]{2008A&ARv..16..155K}
S.~{Krucker}, M.~{Battaglia}, P.~J. {Cargill}, L.~{Fletcher}, H.~S. {Hudson},
  A.~L. {MacKinnon}, S.~{Masuda}, L.~{Sui}, M.~{Tomczak}, A.~L. {Veronig},
  L.~{Vlahos}, S.~M. {White}, \aapr {\bf 16\/}, 155 (2008{\natexlab{a}}),
  doi:10.1007/s00159-008-0014-9

\bibitem[{{Krucker} et~al.(2008{\natexlab{b}}){Krucker}, {Hurford},
  {MacKinnon}, {Shih}, \& {Lin}}]{2008ApJ...678L..63K}
S.~{Krucker}, G.~J. {Hurford}, A.~L. {MacKinnon}, A.~Y. {Shih}, R.~P. {Lin},
  \apjl {\bf 678\/}, L63 (2008{\natexlab{b}}), doi:10.1086/588381

\bibitem[{{Kudela} et~al.(2003){Kudela}, {Kuznetsov}, {Myagkova}, \&
  {Yushkov}}]{2003ICRC....6.3183K}
K.~{Kudela}, S.~N. {Kuznetsov}, N.~{Myagkova}, B.~Y. {Yushkov}, in {\em
  International Cosmic Ray Conference\/} (2003), volume~6 of {\em International
  Cosmic Ray Conference\/}, pp. 3183--+

\bibitem[{{Kuznetsov} et~al.(2006){Kuznetsov}, {Kurt}, {Myagkova}, {Yushkov},
  \& {Kudela}}]{2006SoSyR..40..104K}
S.~N. {Kuznetsov}, V.~G. {Kurt}, I.~N. {Myagkova}, B.~Y. {Yushkov},
  K.~{Kudela}, Solar System Research {\bf 40\/}, 104 (2006),
  doi:10.1134/S0038094606020031

\bibitem[{{Kuznetsov} et~al.(2008){Kuznetsov}, {Kurt}, {Yushkov}, \& {et
  al.}}]{2008ICRC....1..121K}
S.~N. {Kuznetsov}, V.~G. {Kurt}, B.~Y. {Yushkov}, {et al.}, in {\em
  International Cosmic Ray Conference\/} (2008), volume~1 of {\em International
  Cosmic Ray Conference\/}, pp. 121--124

\bibitem[{{Lang} \& {Werntz}(1991)}]{1991AIPC..232..445L}
F.~L. {Lang}, C.~W. {Werntz}, in {\em Gamma-Ray Line Astrophysics\/}, ed. by
  {P.~Durouchoux \& N.~Prantzos} (1991), volume 232 of {\em American Institute
  of Physics Conference Series\/}, pp. 445--450, doi:10.1063/1.40919

\bibitem[{{Leikov} et~al.(1993){Leikov}, {Akimov}, {Volzhenskaia}, {Kalinkin},
  {Nesterov}, {Gal'Per}, {Zemskov}, {Ozerov}, {Topchiev}, \&
  {Fradkin}}]{1993A&AS...97..345L}
N.~G. {Leikov}, V.~V. {Akimov}, V.~A. {Volzhenskaia}, L.~F. {Kalinkin}, V.~E.
  {Nesterov}, A.~M. {Gal'Per}, V.~M. {Zemskov}, I.~V. {Ozerov}, N.~P.
  {Topchiev}, M.~I. {Fradkin}, \aaps {\bf 97\/}, 345 (1993)

\bibitem[{{Lin} et~al.(2002){Lin}, {Dennis}, {Hurford}, {Smith}, {Zehnder},
  {Harvey}, {Curtis}, {Pankow}, {Turin}, {Bester}, {Csillaghy}, {Lewis},
  {Madden}, {van Beek}, {Appleby}, {Raudorf}, {McTiernan}, {Ramaty}, {Schmahl},
  {Schwartz}, {Krucker}, {Abiad}, {Quinn}, {Berg}, {Hashii}, {Sterling},
  {Jackson}, {Pratt}, {Campbell}, {Malone}, {Landis}, {Barrington-Leigh},
  {Slassi-Sennou}, {Cork}, {Clark}, {Amato}, {Orwig}, {Boyle}, {Banks},
  {Shirey}, {Tolbert}, {Zarro}, {Snow}, {Thomsen}, {Henneck}, {McHedlishvili},
  {Ming}, {Fivian}, {Jordan}, {Wanner}, {Crubb}, {Preble}, {Matranga}, {Benz},
  {Hudson}, {Canfield}, {Holman}, {Crannell}, {Kosugi}, {Emslie}, {Vilmer},
  {Brown}, {Johns-Krull}, {Aschwanden}, {Metcalf}, \&
  {Conway}}]{2002SoPh..210....3L}
R.~P. {Lin}, B.~R. {Dennis}, G.~J. {Hurford}, D.~M. {Smith}, A.~{Zehnder},
  P.~R. {Harvey}, D.~W. {Curtis}, D.~{Pankow}, P.~{Turin}, M.~{Bester},
  A.~{Csillaghy}, M.~{Lewis}, N.~{Madden}, H.~F. {van Beek}, M.~{Appleby},
  T.~{Raudorf}, J.~{McTiernan}, R.~{Ramaty}, E.~{Schmahl}, R.~{Schwartz},
  S.~{Krucker}, R.~{Abiad}, T.~{Quinn}, P.~{Berg}, M.~{Hashii}, R.~{Sterling},
  R.~{Jackson}, R.~{Pratt}, R.~D. {Campbell}, D.~{Malone}, D.~{Landis}, C.~P.
  {Barrington-Leigh}, S.~{Slassi-Sennou}, C.~{Cork}, D.~{Clark}, D.~{Amato},
  L.~{Orwig}, R.~{Boyle}, I.~S. {Banks}, K.~{Shirey}, A.~K. {Tolbert},
  D.~{Zarro}, F.~{Snow}, K.~{Thomsen}, R.~{Henneck}, A.~{McHedlishvili},
  P.~{Ming}, M.~{Fivian}, J.~{Jordan}, R.~{Wanner}, J.~{Crubb}, J.~{Preble},
  M.~{Matranga}, A.~{Benz}, H.~{Hudson}, R.~C. {Canfield}, G.~D. {Holman},
  C.~{Crannell}, T.~{Kosugi}, A.~G. {Emslie}, N.~{Vilmer}, J.~C. {Brown},
  C.~{Johns-Krull}, M.~{Aschwanden}, T.~{Metcalf}, A.~{Conway}, \solphys {\bf
  210\/}, 3 (2002), doi:10.1023/A:1022428818870

\bibitem[{{Lin} et~al.(2003){Lin}, {Krucker}, {Hurford}, {Smith}, {Hudson},
  {Holman}, {Schwartz}, {Dennis}, {Share}, {Murphy}, {Emslie}, {Johns-Krull},
  \& {Vilmer}}]{2003ApJ...595L..69L}
R.~P. {Lin}, S.~{Krucker}, G.~J. {Hurford}, D.~M. {Smith}, H.~S. {Hudson},
  G.~D. {Holman}, R.~A. {Schwartz}, B.~R. {Dennis}, G.~H. {Share}, R.~J.
  {Murphy}, A.~G. {Emslie}, C.~{Johns-Krull}, N.~{Vilmer}, \apjl {\bf 595\/},
  L69 (2003), doi:10.1086/378932

\bibitem[{{Lingenfelter} et~al.(1965{\natexlab{a}}){Lingenfelter}, {Flamm},
  {Canfield}, \& {Kellman}}]{1965JGR....70.4077L}
R.~E. {Lingenfelter}, E.~J. {Flamm}, E.~H. {Canfield}, S.~{Kellman}, \jgr {\bf
  70\/}, 4077 (1965{\natexlab{a}}), doi:10.1029/JZ070i017p04077

\bibitem[{{Lingenfelter} et~al.(1965{\natexlab{b}}){Lingenfelter}, {Flamm},
  {Canfield}, \& {Kellman}}]{1965JGR....70.4087L}
R.~E. {Lingenfelter}, E.~J. {Flamm}, E.~H. {Canfield}, S.~{Kellman}, \jgr {\bf
  70\/}, 4087 (1965{\natexlab{b}}), doi:10.1029/JZ070i017p04087

\bibitem[{{Lingenfelter} \& {Ramaty}(1967{\natexlab{a}})}]{1967henr.book...99L}
R.~E. {Lingenfelter}, R.~{Ramaty}, {\em {High Energy Nuclear Reactions in Solar
  Flares}\/} (1967{\natexlab{a}}), pp. 99--+

\bibitem[{{Lingenfelter} \& {Ramaty}(1967{\natexlab{b}})}]{1967P&SS...15.1303L}
R.~E. {Lingenfelter}, R.~{Ramaty}, \planss {\bf 15\/}, 1303
  (1967{\natexlab{b}}), doi:10.1016/0032-0633(67)90184-5

\bibitem[{{Litvinenko} \& {Somov}(1993)}]{1993SoPh..146..127L}
Y.~E. {Litvinenko}, B.~V. {Somov}, \solphys {\bf 146\/}, 127 (1993),
  doi:10.1007/BF00662174

\bibitem[{{Lockwood} \& {Debrunner}(1999)}]{1999SSRv...88..483L}
J.~A. {Lockwood}, H.~{Debrunner}, Space Science Reviews {\bf 88\/}, 483 (1999),
  doi:10.1023/A:1005159816103

\bibitem[{{Lockwood} et~al.(1997){Lockwood}, {Debrunner}, \&
  {Ryan}}]{1997SoPh..173..151L}
J.~A. {Lockwood}, H.~{Debrunner}, J.~M. {Ryan}, \solphys {\bf 173\/}, 151
  (1997)

\bibitem[{{Lopate}(2006)}]{2006GMS...165..283L}
C.~{Lopate}, Washington DC American Geophysical Union Geophysical Monograph
  Series {\bf 165\/}, 283 (2006)

\bibitem[{{L{\"u}thi} et~al.(2004){L{\"u}thi}, {Magun}, \&
  {Miller}}]{2004A&A...415.1123L}
T.~{L{\"u}thi}, A.~{Magun}, M.~{Miller}, \aap {\bf 415\/}, 1123 (2004),
  doi:10.1051/0004-6361:20034624

\bibitem[{{MacKinnon}(1989)}]{1989A&A...226..284M}
A.~L. {MacKinnon}, \aap {\bf 226\/}, 284 (1989)

\bibitem[{{MacKinnon}(2006)}]{2006GMS...165..157M}
A.~L. {MacKinnon}, Washington DC American Geophysical Union Geophysical
  Monograph Series {\bf 165\/}, 157 (2006)

\bibitem[{{MacKinnon}(2007)}]{2007A&A...462..763M}
A.~L. {MacKinnon}, \aap {\bf 462\/}, 763 (2007), doi:10.1051/0004-6361:20066328

\bibitem[{{MacKinnon} \& {Brown}(1990)}]{1990A&A...232..544M}
A.~L. {MacKinnon}, J.~C. {Brown}, \aap {\bf 232\/}, 544 (1990)

\bibitem[{{MacKinnon} \& {Mallik}(2010)}]{2010A&A...510A..29M}
A.~L. {MacKinnon}, P.~C.~V. {Mallik}, \aap {\bf 510\/}, A29+ (2010),
  \eprint{0908.3903}, doi:10.1051/0004-6361/200913190

\bibitem[{{MacKinnon} \& {Toner}(2003)}]{2003A&A...409..745M}
A.~L. {MacKinnon}, M.~P. {Toner}, \aap {\bf 409\/}, 745 (2003),
  doi:10.1051/0004-6361:20030943

\bibitem[{{Mandzhavidze} \& {Ramaty}(1992)}]{1992ApJ...389..739M}
N.~{Mandzhavidze}, R.~{Ramaty}, \apj {\bf 389\/}, 739 (1992),
  doi:10.1086/171247

\bibitem[{{Mandzhavidze} \& {Ramaty}(2000)}]{2000ASPC..206...64M}
N.~{Mandzhavidze}, R.~{Ramaty}, in {\em High Energy Solar Physics Workshop -
  Anticipating HESSI\/}, ed. by {R.~Ramaty \& N.~Mandzhavidze} (2000), volume
  206 of {\em Astronomical Society of the Pacific Conference Series\/}, pp.
  64--+

\bibitem[{{Mandzhavidze} et~al.(1999){Mandzhavidze}, {Ramaty}, \&
  {Kozlovsky}}]{1999ApJ...518..918M}
N.~{Mandzhavidze}, R.~{Ramaty}, B.~{Kozlovsky}, \apj {\bf 518\/}, 918 (1999),
  doi:10.1086/307321

\bibitem[{{Masson} et~al.(2009){Masson}, {Klein}, {B{\"u}tikofer},
  {Fl{\"u}ckiger}, {Kurt}, {Yushkov}, \& {Krucker}}]{2009SoPh..257..305M}
S.~{Masson}, K.~{Klein}, R.~{B{\"u}tikofer}, E.~{Fl{\"u}ckiger}, V.~{Kurt},
  B.~{Yushkov}, S.~{Krucker}, \solphys {\bf 257\/}, 305 (2009),
  \eprint{0905.1816}, doi:10.1007/s11207-009-9377-y

\bibitem[{{McConnell} et~al.(1997){McConnell}, {Bennett}, {MacKinnon},
  {Miller}, {Rank}, {Ryan}, \& {Sch\"{o}nfelder}}]{1997ICRC...1..13M}
M.~L. {McConnell}, K.~{Bennett}, A.~L. {MacKinnon}, R.~{Miller}, G.~{Rank},
  J.~M. {Ryan}, V.~{Sch\"{o}nfelder}, in {\em Proc. 25th ICRC, Durban (South
  Africa)\/} (1997), volume~1, p.~13

\bibitem[{{Mewaldt} et~al.(2009){Mewaldt}, {Leske}, {Stone}, {Barghouty},
  {Labrador}, {Cohen}, {Cummings}, {Davis}, {von Rosenvinge}, \&
  {Wiedenbeck}}]{2009ApJ...693L..11M}
R.~A. {Mewaldt}, R.~A. {Leske}, E.~C. {Stone}, A.~F. {Barghouty}, A.~W.
  {Labrador}, C.~M.~S. {Cohen}, A.~C. {Cummings}, A.~J. {Davis}, T.~T. {von
  Rosenvinge}, M.~E. {Wiedenbeck}, \apjl {\bf 693\/}, L11 (2009),
  doi:10.1088/0004-637X/693/1/L11

\bibitem[{{Meyer}(1993)}]{1993AdSpR..13..377M}
J.~{Meyer}, Advances in Space Research {\bf 13\/}, 377 (1993),
  doi:10.1016/0273-1177(93)90509-A

\bibitem[{{Morrison}(1958)}]{1958cora.conf..305M}
P.~{Morrison}, in {\em Fifth International Congress on Cosmic Radiation\/}, ed.
  by {M.~S.~Vallarta} (1958), pp. 305--+

\bibitem[{{Muraki} et~al.(1992){Muraki}, {Murakami}, {Miyazaki}, {Mitsui},
  {Shibata}, {Sakakibara}, {Sakai}, {Takahashi}, {Yamada}, \&
  {Yamaguchi}}]{1992ApJ...400L..75M}
Y.~{Muraki}, K.~{Murakami}, M.~{Miyazaki}, K.~{Mitsui}, S.~{Shibata},
  S.~{Sakakibara}, T.~{Sakai}, T.~{Takahashi}, T.~{Yamada}, K.~{Yamaguchi},
  \apjl {\bf 400\/}, L75 (1992), doi:10.1086/186653

\bibitem[{{Murphy} et~al.(1987){Murphy}, {Dermer}, \&
  {Ramaty}}]{1987ApJS...63..721M}
R.~J. {Murphy}, C.~D. {Dermer}, R.~{Ramaty}, \apjs {\bf 63\/}, 721 (1987),
  doi:10.1086/191180

\bibitem[{{Murphy} et~al.(2009){Murphy}, {Kozlovsky}, {Kiener}, \&
  {Share}}]{2009ApJS..183..142M}
R.~J. {Murphy}, B.~{Kozlovsky}, J.~{Kiener}, G.~H. {Share}, \apjs {\bf 183\/},
  142 (2009), doi:10.1088/0067-0049/183/1/142

\bibitem[{{Murphy} et~al.(1988){Murphy}, {Kozlovsky}, \&
  {Ramaty}}]{1988ApJ...331.1029M}
R.~J. {Murphy}, B.~{Kozlovsky}, R.~{Ramaty}, \apj {\bf 331\/}, 1029 (1988),
  doi:10.1086/166619

\bibitem[{{Murphy} et~al.(2007){Murphy}, {Kozlovsky}, {Share}, {Hua}, \&
  {Lingenfelter}}]{2007ApJS..168..167M}
R.~J. {Murphy}, B.~{Kozlovsky}, G.~H. {Share}, X.-M. {Hua}, R.~E.
  {Lingenfelter}, \apjs {\bf 168\/}, 167 (2007), doi:10.1086/509637

\bibitem[{{Murphy} et~al.(1991){Murphy}, {Ramaty}, {Reames}, \&
  {Kozlovsky}}]{1991ApJ...371..793M}
R.~J. {Murphy}, R.~{Ramaty}, D.~V. {Reames}, B.~{Kozlovsky}, \apj {\bf 371\/},
  793 (1991), doi:10.1086/169944

\bibitem[{{Murphy} et~al.(1999){Murphy}, {Share}, {Delsignore}, \&
  {Hua}}]{1999ApJ...510.1011M}
R.~J. {Murphy}, G.~H. {Share}, K.~W. {Delsignore}, X.-M. {Hua}, \apj {\bf
  510\/}, 1011 (1999), doi:10.1086/306614

\bibitem[{{Murphy} et~al.(1997){Murphy}, {Share}, {Grove}, {Johnson}, {Kinzer},
  {Kurfess}, {Strickman}, \& {Jung}}]{1997ApJ...490..883M}
R.~J. {Murphy}, G.~H. {Share}, J.~E. {Grove}, W.~N. {Johnson}, R.~L. {Kinzer},
  J.~D. {Kurfess}, M.~S. {Strickman}, G.~V. {Jung}, \apj {\bf 490\/}, 883
  (1997), doi:10.1086/304902

\bibitem[{{Murphy} et~al.(2003){Murphy}, {Share}, {Hua}, {Lin}, {Smith}, \&
  {Schwartz}}]{2003ApJ...595..L93M}
R.~J. {Murphy}, G.~H. {Share}, X.-M. {Hua}, R.~P. {Lin}, D.~M. {Smith}, R.~A.
  {Schwartz}, \apj {\bf 595\/}, L93 (2003)

\bibitem[{{Murphy} et~al.(2005){Murphy}, {Share}, {Skibo}, \&
  {Kozlovsky}}]{2005ApJS..161..495M}
R.~J. {Murphy}, G.~H. {Share}, J.~G. {Skibo}, B.~{Kozlovsky}, \apjs {\bf
  161\/}, 495 (2005), doi:10.1086/452634

\bibitem[{{Myagkova} et~al.(2007){Myagkova}, {Kuznetsov}, {Kurt}, {Yuskov},
  {Galkin}, {Muravieva}, \& {Kudela}}]{2007AdSpR..40.1929M}
I.~N. {Myagkova}, S.~N. {Kuznetsov}, V.~G. {Kurt}, B.~Y. {Yuskov}, V.~I.
  {Galkin}, E.~A. {Muravieva}, K.~{Kudela}, Advances in Space Research {\bf
  40\/}, 1929 (2007), doi:10.1016/j.asr.2007.01.091

\bibitem[{{Myagkova} et~al.(2004){Myagkova}, {Kuznetsov}, {Yushkov}, \&
  {Kudela}}]{2004cosp...35.1511M}
I.~N. {Myagkova}, S.~N. {Kuznetsov}, B.~Y. {Yushkov}, K.~{Kudela}, in {\em 35th
  COSPAR Scientific Assembly\/} (2004), volume~35 of {\em COSPAR, Plenary
  Meeting\/}, pp. 1511--+

\bibitem[{{Naiman} et~al.(2008){Naiman}, {Smith}, {Murphy}, {Share}, {Shih}, \&
  {Kiener}}]{2008AGUFMSH31B1670N}
J.~P. {Naiman}, D.~M. {Smith}, R.~J. {Murphy}, G.~H. {Share}, A.~Y. {Shih},
  J.~{Kiener}, AGU Fall Meeting Abstracts pp. B1670+ (2008)

\bibitem[{{Pelaez} et~al.(1992){Pelaez}, {Mandrou}, {Niel}, {Mena}, {Vilmer},
  {Trottet}, {Lebrun}, {Paul}, {Terekhov}, \& {Siuniaev}}]{1992SoPh..140..121P}
F.~{Pelaez}, P.~{Mandrou}, M.~{Niel}, B.~{Mena}, N.~{Vilmer}, G.~{Trottet},
  F.~{Lebrun}, J.~{Paul}, O.~{Terekhov}, R.~{Siuniaev}, \solphys {\bf 140\/},
  121 (1992), doi:10.1007/BF00148433

\bibitem[{{Pirard} et~al.(2009){Pirard}, {Woolf}, {Bravar}, {Bruillard},
  {Fl{\"u}ckiger}, {Legere}, {MacKinnon}, {Macri}, {Mallik}, {Moser}, \&
  {Ryan}}]{2009NIMPA.603..406P}
B.~{Pirard}, R.~S. {Woolf}, U.~{Bravar}, P.~J. {Bruillard}, E.~O.
  {Fl{\"u}ckiger}, J.~S. {Legere}, A.~L. {MacKinnon}, J.~R. {Macri}, P.~C.~V.
  {Mallik}, M.~R. {Moser}, J.~M. {Ryan}, Nuclear Instruments and Methods in
  Physics Research A {\bf 603\/}, 406 (2009), doi:10.1016/j.nima.2009.02.012

\bibitem[{{Prince} et~al.(1982){Prince}, {Ling}, {Mahoney}, {Riegler}, \&
  {Jacobson}}]{1982ApJ...255L..81P}
T.~A. {Prince}, J.~C. {Ling}, W.~A. {Mahoney}, G.~R. {Riegler}, A.~S.
  {Jacobson}, \apjl {\bf 255\/}, L81 (1982), doi:10.1086/183773

\bibitem[{{Pyle} \& {Simpson}(1991)}]{1991ICRC....3...53P}
K.~R. {Pyle}, J.~A. {Simpson}, in {\em International Cosmic Ray Conference\/}
  (1991), volume~3 of {\em International Cosmic Ray Conference\/}, pp. 53--+

\bibitem[{{Ramaty}(1986)}]{1986psun....2..291R}
R.~{Ramaty}, in {\em Physics of the Sun. Volume 2\/} (1986), volume~2, pp.
  291--323

\bibitem[{{Ramaty} \& {Crannell}(1976)}]{1976ApJ...203..766R}
R.~{Ramaty}, C.~J. {Crannell}, \apj {\bf 203\/}, 766 (1976), doi:10.1086/154139

\bibitem[{{Ramaty} et~al.(1979){Ramaty}, {Kozlovsky}, \&
  {Lingenfelter}}]{1979ApJS...40..487R}
R.~{Ramaty}, B.~{Kozlovsky}, R.~E. {Lingenfelter}, \apjs {\bf 40\/}, 487
  (1979), doi:10.1086/190596

\bibitem[{{Ramaty} \& {Mandzhavidze}(2000)}]{2000IAUS..195..123R}
R.~{Ramaty}, N.~{Mandzhavidze}, in {\em Highly Energetic Physical Processes and
  Mechanisms for Emission from Astrophysical Plasmas\/}, ed. by
  {P.~C.~H.~Martens, S.~Tsuruta, \& M.~A.~Weber} (2000), volume 195 of {\em IAU
  Symposium\/}, pp. 123--+

\bibitem[{{Ramaty} et~al.(1997){Ramaty}, {Mandzhavidze}, {Barat}, \&
  {Trottet}}]{1997ApJ...479..458R}
R.~{Ramaty}, N.~{Mandzhavidze}, C.~{Barat}, G.~{Trottet}, \apj {\bf 479\/}, 458
  (1997), doi:10.1086/303878

\bibitem[{{Ramaty} et~al.(1996){Ramaty}, {Mandzhavidze}, \&
  {Kozlovsky}}]{1996AIPC..374..172R}
R.~{Ramaty}, N.~{Mandzhavidze}, B.~{Kozlovsky}, in {\em American Institute of
  Physics Conference Series\/}, ed. by {R.~Ramaty, N.~Mandzhavidze, \&
  X.-M.~Hua} (1996), volume 374 of {\em American Institute of Physics
  Conference Series\/}, pp. 172--183, doi:10.1063/1.50953

\bibitem[{{Ramaty} et~al.(1995){Ramaty}, {Mandzhavidze}, {Kozlovsky}, \&
  {Murphy}}]{1995ApJ...455L.193R}
R.~{Ramaty}, N.~{Mandzhavidze}, B.~{Kozlovsky}, R.~J. {Murphy}, \apjl {\bf
  455\/}, L193+ (1995), doi:10.1086/309841

\bibitem[{{Rank} et~al.(2001){Rank}, {Ryan}, {Debrunner}, {McConnell}, \&
  {Sch{\"o}nfelder}}]{2001A&A...378.1046R}
G.~{Rank}, J.~{Ryan}, H.~{Debrunner}, M.~{McConnell}, V.~{Sch{\"o}nfelder},
  \aap {\bf 378\/}, 1046 (2001), doi:10.1051/0004-6361:20011060

\bibitem[{{Raymond} et~al.(2007){Raymond}, {Holman}, {Ciaravella}, {Panasyuk},
  {Ko}, \& {Kohl}}]{2007ApJ...659..750R}
J.~C. {Raymond}, G.~{Holman}, A.~{Ciaravella}, A.~{Panasyuk}, Y.-K. {Ko},
  J.~{Kohl}, \apj {\bf 659\/}, 750 (2007), \eprint{arXiv:astro-ph/0701359},
  doi:10.1086/512604

\bibitem[{{Reames}(1999)}]{1999SSRv...90..413R}
D.~V. {Reames}, Space Science Reviews {\bf 90\/}, 413 (1999),
  doi:10.1023/A:1005105831781

\bibitem[{{Reames} et~al.(1997){Reames}, {Barbier}, {von Rosenvinge}, {Mason},
  {Mazur}, \& {Dwyer}}]{1997ApJ...483..515R}
D.~V. {Reames}, L.~M. {Barbier}, T.~T. {von Rosenvinge}, G.~M. {Mason}, J.~E.
  {Mazur}, J.~R. {Dwyer}, \apj {\bf 483\/}, 515 (1997), doi:10.1086/304229

\bibitem[{{Rieger} et~al.(1998){Rieger}, {Gan}, \&
  {Marschh{\"a}user}}]{1998SoPh..183..123R}
E.~{Rieger}, W.~Q. {Gan}, H.~{Marschh{\"a}user}, \solphys {\bf 183\/}, 123
  (1998)

\bibitem[{{Rieger} \& {Marschh{\"a}user}(1991)}]{1991max..conf...68R}
E.~{Rieger}, H.~{Marschh{\"a}user}, in {\em Max '91/SMM Solar Flares:
  Observations and Theory\/}, ed. by {R.~M.~Winglee \& A.~L.~Kiplinger} (1991),
  pp. 68--+

\bibitem[{{Ruffolo}(1991)}]{1991ApJ...382..688R}
D.~{Ruffolo}, \apj {\bf 382\/}, 688 (1991), doi:10.1086/170756

\bibitem[{{Ryan} et~al.(1993){Ryan}, {Bennett}, {Debrunner}, {Forrest},
  {Lockwood}, {Loomis}, {McConnell}, {Morris}, {Sch{\"o}nfelder}, {Swanenburg},
  \& {Webber}}]{1993AdSpR..13..255R}
J.~{Ryan}, K.~{Bennett}, H.~{Debrunner}, D.~{Forrest}, J.~{Lockwood},
  M.~{Loomis}, M.~{McConnell}, D.~{Morris}, V.~{Sch{\"o}nfelder}, B.~N.
  {Swanenburg}, W.~{Webber}, Advances in Space Research {\bf 13\/}, 255 (1993),
  doi:10.1016/0273-1177(93)90487-V

\bibitem[{{Ryan} et~al.(1994){Ryan}, {Forrest}, {Lockwood}, {Loomis},
  {McConnell}, {Morris}, {Webber}, {Bennett}, {Hanlon}, {Winkler}, {Debrunner},
  {Rank}, {Sch{\"o}nfelder}, \& {Swanenburg}}]{1994AIPC..294...89R}
J.~{Ryan}, D.~{Forrest}, J.~{Lockwood}, M.~{Loomis}, M.~{McConnell},
  D.~{Morris}, W.~{Webber}, K.~{Bennett}, L.~{Hanlon}, C.~{Winkler},
  H.~{Debrunner}, G.~{Rank}, V.~{Sch{\"o}nfelder}, B.~N. {Swanenburg}, in {\em
  High-Energy Solar Phenomena - a New Era of Spacecraft Measurements\/}, ed. by
  {J.~Ryan \& W.~T.~Vestrand} (1994), volume 294 of {\em American Institute of
  Physics Conference Series\/}, pp. 89--93, doi:10.1063/1.45205

\bibitem[{{Ryan}(2000)}]{2000SSRv...93..581R}
J.~M. {Ryan}, Space Science Reviews {\bf 93\/}, 581 (2000)

\bibitem[{{Sakai} \& {Nagasugi}(2007)}]{2007A&A...474L..33S}
J.~I. {Sakai}, Y.~{Nagasugi}, \aap {\bf 474\/}, L33 (2007),
  doi:10.1051/0004-6361:20078471

\bibitem[{{Sako} et~al.(2006){Sako}, {Watanabe}, {Muraki}, {Matsubara},
  {Tsujihara}, {Yamashita}, {Sakai}, {Shibata}, {Vald{\'e}s-Galicia},
  {Gonz{\'a}lez}, {Hurtado}, {Musalem}, {Miranda}, {Martinic}, {Ticona},
  {Velarde}, {Kakimoto}, {Ogio}, {Tsunesada}, {Tokuno}, {Tanaka}, {Yoshikawa},
  {Terasawa}, {Saito}, {Mukai}, \& {Gros}}]{2006ApJ...651L..69S}
T.~{Sako}, K.~{Watanabe}, Y.~{Muraki}, Y.~{Matsubara}, H.~{Tsujihara},
  M.~{Yamashita}, T.~{Sakai}, S.~{Shibata}, J.~F. {Vald{\'e}s-Galicia}, L.~X.
  {Gonz{\'a}lez}, A.~{Hurtado}, O.~{Musalem}, P.~{Miranda}, N.~{Martinic},
  R.~{Ticona}, A.~{Velarde}, F.~{Kakimoto}, S.~{Ogio}, Y.~{Tsunesada},
  H.~{Tokuno}, Y.~T. {Tanaka}, I.~{Yoshikawa}, T.~{Terasawa}, Y.~{Saito},
  T.~{Mukai}, M.~{Gros}, \apjl {\bf 651\/}, L69 (2006), doi:10.1086/509145

\bibitem[{{Schmelz} et~al.(2005){Schmelz}, {Nasraoui}, {Roames}, {Lippner}, \&
  {Garst}}]{2005ApJ...634L.197S}
J.~T. {Schmelz}, K.~{Nasraoui}, J.~K. {Roames}, L.~A. {Lippner}, J.~W. {Garst},
  \apjl {\bf 634\/}, L197 (2005), \eprint{arXiv:astro-ph/0510230},
  doi:10.1086/499051

\bibitem[{{Schrijver} et~al.(2006){Schrijver}, {Hudson}, {Murphy}, {Share}, \&
  {Tarbell}}]{2006ApJ...650.1184S}
C.~J. {Schrijver}, H.~S. {Hudson}, R.~J. {Murphy}, G.~H. {Share}, T.~D.
  {Tarbell}, \apj {\bf 650\/}, 1184 (2006), doi:10.1086/506583

\bibitem[{{Share} \& {Murphy}(1995)}]{1995ApJ...452..933S}
G.~H. {Share}, R.~J. {Murphy}, \apj {\bf 452\/}, 933 (1995), doi:10.1086/176360

\bibitem[{{Share} \& {Murphy}(1998)}]{1998ApJ...508..876S}
G.~H. {Share}, R.~J. {Murphy}, \apj {\bf 508\/}, 876 (1998)

\bibitem[{{Share} \& {Murphy}(2000)}]{2000ASPC..206..377S}
G.~H. {Share}, R.~J. {Murphy}, in {\em High Energy Solar Physics Workshop -
  Anticipating Hess!\/}, ed. by {R.~Ramaty \& N.~Mandzhavidze} (2000), volume
  206 of {\em Astronomical Society of the Pacific Conference Series\/}, pp.
  377--+

\bibitem[{{Share} \& {Murphy}(2006)}]{2006GMS...165..177S}
G.~H. {Share}, R.~J. {Murphy}, Washington DC American Geophysical Union
  Geophysical Monograph Series {\bf 165\/}, 177 (2006)

\bibitem[{{Share} et~al.(2002){Share}, {Murphy}, {Kiener}, \& {de
  S{\'e}r{\'e}ville}}]{2002ApJ...573..464S}
G.~H. {Share}, R.~J. {Murphy}, J.~{Kiener}, N.~{de S{\'e}r{\'e}ville}, \apj
  {\bf 573\/}, 464 (2002), \eprint{arXiv:astro-ph/0203215}, doi:10.1086/340595

\bibitem[{{Share} et~al.(2001){Share}, {Murphy}, \&
  {Newton}}]{2001SoPh..201..191S}
G.~H. {Share}, R.~J. {Murphy}, E.~K. {Newton}, \solphys {\bf 201\/}, 191 (2001)

\bibitem[{{Share} et~al.(2003{\natexlab{a}}){Share}, {Murphy}, {Skibo},
  {Smith}, {Hudson}, {Lin}, {Shih}, {Dennis}, {Schwartz}, \&
  {Kozlovsky}}]{2003ApJ...595L..85S}
G.~H. {Share}, R.~J. {Murphy}, J.~G. {Skibo}, D.~M. {Smith}, H.~S. {Hudson},
  R.~P. {Lin}, A.~Y. {Shih}, B.~R. {Dennis}, R.~A. {Schwartz}, B.~{Kozlovsky},
  \apjl {\bf 595\/}, L85 (2003{\natexlab{a}}), doi:10.1086/378174

\bibitem[{{Share} et~al.(2003{\natexlab{b}}){Share}, {Murphy}, {Smith}, {Lin},
  {Dennis}, \& {Schwartz}}]{2003ApJ...595L..89S}
G.~H. {Share}, R.~J. {Murphy}, D.~M. {Smith}, R.~P. {Lin}, B.~R. {Dennis},
  R.~A. {Schwartz}, \apjl {\bf 595\/}, L89 (2003{\natexlab{b}}),
  doi:10.1086/378176

\bibitem[{{Share} et~al.(2004){Share}, {Murphy}, {Smith}, {Schwartz}, \&
  {Lin}}]{2004ApJ...615L.169S}
G.~H. {Share}, R.~J. {Murphy}, D.~M. {Smith}, R.~A. {Schwartz}, R.~P. {Lin},
  \apjl {\bf 615\/}, L169 (2004), doi:10.1086/426478

\bibitem[{{Shea} et~al.(1991){Shea}, {Smart}, \& {Pyle}}]{1991GeoRL..18.1655S}
M.~A. {Shea}, D.~F. {Smart}, K.~R. {Pyle}, \grl {\bf 18\/}, 1655 (1991),
  doi:10.1029/91GL02001

\bibitem[{{Shibata}(1994)}]{1994JGR....99.6651S}
S.~{Shibata}, \jgr {\bf 99\/}, 6651 (1994), doi:10.1029/93JA03175

\bibitem[{{Shih} et~al.(2009{\natexlab{a}}){Shih}, {Lin}, {Hurford}, {Boggs},
  {Zoglauer}, {Wunderer}, {Sample}, {Turin}, {McBride}, {Smith}, {Tajima},
  {Luke}, \& {Amman}}]{2009SPD....40.1810S}
A.~Y. {Shih}, R.~P. {Lin}, G.~J. {Hurford}, S.~E. {Boggs}, A.~C. {Zoglauer},
  C.~B. {Wunderer}, J.~G. {Sample}, P.~{Turin}, S.~{McBride}, D.~M. {Smith},
  H.~{Tajima}, P.~N. {Luke}, M.~S. {Amman}, in {\em AAS/Solar Physics Division
  Meeting\/} (2009{\natexlab{a}}), volume~40 of {\em AAS/Solar Physics Division
  Meeting\/}, pp. \#18.10--+

\bibitem[{{Shih} et~al.(2009{\natexlab{b}}){Shih}, {Lin}, \&
  {Smith}}]{2009ApJ...698L.152S}
A.~Y. {Shih}, R.~P. {Lin}, D.~M. {Smith}, \apjl {\bf 698\/}, L152
  (2009{\natexlab{b}}), doi:10.1088/0004-637X/698/2/L152

\bibitem[{{Shih} et~al.(2007){Shih}, {Smith}, {Lin}, \&
  {Hurford}}]{2007AAS...210.6804S}
A.~Y. {Shih}, D.~M. {Smith}, R.~P. {Lin}, G.~J. {Hurford}, in {\em Bulletin of
  the American Astronomical Society\/} (2007), volume~38 of {\em Bulletin of
  the American Astronomical Society\/}, pp. 175--+

\bibitem[{{Shih} et~al.(2003){Shih}, {Smith}, {Lin}, {Hurford}, {Krucker},
  {Schwartz}, {Share}, \& {Murphy}}]{2003AGUFMSH11D1132S}
A.~Y. {Shih}, D.~M. {Smith}, R.~P. {Lin}, G.~J. {Hurford}, S.~{Krucker}, R.~A.
  {Schwartz}, G.~H. {Share}, R.~J. {Murphy}, AGU Fall Meeting Abstracts pp.
  D1132+ (2003)

\bibitem[{{Silva} et~al.(2007){Silva}, {Share}, {Murphy}, {Costa}, {de Castro},
  {Raulin}, \& {Kaufmann}}]{2007SoPh..245..311S}
A.~V.~R. {Silva}, G.~H. {Share}, R.~J. {Murphy}, J.~E.~R. {Costa}, C.~G.~G. {de
  Castro}, J.~{Raulin}, P.~{Kaufmann}, \solphys {\bf 245\/}, 311 (2007),
  doi:10.1007/s11207-007-9044-0

\bibitem[{{Smart} et~al.(1995){Smart}, {Shea}, \&
  {O'Brien}}]{1995ICRC....4..171S}
D.~F. {Smart}, M.~A. {Shea}, K.~{O'Brien}, in {\em International Cosmic Ray
  Conference\/} (1995), volume~4 of {\em International Cosmic Ray
  Conference\/}, pp. 171--+

\bibitem[{{Smith} et~al.(2002){Smith}, {Lin}, {Turin}, {Curtis}, {Primbsch},
  {Campbell}, {Abiad}, {Schroeder}, {Cork}, {Hull}, {Landis}, {Madden},
  {Malone}, {Pehl}, {Raudorf}, {Sangsingkeow}, {Boyle}, {Banks}, {Shirey}, \&
  {Schwartz}}]{2002SoPh..210...33S}
D.~M. {Smith}, R.~P. {Lin}, P.~{Turin}, D.~W. {Curtis}, J.~H. {Primbsch}, R.~D.
  {Campbell}, R.~{Abiad}, P.~{Schroeder}, C.~P. {Cork}, E.~L. {Hull}, D.~A.
  {Landis}, N.~W. {Madden}, D.~{Malone}, R.~H. {Pehl}, T.~{Raudorf},
  P.~{Sangsingkeow}, R.~{Boyle}, I.~S. {Banks}, K.~{Shirey}, R.~{Schwartz},
  \solphys {\bf 210\/}, 33 (2002), doi:10.1023/A:1022400716414

\bibitem[{{Smith} et~al.(2003){Smith}, {Share}, {Murphy}, {Schwartz}, {Shih},
  \& {Lin}}]{2003ApJ...595L..81S}
D.~M. {Smith}, G.~H. {Share}, R.~J. {Murphy}, R.~A. {Schwartz}, A.~Y. {Shih},
  R.~P. {Lin}, \apjl {\bf 595\/}, L81 (2003), \eprint{arXiv:astro-ph/0306292},
  doi:10.1086/378173

\bibitem[{{Stecker}(1970)}]{1970Ap&SS...6..377S}
F.~W. {Stecker}, \apss {\bf 6\/}, 377 (1970), doi:10.1007/BF00653856

\bibitem[{{Struminsky} et~al.(1994){Struminsky}, {Matsuoka}, \&
  {Takahashi}}]{1994ApJ...429..400S}
A.~{Struminsky}, M.~{Matsuoka}, K.~{Takahashi}, \apj {\bf 429\/}, 400 (1994),
  doi:10.1086/174330

\bibitem[{{Suri} et~al.(1975){Suri}, {Chupp}, {Forrest}, \&
  {Reppin}}]{1975SoPh...43..415S}
A.~N. {Suri}, E.~L. {Chupp}, D.~J. {Forrest}, C.~{Reppin}, \solphys {\bf 43\/},
  415 (1975), doi:10.1007/BF00152365

\bibitem[{{Talon} et~al.(1975){Talon}, {Vedrenne}, {Melioranskii},
  {Pissarenko}, {Shamolin}, \& {Likin}}]{1975IAUS...68..315T}
R.~{Talon}, G.~{Vedrenne}, A.~S. {Melioranskii}, N.~F. {Pissarenko}, V.~M.
  {Shamolin}, O.~B. {Likin}, in {\em Solar Gamma-, X-, and EUV Radiation\/},
  ed. by {S.~R.~Kane} (1975), volume~68 of {\em IAU Symposium\/}, pp. 315--339

\bibitem[{{Tamres} et~al.(1989){Tamres}, {Melrose}, \&
  {Canfield}}]{1989ApJ...342..576T}
D.~H. {Tamres}, D.~B. {Melrose}, R.~C. {Canfield}, \apj {\bf 342\/}, 576
  (1989), doi:10.1086/167617

\bibitem[{{Tatischeff} et~al.(2005){Tatischeff}, {Kiener}, \&
  {Gros}}]{2005astro.ph..1121T}
V.~{Tatischeff}, J.~{Kiener}, M.~{Gros}, ArXiv Astrophysics e-prints:
  Rencontres du Vietnam, New Views on the Universe, Hanoi, Vietnam  (2005),
  \eprint{arXiv:astro-ph/0501121}

\bibitem[{{Tatischeff} et~al.(2006){Tatischeff}, {Kozlovsky}, {Kiener}, \&
  {Murphy}}]{2006ApJS..165..606T}
V.~{Tatischeff}, B.~{Kozlovsky}, J.~{Kiener}, R.~J. {Murphy}, \apjs {\bf
  165\/}, 606 (2006), \eprint{arXiv:astro-ph/0604325}, doi:10.1086/505112

\bibitem[{{Toner}(2004)}]{tonerphd}
M.~P. {Toner}, {\em {High-energy Neutral Emissions from Solar Flares}\/}, Ph.D.
  thesis, {University of Glasgow} (2004)

\bibitem[{{Toner} \& {MacKinnon}(2004)}]{2004SoPh..223..155T}
M.~P. {Toner}, A.~L. {MacKinnon}, \solphys {\bf 223\/}, 155 (2004),
  doi:10.1007/s11207-004-5168-7

\bibitem[{{Trottet} et~al.(1996){Trottet}, {Barat}, {Ramaty}, {Vilmer},
  {Dezalay}, {Kuznetsov}, {Mandzhavidze}, {Sunyaev}, {Talon}, \&
  {Terekhov}}]{1996AIPC..374..153T}
G.~{Trottet}, C.~{Barat}, R.~{Ramaty}, N.~{Vilmer}, J.~P. {Dezalay},
  A.~{Kuznetsov}, N.~{Mandzhavidze}, R.~{Sunyaev}, R.~{Talon}, O.~{Terekhov},
  in {\em American Institute of Physics Conference Series\/}, ed. by
  {R.~Ramaty, N.~Mandzhavidze, \& X.-M.~Hua} (1996), volume 374 of {\em
  American Institute of Physics Conference Series\/}, pp. 153--161,
  doi:10.1063/1.50951

\bibitem[{{Trottet} et~al.(2008){Trottet}, {Krucker}, {L{\"u}thi}, \&
  {Magun}}]{2008ApJ...678..509T}
G.~{Trottet}, S.~{Krucker}, T.~{L{\"u}thi}, A.~{Magun}, \apj {\bf 678\/}, 509
  (2008), doi:10.1086/528787

\bibitem[{{Trottet} \& {Vilmer}(1997)}]{1997LNP...489..219T}
G.~{Trottet}, N.~{Vilmer}, in {\em European Meeting on Solar Physics\/}, ed. by
  {G.~M.~Simnett, C.~E.~Alissandrakis, \& L.~Vlahos} (1997), volume 489 of {\em
  Lecture Notes in Physics, Berlin Springer Verlag\/}, pp. 219--+

\bibitem[{{Trottet} et~al.(1998){Trottet}, {Vilmer}, {Barat}, {Benz}, {Magun},
  {Kuznetsov}, {Sunyaev}, \& {Terekhov}}]{1998A&A...334.1099T}
G.~{Trottet}, N.~{Vilmer}, C.~{Barat}, A.~{Benz}, A.~{Magun}, A.~{Kuznetsov},
  R.~{Sunyaev}, O.~{Terekhov}, \aap {\bf 334\/}, 1099 (1998)

\bibitem[{{Trottet} et~al.(1993){Trottet}, {Vilmer}, {Barat}, {Dezalay},
  {Talon}, {Siuniaev}, {Kuznetsov}, \& {Terekhov}}]{1993A&AS...97..337T}
G.~{Trottet}, N.~{Vilmer}, C.~{Barat}, J.~P. {Dezalay}, R.~{Talon}, R.~A.
  {Siuniaev}, A.~{Kuznetsov}, O.~{Terekhov}, \aaps {\bf 97\/}, 337 (1993)

\bibitem[{{Tsuchiya} et~al.(2001){Tsuchiya}, {Muraki}, {Masuda}, {Matsubara},
  {Koi}, {Sako}, {Ohno}, {Hoshida}, {Shibata}, {Munakata}, {Hatanaka},
  {Wakasa}, \& {Sakai}}]{2001NIMPA.463..183T}
H.~{Tsuchiya}, Y.~{Muraki}, K.~{Masuda}, Y.~{Matsubara}, T.~{Koi}, T.~{Sako},
  S.~{Ohno}, T.~{Hoshida}, S.~{Shibata}, Y.~{Munakata}, K.~{Hatanaka},
  T.~{Wakasa}, H.~{Sakai}, Nuclear Instruments and Methods in Physics Research
  A {\bf 463\/}, 183 (2001), doi:10.1016/S0168-9002(01)00290-X

\bibitem[{{Usoskin} et~al.(1997){Usoskin}, {Kovaltsov}, {Kananen}, \&
  {Tanskanen}}]{1997AnGeo..15..375U}
I.~G. {Usoskin}, G.~A. {Kovaltsov}, H.~{Kananen}, P.~{Tanskanen}, Annales
  Geophysicae {\bf 15\/}, 375 (1997), doi:10.1007/s005850050451

\bibitem[{{Vald{\'e}s-Galicia} et~al.(2009){Vald{\'e}s-Galicia}, {Muraki},
  {Watanabe}, {Matsubara}, {Sako}, {Gonzalez}, {Musalem}, \&
  {Hurtado}}]{2009AdSpR..43..565V}
J.~F. {Vald{\'e}s-Galicia}, Y.~{Muraki}, K.~{Watanabe}, Y.~{Matsubara},
  T.~{Sako}, L.~X. {Gonzalez}, O.~{Musalem}, A.~{Hurtado}, Advances in Space
  Research {\bf 43\/}, 565 (2009), doi:10.1016/j.asr.2008.09.023

\bibitem[{{Vedrenne} et~al.(2003){Vedrenne}, {Roques}, {Sch{\"o}nfelder},
  {Mandrou}, {Lichti}, {von Kienlin}, {Cordier}, {Schanne}, {Kn{\"o}dlseder},
  {Skinner}, {Jean}, {Sanchez}, {Caraveo}, {Teegarden}, {von Ballmoos},
  {Bouchet}, {Paul}, {Matteson}, {Boggs}, {Wunderer}, {Leleux},
  {Weidenspointner}, {Durouchoux}, {Diehl}, {Strong}, {Cass{\'e}}, {Clair}, \&
  {Andr{\'e}}}]{2003A&A...411L..63V}
G.~{Vedrenne}, J.~{Roques}, V.~{Sch{\"o}nfelder}, P.~{Mandrou}, G.~G. {Lichti},
  A.~{von Kienlin}, B.~{Cordier}, S.~{Schanne}, J.~{Kn{\"o}dlseder},
  G.~{Skinner}, P.~{Jean}, F.~{Sanchez}, P.~{Caraveo}, B.~{Teegarden}, P.~{von
  Ballmoos}, L.~{Bouchet}, P.~{Paul}, J.~{Matteson}, S.~{Boggs}, C.~{Wunderer},
  P.~{Leleux}, G.~{Weidenspointner}, P.~{Durouchoux}, R.~{Diehl}, A.~{Strong},
  M.~{Cass{\'e}}, M.~A. {Clair}, Y.~{Andr{\'e}}, \aap {\bf 411\/}, L63 (2003),
  doi:10.1051/0004-6361:20031482

\bibitem[{{Vernazza} et~al.(1981){Vernazza}, {Avrett}, \&
  {Loeser}}]{1981ApJS...45..635V}
J.~E. {Vernazza}, E.~H. {Avrett}, R.~{Loeser}, \apjs {\bf 45\/}, 635 (1981),
  doi:10.1086/190731

\bibitem[{{Vestrand} \& {Forrest}(1993)}]{1993ApJ...409L..69V}
W.~T. {Vestrand}, D.~J. {Forrest}, \apjl {\bf 409\/}, L69 (1993),
  doi:10.1086/186862

\bibitem[{{Vilmer} et~al.(1982){Vilmer}, {Kane}, \&
  {Trottet}}]{1982A&A...108..306V}
N.~{Vilmer}, S.~R. {Kane}, G.~{Trottet}, \aap {\bf 108\/}, 306 (1982)

\bibitem[{{Vilmer} \& {MacKinnon}(2003)}]{2003LNP...612..127V}
N.~{Vilmer}, A.~L. {MacKinnon}, in {\em Energy Conversion and Particle
  Acceleration in the Solar Corona\/}, ed. by {L.~Klein} (2003), volume 612 of
  {\em Lecture Notes in Physics, Berlin Springer Verlag\/}, pp. 127--160

\bibitem[{{Vilmer} et~al.(2003){Vilmer}, {MacKinnon}, {Trottet}, \&
  {Barat}}]{2003A&A...412..865V}
N.~{Vilmer}, A.~L. {MacKinnon}, G.~{Trottet}, C.~{Barat}, \aap {\bf 412\/}, 865
  (2003), doi:10.1051/0004-6361:20031488

\bibitem[{{Vilmer} et~al.(2001){Vilmer}, {Maksimovic}, {Lin}, \&
  {Trottet}}]{2001ESASP.493..405V}
N.~{Vilmer}, M.~{Maksimovic}, R.~P. {Lin}, G.~{Trottet}, in {\em Solar
  encounter. Proceedings of the First Solar Orbiter Workshop\/}, ed. by
  B.~{Battrick}, H.~{Sawaya-Lacoste}, E.~{Marsch}, V.~{Martinez Pillet},
  B.~{Fleck}, R.~{Marsden} (2001), volume 493 of {\em ESA Special
  Publication\/}, pp. 405--410

\bibitem[{{Vilmer} et~al.(1994){Vilmer}, {Trottet}, {Barat}, {Dezalay},
  {Talon}, {Sunyaev}, {Terekhov}, \& {Kuznetsov}}]{1994LNP...432..197V}
N.~{Vilmer}, G.~{Trottet}, C.~{Barat}, J.~P. {Dezalay}, R.~{Talon},
  R.~{Sunyaev}, O.~{Terekhov}, A.~{Kuznetsov}, in {\em Advances in Solar
  Physics\/}, ed. by {G.~Belvedere, M.~Rodono, \& G.~M.~Simnett} (1994), volume
  432 of {\em Lecture Notes in Physics, Berlin Springer Verlag\/}, pp.
  197--202, doi:10.1007/3-540-58041-7\_219

\bibitem[{{Vilmer} et~al.(1999){Vilmer}, {Trottet}, {Barat}, {Schwartz},
  {Enome}, {Kuznetsov}, {Sunyaev}, \& {Terekhov}}]{1999A&A...342..575V}
N.~{Vilmer}, G.~{Trottet}, C.~{Barat}, R.~A. {Schwartz}, S.~{Enome},
  A.~{Kuznetsov}, R.~{Sunyaev}, O.~{Terekhov}, \aap {\bf 342\/}, 575 (1999)

\bibitem[{{Vogt} \& {H{\'e}noux}(1996)}]{1996SoPh..164..345V}
E.~{Vogt}, J.~C. {H{\'e}noux}, \solphys {\bf 164\/}, 345 (1996),
  doi:10.1007/BF00146646

\bibitem[{{Wang} \& {Ramaty}(1974)}]{1974SoPh...36..129W}
H.~T. {Wang}, R.~{Ramaty}, \solphys {\bf 36\/}, 129 (1974),
  doi:10.1007/BF00151553

\bibitem[{{Watanabe} et~al.(2006){Watanabe}, {Gros}, {Stoker}, {Kudela},
  {Lopate}, {Vald{\'e}s-Galicia}, {Hurtado}, {Musalem}, {Ogasawara},
  {Mizumoto}, {Nakagiri}, {Miyashita}, {Matsubara}, {Sako}, {Muraki}, {Sakai},
  \& {Shibata}}]{2006ApJ...636.1135W}
K.~{Watanabe}, M.~{Gros}, P.~H. {Stoker}, K.~{Kudela}, C.~{Lopate}, J.~F.
  {Vald{\'e}s-Galicia}, A.~{Hurtado}, O.~{Musalem}, R.~{Ogasawara},
  Y.~{Mizumoto}, M.~{Nakagiri}, A.~{Miyashita}, Y.~{Matsubara}, T.~{Sako},
  Y.~{Muraki}, T.~{Sakai}, S.~{Shibata}, \apj {\bf 636\/}, 1135 (2006),
  \eprint{arXiv:astro-ph/0509527}, doi:10.1086/498086

\bibitem[{{Watanabe} et~al.(2003){Watanabe}, {Muraki}, {Matsubara}, {Murakami},
  {Sako}, {Tsuchiya}, {Masuda}, {Yoshimori}, {Ohmori}, {Miranda}, {Martinic},
  {Ticona}, {Velarde}, {Kakimoto}, {Ogio}, {Tsunesada}, {Tokuno}, \&
  {Shirasaki}}]{2003ApJ...592..590W}
K.~{Watanabe}, Y.~{Muraki}, Y.~{Matsubara}, K.~{Murakami}, T.~{Sako},
  H.~{Tsuchiya}, S.~{Masuda}, M.~{Yoshimori}, N.~{Ohmori}, P.~{Miranda},
  N.~{Martinic}, R.~{Ticona}, A.~{Velarde}, F.~{Kakimoto}, S.~{Ogio},
  Y.~{Tsunesada}, H.~{Tokuno}, Y.~{Shirasaki}, \apj {\bf 592\/}, 590 (2003),
  \eprint{arXiv:astro-ph/0304067}, doi:10.1086/375685

\bibitem[{{Watanabe} et~al.(2005){Watanabe}, {Muraki}, {Matsubara}, {Sako},
  {Sakai}, {Tsuchiya}, \& {et al.}}]{2005ICRC....1...37W}
K.~{Watanabe}, Y.~{Muraki}, Y.~{Matsubara}, T.~{Sako}, T.~{Sakai},
  H.~{Tsuchiya}, {et al.}, in {\em International Cosmic Ray Conference\/}
  (2005), volume~1 of {\em International Cosmic Ray Conference\/}, pp. 37--+

\bibitem[{{Werntz} et~al.(1990){Werntz}, {Kim}, \&
  {Lang}}]{1990ApJS...73..349W}
C.~{Werntz}, Y.~E. {Kim}, F.~L. {Lang}, \apjs {\bf 73\/}, 349 (1990),
  doi:10.1086/191471

\bibitem[{{Wiehr}(1978)}]{1978A&A....69..279W}
E.~{Wiehr}, \aap {\bf 69\/}, 279 (1978)

\bibitem[{{Winglee}(1989)}]{1989ApJ...343..511W}
R.~M. {Winglee}, \apj {\bf 343\/}, 511 (1989), doi:10.1086/167726

\bibitem[{{Woodgate} et~al.(1992){Woodgate}, {Robinson}, {Carpenter}, {Maran},
  \& {Shore}}]{1992ApJ...397L..95W}
B.~E. {Woodgate}, R.~D. {Robinson}, K.~G. {Carpenter}, S.~P. {Maran}, S.~N.
  {Shore}, \apjl {\bf 397\/}, L95 (1992), doi:10.1086/186553

\bibitem[{{Woolf} et~al.(2009){Woolf}, {Ryan}, {Bloser}, {Bravar},
  {Fl{\"u}ckiger}, {Legere}, {MacKinnon}, {Mallik}, {McConnell}, \&
  {Pirard}}]{2009SPIE.7438E..19W}
R.~S. {Woolf}, J.~M. {Ryan}, P.~F. {Bloser}, U.~{Bravar}, E.~O.
  {Fl{\"u}ckiger}, J.~S. {Legere}, A.~{MacKinnon}, P.~C. {Mallik}, M.~L.
  {McConnell}, B.~{Pirard}, in {\em Society of Photo-Optical Instrumentation
  Engineers (SPIE) Conference Series\/} (2009), volume 7438 of {\em Society of
  Photo-Optical Instrumentation Engineers (SPIE) Conference Series\/},
  doi:10.1117/12.826425

\bibitem[{{Xu} et~al.(2005){Xu}, {H{\'e}noux}, {Chambe}, {Karlick{\'y}}, \&
  {Fang}}]{2005ApJ...631..618X}
Z.~{Xu}, J.-C. {H{\'e}noux}, G.~{Chambe}, M.~{Karlick{\'y}}, C.~{Fang}, \apj
  {\bf 631\/}, 618 (2005), doi:10.1086/432505

\bibitem[{{Xu} et~al.(2006){Xu}, {Henoux}, {Chambe}, {Petrashen}, \&
  {Fang}}]{2006ApJ...650.1193X}
Z.~{Xu}, J.~C. {Henoux}, G.~{Chambe}, A.~G. {Petrashen}, C.~{Fang}, \apj {\bf
  650\/}, 1193 (2006), doi:10.1086/507477

\bibitem[{{Yoshimori}(1989)}]{1989SSRv...51...85Y}
M.~{Yoshimori}, Space Science Reviews {\bf 51\/}, 85 (1989),
  doi:10.1007/BF00226269

\bibitem[{{Yoshimori}(1999{\natexlab{a}})}]{1999ICRC....6....5Y}
Y.~{Yoshimori}, in {\em International Cosmic Ray Conference\/}
  (1999{\natexlab{a}}), volume~6 of {\em International Cosmic Ray
  Conference\/}, pp. 5--+

\bibitem[{{Yoshimori}(1999{\natexlab{b}})}]{1999ICRC....6....1Y}
Y.~{Yoshimori}, in {\em International Cosmic Ray Conference\/}
  (1999{\natexlab{b}}), volume~6 of {\em International Cosmic Ray
  Conference\/}, pp. 1--+

\bibitem[{{Zharkova} \& {Gordovskyy}(2004)}]{2004ApJ...604..884Z}
V.~V. {Zharkova}, M.~{Gordovskyy}, \apj {\bf 604\/}, 884 (2004),
  doi:10.1086/381966

\bibitem[{{Zweibel} \& {Haber}(1983)}]{1983ApJ...264..648Z}
E.~G. {Zweibel}, D.~A. {Haber}, \apj {\bf 264\/}, 648 (1983),
  doi:10.1086/160638

\end{thebibliography}

\printindex

\end{document}